\documentclass[a4paper,11pt]{article}
\pdfoutput=1 

\usepackage{jcappub} 

\usepackage[T1]{fontenc} 

\usepackage{subfigure}
\usepackage{enumitem}
\usepackage[normalem]{ulem}
\usepackage{graphicx}
\usepackage{upgreek} 
\usepackage{subeqnarray}

\usepackage{cases}
\usepackage{subfigure}
\usepackage{latexsym}
\usepackage{mathrsfs}
\usepackage{color}
\usepackage{dcolumn}
\usepackage{bm}
\usepackage{verbatim} 

\RequirePackage{graphicx}
\definecolor{dyellow}{rgb}{1.,0.8,.0}
\definecolor{myblue}{rgb}{.1,.1,.7}
\definecolor{dcyan}{rgb}{.0,.6,.6}
\definecolor{dmagenta}{rgb}{0.6,0.0,0.6}
\definecolor{brown}{rgb}{0.6,0.2,0.}
\definecolor{darkblue}{rgb}{.0,.0,0.5}
\definecolor{darkred}{rgb}{0.75,0.0,0.0}
\definecolor{orange}{rgb}{1.,.6,.0}
\definecolor{dorange}{rgb}{0.8,.4,.0}
\definecolor{darkgreen}{rgb}{0.0,0.6,0.0}
\definecolor{purple}{rgb}{.4,.0,.4}
\definecolor{grey}{rgb}{0.5,0.5,0.5}

\title{\boldmath Multipole expansion of the gravitational field in a general class of fourth-order theories of gravity and the application in gyroscopic precession}

\author{Bofeng Wu
}
\author{and En-Wei Liang
}

\affiliation{Guangxi Key Laboratory for Relativistic Astrophysics, School of Physical Science and Technology, Guangxi University, Nanning 530004, China}

\abstract{A viable weak-field and slow-motion approximation method is constructed in
$F(R,R_{\mu\nu}R^{\mu\nu}, R_{\mu\nu\rho\sigma}R^{\mu\nu\rho\sigma})$ gravity, a general class of fourth-order theories of gravity. By applying this method,  the metric, presented in the form of the multipole expansion, outside a spatially compact source up to $1/c^3$ order is provided, and the closed-form expressions for the source multipole moments are all presented explicitly. The metric consists of the massless tensor part, the massive scalar part, and the massive tensor part, where the former is exactly the metric in General Relativity, and the latter two are the corrections to it. It is shown that the corrections bear the Yukawa-like dependence on the two massive parameters and predict the appearance of six additional sets of source multipole moments, which indicates that up to $1/c^3$ order, there exist six degrees of freedom beyond General Relativity within $F(R,R_{\mu\nu}R^{\mu\nu}, R_{\mu\nu\rho\sigma}R^{\mu\nu\rho\sigma})$ gravity. By means of the metric, for a gyroscope moving around the source without experiencing any torque, the multipole expansions of its spin's angular velocities of the Thomas precession, the geodetic precession, and the Lense-Thirring precession  are derived, and from them, the
corrections to the angular velocities of the three types of precession in General Relativity can be read off. These results indicate that differently from $f(R)$ or $f(R,\mathcal{G})$ gravity, the most salient feature of the general $F(R,R_{\mu\nu}R^{\mu\nu}, R_{\mu\nu\rho\sigma}R^{\mu\nu\rho\sigma})$ gravity is that it gives the nonvanishing correction to the gyroscopic spin's angular velocity of the Lense-Thirring precession in General Relativity.}

\begin{document}
\maketitle
\flushbottom
\section{Introduction}
\label{sec:first}
General Relativity (GR) is a great theory of gravity~\cite{Clifford2018,eric2018,TheLIGOScientific:2016agk1,TheLIGOScientific:2016agk2,TheLIGOScientific:2016agk3,
TheLIGOScientific:2016agk4,TheLIGOScientific:2016agk5,TheLIGOScientific:2016agk6,TheLIGOScientific:2016agk7,TheLIGOScientific:2016agk8}, but in order to give a physical explanation to many observed data at astrophysical and cosmic scales~\cite{cosmicacceleration1,cosmicacceleration2,cosmicacceleration3,cosmicacceleration4,cosmicacceleration5,cosmicacceleration6,
cosmicacceleration7},  the concepts of dark matter and dark energy need to be introduced~\cite{Naf:2010zy}. As a comparison, these data can also be interpreted within the framework of many alternative theories of gravity by extending the geometric sector of field equations~\cite{Sotiriou2010,Gibbons1977,Starobinsky:1980te,Nojiri:2010wj,DNojiri:2017ncd,Capozziello:2011et,Odintsov:2023weg}. In this paper, our attention is focused on the fourth-order theories of gravity (FOTGs)~\cite{Clifton:2011jh}.
$F(X,Y,Z)$ gravity, being a general class of FOTGs, is constructed by replacing the Einstein-Hilbert action by a general function $F$ of curvature invariants $X,Y,$ and $Z$ in the gravitational
Lagrangian~\cite{Stabile:2010mz,stabile2015,Bogdanos:2009tn}, where $X:=R$, $Y:=R_{\mu\nu}R^{\mu\nu}$, and
$Z:=R_{\mu\nu\rho\sigma}R^{\mu\nu\rho\sigma}$ with $R_{\mu\nu}$ Ricci tensor and $R_{\mu\nu\rho\sigma}$ Riemann tensor.  Numerous FOTGs, such as GR, Starobinsky gravity~\cite{Starobinsky:1980te,Alexandre2014}, $f(R)$ gravity~\cite{Gibbons1977,Starobinsky:1980te,Nojiri:2010wj,DNojiri:2017ncd}, and $f(R,\mathcal{G})$ gravity~\cite{Cognola:2006eg,Alimohammadi:2008fq,Bamba:2009uf,Wu:2015maa,Shamir:2017ndy,Odintsov:2018nch} ($\mathcal{G}$ is the Gauss-Bonnet invariant), etc.,
are sub-models of $F(X,Y,Z)$ gravity, so in essence, it could be regarded as a general theoretical framework for FOTGs.

When it comes to FOTGs, it is known that they are plagued with ghost degrees of freedom~\cite{Clifton:2011jh,Stelle:1977ry}, which
give rise to physical states of negative energy or negative norm upon quantization due to the presence of fourth-order derivatives.
According to refs.~\cite{Johnson2022,Tomboulis:1983sw,Antoniadis:1986tu},  it may be misleading that the general quadratic gravity, equivalent to $F(X,Y,Z)$ gravity in the present paper (cf.~eq.~(\ref{equ3.4})), has traditionally been rejected because the fourth-order derivatives lead to ghosts in the perturbation series about the linearized theory.
It is shown that the unitarity picture presented by the linearized theory is simply too naive, and is substantially modified by nonperturbative effects~\cite{Tomboulis:1983sw}. In addition, some attempts to avoid the ghost instabilities by reducing the phase space are described in refs.~\cite{Chen:2012au,Chen:2013aha,Akita:2015dda,Morales:2018imi}. To say the least, as shown in refs.~\cite{DeFelice:2006pg,Moldenhauer:2009kc,Biswas:2011ar,Berti:2015itd}, even though the ghosts are malicious, one can also construct the ghost-free models by means of some particular methods. In view of the above, $F(X,Y,Z)$ gravity (or some of its sub-models) could serve as a potential candidate for quantum gravity theories, which implies that this model is worthy of being examined~\cite{Capozziello:2011et}.

It is very meaningful to deduce the metric for the gravitational field of a source in $F(X,Y,Z)$ gravity, because   $F(X,Y,Z)$ gravity is a comprehensive model of FOTGs. The gravitational field equations of $F(X,Y,Z)$ gravity are exceedingly complicated, and they are still hard to solve even after simplified by adopting the weak-field and slow-motion (WFSM) approximation. Under the WFSM approximation, the expression of the linearized Ricci scalar $X^{(1)}$ up to $1/c^2$ order is given in refs.~\cite{stabile2015,Capozziello:2009ss,Stabile:2010mz},
\begin{eqnarray}\label{equ1.1}
X^{(1)}(t,\boldsymbol{x})=\frac{1}{4\pi}\int\frac{\text{e}^{-m_{1}|\boldsymbol{x}-\boldsymbol{x}'|}}{|\boldsymbol{x}-\boldsymbol{x}'|}
\big(-m_{1}^2\kappa T(t,\boldsymbol{x}')\big)\text{d}^{3}x',
\end{eqnarray}
where $1/c$ is the WFSM parameter~\cite{Blanchet:2013haa} with $c$ as the speed of light in vacuum, $m_{1}$ is a massive parameter in $F(X,Y,Z)$ gravity, $\kappa=8\pi G/c^{4}$ with $G$ as the gravitational constant, and $T(t,\boldsymbol{x}')$ is the trace of the energy-momentum tensor of a source living in Minkowski spacetime. With $X^{(1)}$, the metric up to $1/c^2$ order for the source is given by imposing the harmonic gauge condition in refs.~\cite{stabile2015,Capozziello:2009ss,Stabile:2010mz}. In the expressions of the components of the metric, infinite integrals like
\begin{eqnarray}\label{equ1.2}
\int\text{d}^{3}x'\text{d}^{3}x''\frac{\text{e}^{-m_{2}|\boldsymbol{x}-\boldsymbol{x}'|}}{|\boldsymbol{x}-\boldsymbol{x}'|^{2}}X^{(1)}(t,\boldsymbol{x}'')\quad \text{or}\quad
\int\text{d}^{3}x'\text{d}^{3}x''\frac{\text{e}^{-m_{2}|\boldsymbol{x}-\boldsymbol{x}'|}}{|\boldsymbol{x}-\boldsymbol{x}'|^{2}}\frac{\partial^{2}}{\partial x''_{i}x''_{j}}X^{(1)}(t,\boldsymbol{x}'')\quad
\end{eqnarray}
are always involved, where $m_{2}$ is another massive parameter in $F(X,Y,Z)$ gravity.  $X^{(1)}(t,\boldsymbol{x}'')$ is non-zero in the whole space in general, which means that even for a spatially compact source the above two complicated integrals need to be performed over the whole space, and as a consequence, it is almost impossible to gain the analytic expression of the metric by performing them.

\subsection{How to derive the external metric for a spatially compact source in $F(X,Y,Z)$ gravity?}

In the harmonic gauge, the linearized gravitational field equations  are not fully simplified, so under this case, if one directly uses the Green's function method to find their solutions, the infinite integrals like those in eqs.~(\ref{equ1.2}) will appear in the metric. The key to overcoming this difficulty is to seek appropriate gauge condition.  In ref.~\cite{Wu:2022mna}, by imposing a new type of gauge condition, a weak-field approximation framework of $F(X,Y,Z)$ gravity is developed. As one of the tasks of the present paper, a viable WFSM approximation method is constructed in $F(X,Y,Z)$ gravity based on this framework by further adopting the slow-motion approximation.
After applying this method, the metric is decomposed into three separated parts relevant to a massless tensor field $\tilde{h} ^{\mu\nu}$, a massive scalar field $X^{(1)}$, and a massive tensor field $P^{\mu\nu}$, respectively, and the complicated linearized gravitational field equations of $F(X,Y,Z)$ gravity up to $1/c^{3}$ order are converted to an easy-to-handle system of equations consisting of a Poisson equation satisfied by $\tilde{h} ^{\mu\nu}$
and two screened Poisson equations satisfied by $X^{(1)}$ and  $P^{\mu\nu}$. These equations can be resolved by use of the Green's function method, and the metric up to $1/c^3$ order for a source is then acquired in this way. However, the metric obtained in such manner is still inconvenient to be employed in practical application because the two tensor fields $\tilde{h} ^{\mu\nu}$ and $P^{\mu\nu}$ also fulfill some additional conditions, and these conditions need
to be considered separately.

The main task of this paper is to derive the external metric for a spatially compact source in $F(X,Y,Z)$ gravity by following the above method, but how to solve the problem mentioned above is the key point. Starting from the three equations fulfilled by $\tilde{h} ^{\mu\nu}$, $X^{(1)}$, and $P^{\mu\nu}$, by applying the symmetric and trace-free (STF) formalism in terms of the irreducible Cartesian tensors~\cite{Damour:1990gj,Thorne:1980ru,Blanchet:1985sp,Blanchet:1989ki},
these fields are expressed in the form of the STF-tensor spherical harmonics expansions outside the source. Those additional conditions fulfilled by the tensor fields $\tilde{h}^{\mu\nu}$ and $P^{\mu\nu}$ can be used to eliminate the redundant multipole tensor
coefficients in their expansions. By such approach, the multipole expansions of $\tilde{h} ^{\mu\nu}$, $X^{(1)}$, and $P^{\mu\nu}$ up to $1/c^3$ order are achieved. With these results, the multipole expansion of the metric outside the source up to $1/c^3$ order is also derived, and the closed-form expressions for the source multipole moments are all presented explicitly. It is shown that all the integrals involved in the multipole moments are carried out only over the actual source distribution, and therefore, compared with previous expressions of the metric given in refs.~\cite{stabile2015,Capozziello:2009ss,Stabile:2010mz}, our result is not plagued by the infinite integrals over the whole space like those in eqs.~(\ref{equ1.2}) and thus yields a ready-to-use form of the metric.

In the expansion of the metric, the three parts relevant to $\tilde{h} ^{\mu\nu}$, $X^{(1)}$, and $P^{\mu\nu}$ could be referred to as the massless tensor part, the massive scalar part, and the massive tensor part, respectively. When $F(X,Y,Z)=R$, the massless tensor part is exactly the result in GR,
and it shows the Coulomb-like dependence and is characterized by the mass and spin multipole moments. The Lense-Thirring metric in the isotropic coordinate can easily be found from the massless tensor part at the leading pole order in the stationary spacetime. When $F(X,Y,Z)\rightarrow f(R)$ or $f(R,\mathcal{G})$, it is proved that the massless tensor part plus the massive scalar part recovers the metric in $f(R)$ or $f(R,\mathcal{G})$ gravity~\cite{Wu:2021uws}, so in either of these two models, the massive scalar part actually provides the correction to the metric in GR, where one needs to note that the Gauss-Bonnet scalar $\mathcal{G}$ has no contribution to the gravitational field dynamics because it is a topological invariant. Thus, in $F(X,Y,Z)$ gravity, the massive tensor part should be identified as the further correction to the metric in GR. The expansions of the massive scalar and tensor parts display that they bear the Yukawa-like dependence on the two massive parameters and predict the appearance of six additional sets of source multipole moments. Five of these additional sets of multipole moments come from the massive tensor part, and among them besides the counterparts of the mass and spin multipole moments, the remaining three, unlike the situation encountered in the massless tensor part, can not be transformed away by the gauge symmetry.

\subsection{How to study gyroscopic precession in $F(X,Y,Z)$ gravity?}

The previous discussions indicate that up to $1/c^3$ order, there exist a total of six degrees of freedom beyond GR in $F(X,Y,Z)$ gravity, whereas in $f(R)$ or $f(R,\mathcal{G})$ gravity the number of such degrees of freedom is only one.
In order to study the effects of these new degrees of freedom,  the third task of this paper is to make use of the metric to deal with gyroscopic precession. According to refs.~\cite{Wu:2021uws,MTW1973,Dass:2019kon,Dass:2019hnb}, the metric up to $1/c^3$ order under the WFSM approximation is sufficient to be used to analyze the gyroscopic precession, and the condition ``up to $1/c^3$ order'' can greatly simplify the related derivations. The basic method to analyze gyroscopic precession in the stationary spacetime is presented in ref.~\cite{MTW1973}, but since the metric obtained in this paper for the external gravitational field of the source is time-dependent in general, this basic method needs to be extended. In addition, based on the method in ref.~\cite{MTW1973}, a local orthonormal tetrad at rest in the coordinate frame needs to be constructed in the derivation, and its three spatial vectors should be on an equal footing. In $F(X,Y,Z)$ gravity, the construction of such a tetrad  is a bit subtle because the off-diagonal space-space components of the metric are nonvanishing. In this paper, after solving these difficulties, for a gyroscope moving around the source without experiencing any torque, the multipole expansion of the precessional angular velocity of its spin up to $1/c^3$ order is derived.

From the result, the gyroscopic spin's angular velocities of the Thomas precession, the geodetic precession, and the Lense-Thirring precession in $F(X,Y,Z)$ gravity are provided. In their multipole expansions, those terms associated with the massless tensor part of the metric are the corresponding angular velocities of the gyroscopic spin in GR~\cite{Wu:2021uws}, and the remaining terms yield the corrections to them. These corrections stem from the massive scalar and tensor parts of the metric and are characterized by the additional multipole moments, so from them, the effects of the degrees of freedom beyond GR appearing in $F(X,Y,Z)$ gravity can be read off. The multipole expansions of these corrections are a bit lengthy, but since one only needs to consider their leading and next-leading pole terms in the stationary spacetime in general applications
the expressions of these corrections can be reduced to a sufficiently simple form, which greatly helps to make an analysis on those degrees of freedom beyond GR. The corrections in $f(R)$ gravity~\cite{Wu:2021uws} or $f(R,\mathcal{G})$ gravity are recovered when $F(X,Y,Z)\rightarrow f(R)$ or $f(R,\mathcal{G})$, which shows that these two models do not give the correction to the gyroscopic spin's angular velocity of the Lense-Thirring precession in GR. However, in the general $F(X,Y,Z)$ gravity, due to the presence of the massive tensor part of the metric, this angular velocity in GR is corrected. Moreover, for the Lense-Thirring precession, one interesting fact is that
the angular velocity of the gyroscopic spin is independent of the nonvanishing time-space components of the metric at the monopole order, which results in that its monopole term disappears.

The multipole expansions of the gyroscopic spin's angular velocities of the three types of precession describe all the effects of the external gravitational field of the source up to $1/c^3$ order on the gyroscopic spin. In order to apply these results to gyroscope experiments, e.g., Gravity Probe B (GP-B), the Thomas-monopole term, the geodetic-monopole term, and the Lense-Thirring-dipole term in the stationary spacetime are gained, and from them, the most main corrections to the classical angular velocities of the three types of precession in GR by the new degrees of freedom in $F(X,Y,Z)$ gravity can be read off. As stated above, differently from $f(R)$ gravity, the gyroscopic spin's angular velocity of the Lense-Thirring precession in GR is corrected by the massive tensor part of the metric in the general $F(X,Y,Z)$ gravity, and therefore, if one employs the Lense-Thirring-dipole term to explain the corresponding data in gyroscope experiment GP-B, the constraint on the massive parameter $m_{2}$ relevant to the massive tensor part will be obtained. In this manner, one can identify whether the degrees of freedom beyond $f(R)$ gravity exist or not in principle.  After the massive parameter $m_{2}$ is determined, by further comparing the geodetic-monopole term with the measurement in gyroscope experiment GP-B, the constraint on the massive parameter $m_{1}$ relevant to the massive scalar part will also be obtained. If both the massive parameters can be fixed with the help of the gyroscope experiment GP-B, the effects of the degrees of freedom beyond GR in $F(X,Y,Z)$ gravity are clear in gyroscopic precession. Since the two massive parameters are dependent on the related coefficients appearing in the Taylor expansion of $F(X,Y,Z)$ at $X=Y=Z=0$
in the Lagrangian density, once the constraints on the two massive parameters are given the constraints on these coefficients could also be acquired.  Although the leading-pole terms of the angular velocities for the three types of precession are the most significant, if one intends to probe the influence of the scale and shape of the source on gyroscopic spin those terms at the next-leading and higher pole order need to be taken into account. In those circumstances, more effects brought about by the degrees of freedom beyond GR will emerge.

\subsection{The purpose of this paper}

The main purpose of this paper is to provide the metric, presented in the form of the multipole expansion, outside
a spatially compact source up to $1/c^{3}$ order in $F(X,Y,Z)$ gravity,  and the derivations of the precessional angular velocities of the gyroscopic spin could be treated as a typical application example of the metric. The first advantage of the metric is that all the integrals involved in the source multipole moments only need to be performed over the actual source distributions, so the metric can be conveniently employed to explore astrophysical phenomena happening  in the gravitational field outside a \emph{realistic} source. The second advantage of the metric is that it is expressed in a structurally transparent way, which enables us to easily recognize the GR-like metric and the massive scalar and tensor corrections to it in $F(X,Y,Z)$ gravity. Thus, while the metric is utilized to study some phenomenon, the corresponding results can also be presented in such way, which will greatly facilitate the analysis on the effects of the new degrees of freedom appearing in $F(X,Y,Z)$ gravity. Besides gyroscopic precession, one is also able to make use of the metric in this paper to explore other astrophysical phenomena like the gravitational redshift of light and the light bending, and it is expected that more effects beyond GR will be predicted by those new degrees of freedom in $F(X,Y,Z)$ gravity.

This paper is organized as follows. In Sec.~\ref{Sec:second}, notation and related formulas in the STF formalism are briefly reviewed. In Sec.~\ref{Sec:third}, a weak-field and slow-motion approximation method is constructed within $F(X,Y,Z)$ gravity, and by applying this method, the multipole expansion of the metric for the external gravitational field of a spatially compact source up to $1/c^3$ order is provided in Sec.~\ref{Sec:fourth}. In Sec.~\ref{Sec:fifth}, by applying the metric, for a gyroscope moving around the source without experiencing any torque,
the multipole expansion of  the precessional angular velocity of its spin is derived. In Sec.~\ref{Sec:sixth}, the summary and the related discussions are presented.  In the Appendices, we provide the detailed derivations for eqs.~(\ref{equ4.67})--(\ref{equ4.77}) and (\ref{equ5.39}).
\section{Notation and formulas in the STF formalism~\label{Sec:second}}
Throughout this paper, the international system of units is used.  Greek indices $\alpha,\beta,\gamma,\cdots$ run from 0 to 3 and mark spacetime indices. Latin indices $i,j,k,\cdots$ run from 1 to 3 and denote space indices. The repeated indices within a term represent that the sum should be taken over.

When the discussion is focused on the linearized gravitational theory, the coordinates
$(x^{\mu})=(ct,x_{i})$ behave as if they were Minkowskian coordinates~\cite{Wu:2017huang}. With the coordinates $t,x_{i}$, we also use $r=(x_{i}x_{i})^{1/2}$, $n_{i}=x_{i}/r$, $\partial_{t}=\partial/\partial t$, $\partial_{i}=\partial/\partial x_{i}$, $\boldsymbol{x}=x_{i}\partial_{i}$, $\Delta=\partial_{a}\partial_{a}$,
\begin{eqnarray}
\label{equ2.1}X_{I_{l}}&=&X_{i_{1}i_{2}\cdots i_{l}}:= x_{i_{1}}x_{i_{2}}\cdots x_{i_{l}},\\
\label{equ2.2}N_{I_{l}}&=&N_{i_{1}i_{2}\cdots i_{l}}:= n_{i_{1}}n_{i_{2}}\cdots n_{i_{l}},
\end{eqnarray}
\begin{eqnarray}
\label{equ2.3}\partial_{I_{l}}&=&\partial_{i_{1}i_{2}\cdots i_{l}}:=\partial_{i_{1}}\partial_{i_{2}}\cdots\partial_{i_{l}},
\end{eqnarray}
where spatial indices,  $i=1,2,3$, are freely raised or lowered by means of the Cartesian metric
$\delta_{ij}=\delta^{ij}=\text{diag}(+1,+1,+1)$.
The totally antisymmetric Levi-Civita tensor is represented by $\epsilon_{ijk}$ with $\epsilon_{123}=1$.

For a Cartesian tensor $B_{I_{l}}=B_{i_{1}i_{2}\cdots i_{l}}$~\cite{Thorne:1980ru},
its STF part is denoted by
\begin{eqnarray}
\label{equ2.4}\hat{B}_{I_{l}}:=B_{\langle I_{l}\rangle}=B_{\langle i_{1}i_{2}\cdots i_{l}\rangle}
:=\sum_{k=0}^{\left[\frac{l}{2}\right]}b_{k}\delta_{(i_{1}i_{2}}\cdots\delta_{i_{2k-1}i_{2k}}S_{i_{2k+1}\cdots i_{l})a_{1}a_{1}\cdots a_{k}a_{k}},
\end{eqnarray}
where $\left[l/2\right]$ denotes the integer part of $l/2$,
\begin{equation}
\label{equ2.5}b_{k}:=(-1)^{k}\frac{(2l-2k-1)!!}{(2l-1)!!}\frac{l!}{(2k)!!(l-2k)!},
\end{equation}
and
\begin{equation}
\label{equ2.6}S_{I_{l}}:=B_{(I_{l})}=B_{(i_{1}i_{2}\cdots i_{l})}:=\frac{1}{l!}\sum_{\sigma} B_{i_{\sigma(1)}i_{\sigma(2)}\cdots i_{\sigma(l)}}
\end{equation}
is its symmetric part with $\sigma$ running over all permutations of $(12\cdots l)$.

In the STF formalism, a Cartesian tensor $T_{I_{l}}$ can be decomposed into a finite sum of terms of the type $\gamma_{I_{l}J_{k}}\hat{R}_{J_{k}}$, where $\gamma_{I_{l}J_{k}}$ is a tensor invariant under the group of proper rotations $\text{SO(3)}$, and $\hat{R}_{J_{k}}$ is an irreducible STF
tensor~\cite{Thorne:1980ru,Blanchet:1985sp, Damour:1990gj}. The simplest case ($l=1$) reads
\begin{eqnarray}\label{equ2.7}
A_{i}\hat{T}_{I_{l}}&=&\hat{R}^{(+)}_{iI_{l}}+\frac{l}{l+1}\epsilon_{ai\langle i_{l}}\hat{R}^{(0)}_{i_{1}\cdots i_{l-1}\rangle a}+\frac{2l-1}{2l+1}\delta_{i\langle i_{l}}\hat{R}^{(-)}_{i_{1}\cdots i_{l-1}\rangle},
\end{eqnarray}
where
\begin{eqnarray}
\label{equ2.8}&&\hat{R}^{(+)}_{I_{l+1}}:=A_{\langle i_{l+1}}\hat{T}_{i_{1}\cdots i_{l}\rangle},\\
\label{equ2.9}&&\hat{R}^{(0)}_{I_{l}}:=A_{a}\hat{T}_{b\langle i_{1}\cdots i_{l-1}}\epsilon_{i_{l}\rangle ab},\\
\label{equ2.10}&&\hat{R}^{(-)}_{I_{l-1}}:=A_{a}\hat{T}_{ai_{1}\cdots i_{l-1}}.
\end{eqnarray}
For a general tensor $A_{J_{k}}$ with $k>1$, the decomposition can be implemented by repeated use of eqs.~(\ref{equ2.7})--(\ref{equ2.10}). Other related formulas of direct use in this paper are
\begin{eqnarray}
\label{equ2.11}&&n_{i}\hat{N}_{I_{l}}=\hat{N}_{iI_{l}}+\frac{l}{2l+1}\delta_{i\langle i_{l}}\hat{N}_{i_{1}\cdots i_{l-1}\rangle},\\
\label{equ2.12}&&\int\hat{N}_{I_{l}}\text{d}\varOmega=0,\quad l\geqslant1,\\
\label{equ2.13}&&\int\Big(\hat{A}_{I_{l}}\hat{N}_{I_{l}}\Big)\Big(\hat{B}_{J_{l'}}\hat{N}_{J_{l'}}\Big)\text{d}\varOmega=\frac{ 4\pi l!}{(2l+1)!!}\hat{A}_{I_{l}}\hat{B}_{I_{l}}\delta_{l\,l'},\\
\label{equ2.14}&&\hat{\partial}_{I_{l}}\left(\frac{F(r)}{r}\right)=\hat{N}_{I_{l}}\sum_{k=0}^{l}\frac{(l+k)!}{(-2)^{k}k!(l-k)!}
\frac{\partial_{r}^{l-k}F(r)}{r^{k+1}},
\end{eqnarray}
where $\varOmega$ is the solid angle about the radial vector, $\hat{A}_{I_{l}}$ and $\hat{B}_{J_{l'}}$ are any two STF tensors,  and $\partial_r^{l-k}$ is the $(l-k)$-th derivative with respect to $r$.
\section{The weak-field and slow-motion approximation of $F(X,Y,Z)$ gravity~\label{Sec:third}}
The spacetime with metric $g_{\mu\nu}$ is considered in the $(-,+,+,+)$ signature.
The action of $F(X,Y,Z)$ gravity~\cite{Stabile:2010mz,stabile2015,Bogdanos:2009tn} is
\begin{equation}
\label{equ3.1}S=\frac{1}{2\kappa c}\int\text{d}^4x\sqrt{-g}F(X,Y,Z)+S_{M}(g^{\mu\nu},\psi),
\end{equation}
where $g$ is the determinant of metric $g_{\mu\nu}$, and $S_{M}(g^{\mu\nu},\psi)$ is the matter action. Varying this action with respect to metric $g^{\mu\nu}$ yields the gravitational field equations
\begin{equation}
\label{equ3.2}H_{\mu\nu}=\kappa T_{\mu\nu}
\end{equation}
with
\begin{eqnarray}
\label{equ3.3}H_{\mu\nu}&=&-\frac{1}{2}g_{\mu\nu}F+(R_{\mu\nu}+g_{\mu\nu}\square -\nabla_{\mu}\nabla_{\nu}) F_{X}-2\nabla^{\lambda}\nabla_{(\mu}(F_{Y}R_{\nu)\lambda})+g_{\mu\nu}\nabla^{\alpha}\nabla^{\beta}(F_{Y}R_{\alpha\beta})\notag\\
&&+\square(F_{Y}R_{\mu\nu})+
4\nabla^{(\rho}\nabla^{\sigma)}(F_{Z}R_{\mu\rho\nu\sigma})+2F_{Y}R_{\mu\alpha}R^{\alpha}_{\phantom{\alpha}\nu}+2F_{Z}R_{\mu}^{\phantom{\mu}\alpha\beta\gamma}R_{\nu\alpha\beta\gamma},
\end{eqnarray}
where $F_{X}:=\partial F/\partial{X}$, $F_{Y}:=\partial F/\partial{Y}$, $F_{Z}:=\partial F/\partial{Z}$, and $T_{\mu\nu}$
is the energy-momentum tensor of matter fields.
$F(X,Y,Z)$ is assumed to be analytic at $X=0$, $Y=0$, and $Z=0$, and thus, it can be expressed as the polynomial form
\begin{eqnarray}\label{equ3.4}
F(X,Y,Z)&=&X+F_{2}Y+F_{3}Z+\frac{1}{2}(F_{11}X^2+F_{22}Y^2+F_{33}Z^2+2F_{12}XY+2F_{13}XZ\notag\\
&&+2F_{23}YZ)+\cdots,
\end{eqnarray}
where since we adopt an expansion about a flat background spacetime (cf.~(\ref{equ3.22})), a possible cosmological constant is ignored, and to link the Lagrangian~(\ref{equ3.4}) to GR, the coefficient of $X$ is set to be $1$.

In ref.~\cite{Wu:2022mna}, we develop a weak-field approximation framework of $F(X,Y,Z)$ gravity, and within this framework, the gravitational field equations of $F(X,Y,Z)$ gravity are simplified greatly. The following is a brief review. In $F(X,Y,Z)$ gravity, inspired by the Landau-Lifshitz formulation of GR, the gravitational field amplitude $h^{\mu\nu}:=\sqrt{-g}g^{\mu\nu}-\eta^{\mu\nu}$ is introduced with $\eta^{\mu\nu}$ as the Minkowskian metric in a fictitious flat spacetime. If  $h^{\mu\nu}$ is a perturbation, the linearized gravitational field equations of $F(X, Y, Z)$ gravity are obtained from eqs.~(\ref{equ3.2}) and (\ref{equ3.3}),
\begin{eqnarray}
\label{equ3.5}H^{\mu\nu(1)}&=&R^{\mu\nu(1)}+\big(F_{2}+4F_{3}\big)\square_{\eta}R^{\mu\nu(1)}-\frac{1}{2}\eta^{\mu\nu}X^{(1)}+\left(F_{11}+\frac{F_{2}}{2}\right)\eta^{\mu\nu}\square_{\eta}X^{(1)}\notag\\
&&-\big(F_{11}+F_{2}+2F_{3}\big)\partial^{\mu}\partial^{\nu}X^{(1)}=\kappa T^{\mu\nu},
\end{eqnarray}
where $\partial^{\rho}:=\eta^{\rho\sigma}\partial_{\sigma}=\eta^{\rho\sigma}\partial/\partial x^{\sigma}$, $\square_{\eta}:=\eta^{\mu\nu}\partial_{\mu}\partial_{\nu}$, and the superscript (1) represents that the linear part of the corresponding quantity is taken.
The key point of the framework is how to simplify these complicated equations  under gauge condition. It is shown that in the harmonic gauge condition $\partial_{\mu}h^{\mu\nu}=0$, eqs.~(\ref{equ3.5}) are not fully simplified so that in the expressions of the obtained solutions to these equations, the infinite integrals like those in eqs.~(1.2) appear~\cite{stabile2015}. In order to avoid such infinite integrals, we need to find a new type of gauge condition to rehandle eqs.~(\ref{equ3.5}).
In fact, even for $f(R)$ gravity, the above issue also exists. In refs.~\cite{Berry:2011pb,Naf:2011za}, by sorting to the gauge condition
\begin{eqnarray}
\label{equ3.5.1}\partial_{\mu}\tilde{h}_{\text{f}}^{\mu\nu}=0\quad \text{with}\quad \tilde{h}_{\text{f}}^{\mu\nu}:=h_{\text{f}}^{\mu\nu}+f_{11} \eta^{\mu\nu}R^{(1)},
\end{eqnarray}
the linearized gravitational field equations of $f(R)$ gravity are simplified to be
\begin{eqnarray}
\label{equ3.5.2}&&\square_{\eta}\tilde{h}_{\text{f}}^{\mu\nu}=2\kappa T^{\mu\nu},\\
\label{equ3.5.3}&&\square_{\eta}R^{(1)}-\frac{1}{3f_{11}}R^{(1)}=\frac{1}{3f_{11}}\kappa T,
\end{eqnarray}
where $h_{\text{f}}^{\mu\nu}$ is the gravitational field amplitude in $f(R)$ gravity, $f_{11}/2$ is the coefficient of $R^{2}$ in the Lagrangian of $f(R)$ gravity, $T=g^{\mu\nu}T_{\mu\nu}$ is the trace of the energy-momentum tensor $T_{\mu\nu}$. It is easy to recognize that both of the above two equations can be
solved without introducing the infinite integrals like those in eqs.~(1.2).
Motivated by this paradigm, we successfully found the similar gauge condition in the weak-field approximation framework of $F(X,Y,Z)$ gravity by applying the method of undetermined coefficients, and readers who are interested in the process could consult our preceding paper~\cite{Wu:2022mna}.
In $F(X,Y,Z)$ gravity, the required gauge condition is
\begin{eqnarray}
\label{equ3.6}\partial_{\mu}\tilde{h}^{\mu\nu}=0
\end{eqnarray}
with
\begin{eqnarray}
\label{equ3.7}\tilde{h}^{\mu\nu}:&=&h^{\mu\nu}+(F_{11}-2F_{3})\eta^{\mu\nu}X^{(1)}+(2F_{2}+8F_{3})R^{\mu\nu(1)},
\end{eqnarray}
and by imposing it,  the linearized gravitational field equations of $F(X, Y, Z)$ gravity reduce to
\begin{eqnarray}
\label{equ3.8}&&\square_{\eta}\tilde{h}^{\mu\nu}=2\kappa T^{\mu\nu},\\
\label{equ3.9}&&\square_{\eta}X^{(1)}-m_{1}^2 X^{(1)}=m_{1}^2\kappa T,\\
\label{equ3.10}&&\square_{\eta}P^{\mu\nu}-m_{2}^2 P^{\mu\nu}=-m_{2}^2\kappa S^{\mu\nu}
\end{eqnarray}
with
\begin{eqnarray}
\label{equ3.11}T&:=&\eta_{\mu\nu}T^{\mu\nu},\\
\label{equ3.12}P^{\mu\nu}&:=&R^{\mu\nu(1)}-\frac{1}{6}\eta^{\mu\nu}X^{(1)}-\frac{1}{3m_{1}^2}\partial^{\mu}\partial^{\nu}X^{(1)},\\
\label{equ3.13}S^{\mu\nu}&:=&T^{\mu\nu}-\frac{1}{3}\eta^{\mu\nu}T+\frac{1}{3m_{2}^2}\partial^{\mu}\partial^{\nu}T,
\end{eqnarray}
where both parameters $m_{1}$ and $m_{2}$ are defined by
\begin{eqnarray}
\label{equ3.14}m_{1}^2&:=&\frac{1}{3F_{11}+2F_{2}+2F_{3}},\\
\label{equ3.15}m_{2}^2&:=&-\frac{1}{F_{2}+4F_{3}}.
\end{eqnarray}
Explicitly, by means of the gauge condition~(\ref{equ3.6}),  the complicated linearized
gravitational field equations~(\ref{equ3.5}) are converted to an easy-to-handle system of equations consisting of eqs.~(\ref{equ3.8})--(\ref{equ3.10}).  Once a physically meaningful solution to the system of equations~(\ref{equ3.8})--(\ref{equ3.10}) for a source is given, from eqs.~(\ref{equ3.7}) and (\ref{equ3.12}), the gravitational field amplitude $h^{\mu\nu}$ can directly be derived by
\begin{eqnarray}
\label{equ3.16}h^{\mu\nu}&=&\tilde{h}^{\mu\nu}+\frac{2}{3m_{1}^2m_{2}^2}\partial^{\mu}\partial^{\nu}X^{(1)}-\frac{1}{3}\left(\frac{1}{m_{1}^2}+\frac{1}{m_{2}^2}\right)\eta^{\mu\nu}X^{(1)}+\frac{2}{m_{2}^2}P^{\mu\nu},
\end{eqnarray}
and then, the metric for the gravitational field of the source under the weak-field approximation will be provided by~\cite{eric2018,Wu:2021uws}
\begin{eqnarray}
\label{equ3.17}g_{\mu\nu}=\eta_{\mu\nu}-\overline{h}_{\mu\nu}
\end{eqnarray}
with
\begin{eqnarray}
\label{equ3.18}\overline{h}_{\mu\nu}:=h_{\mu\nu}-\frac{1}{2}\eta_{\mu\nu}h,\quad h=\eta_{\mu\nu}h^{\mu\nu}.
\end{eqnarray}

The above is the main content of the weak-field approximation framework of $F(X,Y,Z)$ gravity in our preceding paper~\cite{Wu:2022mna}. Equations~(\ref{equ3.16})--(\ref{equ3.18}) indicate that the metric is decomposed into three separated parts relevant to the fields $\tilde{h}^{\mu\nu}$, $X^{(1)}$, and $P^{\mu\nu}$, and in order to analyze their behaviors, the gauge transformations satisfied by them need to be addressed. A direct calculation shows that under an infinitesimal coordinate transformation $x'^{\mu}= x^{\mu}+\varepsilon^{\mu}$, the gravitational field amplitude $h^{\mu\nu}$, the linearized Ricci scalar $X^{(1)}$, and the linearized Ricci tensor $R^{\mu\nu(1)}$ satisfy~\cite{eric2018}
\begin{eqnarray}
\label{equ3.18.1}&&h'^{\mu\nu}=h^{\mu\nu}+\partial^{\mu}\varepsilon^{\nu}+\partial^{\nu}\varepsilon^{\mu}-\eta^{\mu\nu}\partial_{\alpha}\varepsilon^{\alpha},\\
\label{equ3.18.2}&&X'^{(1)}=X^{(1)},\\
\label{equ3.18.3}&&R'^{\mu\nu(1)}=R^{\mu\nu(1)}.
\end{eqnarray}
With these results, by virtue of eqs.~(\ref{equ3.7})  and  (\ref{equ3.12}), the transformations for the tensor fields $\tilde{h}^{\mu\nu}$ and $P^{\mu\nu}$ are gained,
\begin{eqnarray}
\label{equ3.18.4}&&\tilde{h}'^{\mu\nu}=\tilde{h}^{\mu\nu}+\partial^{\mu}\varepsilon^{\nu}+\partial^{\nu}\varepsilon^{\mu}-\eta^{\mu\nu}\partial_{\alpha}\varepsilon^{\alpha},\\
\label{equ3.18.5}&&P'^{\mu\nu}=P^{\mu\nu}.
\end{eqnarray}
One can readily observe that eqs.~(\ref{equ3.6})  and  (\ref{equ3.8}) fulfilled by the tensor field  $\tilde{h}^{\mu\nu}$ are the same as those fulfilled by the gravitational field amplitude $h_{\text{GR}}^{\mu\nu}$ in GR~\cite{eric2018}, and one can also prove that the gauge condition~(\ref{equ3.6}) is preserved under the transformation~(\ref{equ3.18.4}) by setting $\square_{\eta}\varepsilon^{\alpha}=0$. These discussions suggest that the field $\tilde{h}^{\mu\nu}$ behaves as the gravitational field amplitude $h_{\text{GR}}^{\mu\nu}$ in GR, and therefore, it represents the massless propagation in $F(X,Y,Z)$ gravity. From this perspective, we conclude that the main purpose of the introduction of the gauge condition~(\ref{equ3.6}) is to separate the massless tensor field $\tilde{h}^{\mu\nu}$ from the gravitational field amplitude $h^{\mu\nu}$. In fact, as shown in ref.~\cite{Wu:2022mna}, it is based on the requirement
\begin{eqnarray}
\label{equ3.18.6}&&H^{\mu\nu(1)}=\frac{1}{2}\square_{\eta}\tilde{h}^{\mu\nu}=\kappa T^{\mu\nu}
\end{eqnarray}
that the coefficients $F_{11}-2F_{3}$ and $2F_{2}+8F_{3}$ in the definition~(\ref{equ3.7}) of $\tilde{h}^{\mu\nu}$ are  determined.
As for fields $X^{(1)}$ and $P^{\mu\nu}$, their behaviors depend on the algebraic signs of parameters $m_{1}^2$ and  $m_{2}^2$ in eqs.~(\ref{equ3.9}) and (\ref{equ3.10}). When $m_{1}^2>0$ and  $m_{2}^2>0$, these two equations reduce to the Klein-Gordon (KG) equations with external sources, and $X^{(1)}$ and $P^{\mu\nu}$ should present the massive propagations in $F(X,Y,Z)$ gravity. In this situation, both of them, as massive fields, have significant applications in FOTGs~\cite{Capozziello:2007ms,Stabile:2010mz,Stabile:2010zk,DeMartino:2018yqf,DeLaurentis:2018ahr}, which shows that it is very meaningful to make a multipole analysis on them. On the contrary, perhaps the solutions to eqs.~(\ref{equ3.9}) and (\ref{equ3.10}) with the negative squared parameters are also of potential importance in physics, but they have not received much attention so far. Therefore, in this paper, we always restrict our attention to the case of $m_{1}^2>0$ and  $m_{2}^2>0$, which ensures that our results would have practical implications.

The metric can indeed be derived by solving eqs.~(\ref{equ3.8})--(\ref{equ3.10}) through directly using the Green's function method in the above framework, but even so, it is still not an easy task to achieve a practical expression of the metric in this way. The difficulty lies in that the retarded Green's functions of eqs. (\ref{equ3.9}) and (\ref{equ3.10})~\cite{Wu:2017huang}, namely
\begin{eqnarray}
\label{equ3.19}\mathcal{G}_{i}(t,\boldsymbol{x};t',\boldsymbol{x}')&=&-\frac{\displaystyle\updelta\Big(t-t'-\frac{|\boldsymbol{x}
-\boldsymbol{x}'|}{c}\Big)}{4\pi|\boldsymbol{x}-\boldsymbol{x}'|}+\frac{m_{i}}{4\pi}\frac{\displaystyle\text{J}_{1}\left(m_{i}c\sqrt{(t-t')^{2}-\frac{|\boldsymbol{x}-\boldsymbol{x}'|^{2}}{c^2}}\right)}{\sqrt{(t-t')^{2}
-\displaystyle\frac{|\boldsymbol{x}-\boldsymbol{x}'|^{2}}{c^2}}}\notag\\
&&\times \text{H}\left(t-t'-\frac{|\boldsymbol{x}-\boldsymbol{x}'|}{c}\right),\quad i=1,2,
\end{eqnarray}
are complicated, where $\updelta$ is the Dirac delta function, $\text{J}_{1}$ is the Bessel function of the first order, and $\text{H}$ is the Heaviside's step function, and how to simplify them is the key point. A useful clue is that under the slow-motion approximation,
\begin{eqnarray}
\label{equ3.20}\square_{\eta}=-\frac{\partial^{2}}{c^{2}\partial t^{2}}+\delta^{ij}\frac{\partial^{2}}{\partial x_{i}\partial x_{j}}\approx\delta^{ij}\frac{\partial^{2}}{\partial x_{i}\partial x_{j}}=\Delta
\end{eqnarray}
at the leading order~\cite{Clifford2018,eric2018}, which means that eqs. (\ref{equ3.9}) and (\ref{equ3.10}) may reduce to the screened Poisson equations in this case, and thus, their retarded Green's functions $\mathcal{G}_{i}(t,\boldsymbol{x};t',\boldsymbol{x}')$ are simplified to be
\begin{eqnarray}\label{equ3.21}
\mathcal{G}_{i}(\boldsymbol{x};\boldsymbol{x}')=\frac{1}{4\pi}\frac{\text{e}^{-m_{i}|\boldsymbol{x}-\boldsymbol{x}'|}}{|\boldsymbol{x}-\boldsymbol{x}'|},\quad i=1,2.
\end{eqnarray}
The above discussion indicates that one could further simplify eqs.~(\ref{equ3.8})--(\ref{equ3.10}) by applying the slow-motion approximation, which strongly suggests that in this way, one is able to derive an easy-to-handle expression of the metric.

Next, based upon the framework in ref.~\cite{Wu:2022mna}, by applying the slow-motion approximation,
we shall deal with eqs.~(\ref{equ3.8})--(\ref{equ3.10}) so as to construct a viable WFSM approximation method within $F(X,Y,Z)$ gravity.
A basic fact is that so far, even for a point source, the metric up to $1/c^4$ order has not been obtained in $F(X,Y,Z)$ gravity, so the best option for us at the moment is to simplify
eqs.~(\ref{equ3.8})--(\ref{equ3.10})  under the condition ``up to $1/c^3$ order''.  It will be seen that this condition can greatly facilitate the derivation.

As a generic fourth-order theory of gravity, under the WFSM approximation, the metric of $F(X,Y,Z)$ gravity up to $1/c^3$ order outside a source could be assumed to be
\begin{equation}\label{equ3.22}
\left\{\begin{array}{ll}
\displaystyle g_{00}(t,\boldsymbol{x})&=\displaystyle -1+\frac{2}{c^{2}}U(t,\boldsymbol{x}),\smallskip\\
\displaystyle g_{0i}(t,\boldsymbol{x})&=\displaystyle -\frac{4}{c^{3}}U_{i}(t,\boldsymbol{x}),\smallskip\\
\displaystyle g_{ij}(t,\boldsymbol{x})&=\displaystyle \delta_{ij}+\frac{2}{c^{2}}U_{ij}(t,\boldsymbol{x}),
\end{array}\right.
\end{equation}
where $U(t,\boldsymbol{x})$, $U_{i}(t,\boldsymbol{x})$, and $U_{ij}(t,\boldsymbol{x})$ are the scalar, vector, and tensor potentials, respectively. $U(t,\boldsymbol{x})$ is the gravitational potential at the Newtonian order, and its expressions for stationary sources have been given in refs.~\cite{Stabile:2010mz,Wu:2022mna}. $U_{i}(t,\boldsymbol{x})$, also defined in many theories of gravity, is associated with  gravitomagnetic effect, which plays an important role in astrophysical phenomena like gyroscopic precession. As for $U_{ij}(t,\boldsymbol{x})$, we will see that within $F(X,Y,Z)$ gravity, there is $U_{ij}(t,\boldsymbol{x})\neq\delta_{ij}U(t,\boldsymbol{x})$ in general, which is different from the cases in
GR and $f(R)$ gravity.  Besides the metric, under the WFSM approximation, the components of the energy-momentum tensor up to $1/c^3$ order fulfill~\cite{eric2018,Stabile:2010zk,Damour:1990gj}
\begin{equation}
\label{equ3.23}\kappa T^{00}\sim O\left(\frac{1}{c^2}\right),\quad \kappa T^{0i}\sim O\left(\frac{1}{c^{3}}\right),\quad \kappa T^{ij}=0,
\end{equation}
where $O(1/c^m)$ ($m$ is an integer) is used to stand for the ``order of smallness'' of the corresponding quantity,
and the relation~\cite{Clifford2018}
\begin{equation}\label{equ3.24}
\frac{|\partial/\partial x^{0}|}{|\partial/\partial x^{i}|}\sim O\left(\frac{1}{c}\right)
\end{equation}
holds. With these conclusions, from eqs.~(\ref{equ3.11}) and (\ref{equ3.13}),  it can be inferred that there are
\begin{equation}
\label{equ3.25}\kappa T=-\kappa T^{00}\sim O\left(\frac{1}{c^2}\right)
\end{equation}
and
\begin{equation}\label{equ3.26}
\left\{\begin{array}{ll}
\displaystyle \kappa S^{00}&=\displaystyle \frac{2}{3}\kappa T^{00} \sim O\left(\frac{1}{c^2}\right),\smallskip\\
\displaystyle \kappa S^{0i}&=\displaystyle \kappa T^{0i}+\frac{1}{3m_{2}^{2}}\partial_{0}\partial_{i}\left(\kappa T^{00}\right)\sim O\left(\frac{1}{c^3}\right),\smallskip\\
\displaystyle \kappa S^{ij}&=\displaystyle \frac{1}{3}\delta_{ij}\kappa T^{00}-\frac{1}{3m_{2}^{2}}\partial_{i}\partial_{j}\left(\kappa T^{00}\right)\sim O\left(\frac{1}{c^2}\right).
\end{array}\right.
\end{equation}
These results, together with eqs.~(\ref{equ3.20}) and (\ref{equ3.22})--(\ref{equ3.24}), show that up to $1/c^3$ order, eqs.~(\ref{equ3.8})--(\ref{equ3.10}) are simplified to be
\begin{eqnarray}
\label{equ3.27}&&\Delta\tilde{h}^{\mu\nu}=2\kappa T^{\mu\nu},\\
\label{equ3.28}&&\Delta X^{(1)}-m_{1}^2 X^{(1)}=m_{1}^2\kappa T,\\
\label{equ3.29}&&\Delta P^{\mu\nu}-m_{2}^2 P^{\mu\nu}=-m_{2}^2\kappa S^{\mu\nu},
\end{eqnarray}
respectively. According to eqs.~(\ref{equ3.16})--(\ref{equ3.18}), these three equations are the starting point to derive the metric for a source, and hence, they constitute the most fundamental ingredient in the WFSM approximation method within $F(X,Y,Z)$ gravity.

Now, one could adopt the Green's function method to derive $\tilde{h} ^{\mu\nu}$, $X^{(1)}$, and $P^{\mu\nu}$.
As stated previously, the Green's functions of eqs.~(\ref{equ3.27})--(\ref{equ3.29}), namely
\begin{eqnarray}\label{equ3.30}
\mathcal{G}(\boldsymbol{x};\boldsymbol{x}')=\frac{1}{4\pi}\frac{1}{|\boldsymbol{x}-\boldsymbol{x}'|},
\end{eqnarray}
$\mathcal{G}_{1}(\boldsymbol{x};\boldsymbol{x}')$, and $\mathcal{G}_{2}(\boldsymbol{x};\boldsymbol{x}')$,
are convenient to handle, so using them, the solutions to these three equations can be given in a neat form,
\begin{equation}\label{equ3.31}
\left\{\begin{array}{ll}
\displaystyle \tilde{h}^{00}(t,\boldsymbol{x})&=\displaystyle -\frac{4G}{c^2}\int\frac{1}{|\boldsymbol{x}-\boldsymbol{x}'|}\frac{T^{00}(t,\boldsymbol{x}')}{c^2}
\text{d}^{3}x',\smallskip\\
\displaystyle \tilde{h}^{0i}(t,\boldsymbol{x})&=\displaystyle -\frac{4G}{c^3}\int\frac{1}{|\boldsymbol{x}-\boldsymbol{x}'|}\frac{T^{0i}(t,\boldsymbol{x}')}{c}
\text{d}^{3}x',\smallskip\\
\displaystyle \tilde{h}^{ij}(t,\boldsymbol{x})&=0,
\end{array}\right.
\end{equation}
\begin{eqnarray}
\label{equ3.32}&& X^{(1)}(t,\boldsymbol{x})=\frac{2Gm^{2}_{1}}{c^2}\int\frac{\text{e}^{-m_{1}|\boldsymbol{x}-\boldsymbol{x}'|}}{|\boldsymbol{x}-\boldsymbol{x}'|}\frac{T^{00}(t,\boldsymbol{x}')}{c^2}\text{d}^{3}x',
\end{eqnarray}
and
\begin{equation}\label{equ3.33}
\left\{\begin{array}{ll}
\displaystyle P^{00}(t,\boldsymbol{x})&=\displaystyle \frac{2Gm^{2}_{2}}{c^2}\int\frac{\text{e}^{-m_{2}|\boldsymbol{x}-\boldsymbol{x}'|}}{|\boldsymbol{x}-\boldsymbol{x}'|}\frac{S^{00}(t,\boldsymbol{x}')}{c^2}\text{d}^{3}x'
,\smallskip\\
\displaystyle P^{0i}(t,\boldsymbol{x})&=\displaystyle \frac{2Gm^{2}_{2}}{c^3}\int\frac{\text{e}^{-m_{2}|\boldsymbol{x}-\boldsymbol{x}'|}}{|\boldsymbol{x}-\boldsymbol{x}'|}\frac{S^{0i}(t,\boldsymbol{x}')}{c}\text{d}^{3}x'
,\smallskip\\
\displaystyle P^{ij}(t,\boldsymbol{x})&=\displaystyle \frac{2Gm^{2}_{2}}{c^2}\int\frac{\text{e}^{-m_{2}|\boldsymbol{x}-\boldsymbol{x}'|}}{|\boldsymbol{x}-\boldsymbol{x}'|}\frac{S^{ij}(t,\boldsymbol{x}')}{c^2}\text{d}^{3}x'.
\end{array}\right.
\end{equation}
For tensor fields $\tilde{h}^{\mu\nu}$ and $P^{\mu\nu}$, their components are not independent because the former satisfies the gauge condition~(\ref{equ3.6})  and the latter satisfies
the constraint conditions
\begin{eqnarray}
\label{equ3.34}&&\partial_{\mu}P^{\mu\nu}=-\frac{1}{3}\kappa\partial^{\nu}T,\\
\label{equ3.35}&&\eta_{\mu\nu}P^{\mu\nu}=-\frac{1}{3}\kappa T,
\end{eqnarray}
where these two constraint conditions are given from eqs.~(\ref{equ3.9})  and  (\ref{equ3.12}). It is worth noting that eq.~(\ref{equ3.34}) is completely different from the gauge condition~(\ref{equ3.6}), because it does not originate from the gauge transformation. With eqs.~(\ref{equ3.31})--(\ref{equ3.33}), one is able to further acquire the gravitational field amplitude $h^{\mu\nu}$ by inserting them into eq.~(\ref{equ3.16}), and then, by using eqs.~(\ref{equ3.17}) and (\ref{equ3.18}), the metric up to $1/c^3$ order for the gravitational field of the source is finally derived.

From eqs.~(\ref{equ3.31})--(\ref{equ3.33}), it is expected that in the expression of the metric there is no infinite integral like those in eqs.~(\ref{equ1.2}) because the integrals involved only need to be performed over the actual source distributions, and thus, we think that compared with the previous expressions of the metric in refs.~\cite{stabile2015,Capozziello:2009ss,Stabile:2010mz}, our result is more applicable to exploring the effects of the gravitational field of a realistic source. Therefore, the above WFSM approximation method for $F(X,Y,Z)$ gravity is indeed feasible. But, it is still inconvenient to employ the metric in practical application because the gauge condition~(\ref{equ3.6})  and the constraint conditions~(\ref{equ3.34}) and (\ref{equ3.35}) need to be considered separately. In the next section, combining these conditions,
we will further handle eqs.~(\ref{equ3.31})--(\ref{equ3.33}) by means of the STF formalism outside a source so as to achieve the multipole expansions of $\tilde{h}^{\mu\nu}$, $X^{(1)}$, and $P^{\mu\nu}$, and then, with these results, we will also deduce the multipole expansion of the external metric up to $1/c^3$ order by using eqs.~(\ref{equ3.16})--(\ref{equ3.18}). The salient feature of the metric in the form of the multipole expansion is that the field point and the source point are completely separate, and the information of the source is contained in the source multipole moments. Thus, as originally envisaged, we could acquire a ready-to-use expression of the metric by applying the STF formalism.
\section{Metric for the external gravitational field of a spatially compact source up to $1/c^3$ order in $F(X,Y,Z)$ gravity~\label{Sec:fourth}}
As mentioned above, the metric derived based on the expressions of $\tilde{h} ^{\mu\nu}$, $X^{(1)}$, and $P^{\mu\nu}$ in eqs.~(\ref{equ3.31})--(\ref{equ3.33}) is still inconvenient to be employed in practical application because the conditions~(\ref{equ3.6}), (\ref{equ3.34}), and (\ref{equ3.35}) need to be tackled separately.
In this section, we will solve this problem outside a spatially compact source by applying the STF formalism in terms of the irreducible Cartesian tensors~\cite{Thorne:1980ru,Blanchet:1985sp,Blanchet:1989ki,Damour:1990gj} so as to derive
the external metric, presented in the form of multipole expansion, for the source up to $1/c^3$ order in $F(X,Y,Z)$ gravity.
\subsection{Multipole expansion of  $\tilde{h} ^{\mu\nu}$ outside a spatially compact source up to $1/c^3$ order}
For $\tilde{h}^{\mu\nu}$, it satisfies
\begin{equation}\label{equ4.1}
\left\{\begin{array}{ll}
\displaystyle &\Delta\tilde{h}^{\mu\nu}=2\kappa T^{\mu\nu},\smallskip\\
\displaystyle &\partial_{\mu}\tilde{h}^{\mu\nu}=0
\end{array}\right.
\end{equation}
and the gauge transformation~(\ref{equ3.18.4}).  The Green's function of the d'Alembert equation is
$\mathcal{G}(\boldsymbol{x};\boldsymbol{x}')$ in eq.~(\ref{equ3.30}), and by referring to ref.~\cite{Wu:2017huang},
its multipole expansion is
\begin{eqnarray}
\label{equ4.2}\mathcal{G}(\boldsymbol{x};\boldsymbol{x}')=\sum_{l=0}^{\infty}\frac{(2l-1)!!}{4\pi l!}\frac{(r_< )^l}{(r_> )^{l+1}}\hat{N}_{I_{l}}(\theta',\varphi')\hat{N}_{I_{l}}(\theta,\varphi),
\end{eqnarray}
where $(\theta',\varphi')$ and $(\theta,\varphi)$ are the angle coordinates of $\boldsymbol{x}'$ and $\boldsymbol{x}$, respectively, $r_{<}$ represents the lesser of $r=|\boldsymbol{x}|$ and $r'=|\boldsymbol{x}'|$, and $r_{>}$ the greater.
From $\mathcal{G}(\boldsymbol{x};\boldsymbol{x}')$, outside the source, the initial expansion of $\tilde{h}^{\mu\nu}$ is
\begin{eqnarray}\label{equ4.3}
\tilde{h}^{\mu\nu}(t,\boldsymbol{x})&=& \int\mathcal{G}(\boldsymbol{x};\boldsymbol{x}')\left(-2\kappa T^{\mu\nu}(t,\boldsymbol{x}')\right)\text{d}^{3}x'\notag\\
&=&-\frac{4G}{c^{4}}\sum_{l=0}^{\infty}\frac{(-1)^l}{l!}\int \left(r'^l\hat{N}_{I_{l}}(\theta',\varphi')T^{\mu\nu}(t,\boldsymbol{x}')\frac{(-1)^l(2l-1)!!}{r ^{l+1}}\hat{N}_{I_{l}}(\theta,\varphi)\right)\text{d}^{3}x' \notag\\
&=&-\frac{4G}{c^{4}}\sum_{l=0}^{\infty}\frac{(-1)^l}{l!}\int r'^l\hat{N}_{I_{l}}(\theta',\varphi')T^{\mu\nu}(t,\boldsymbol{x}')\text{d}^{3}x'\hat{\partial}_{I_{l}}\left(\frac{1}{r}\right)\notag\\
&=&-\frac{4G}{c^{4}}\sum_{l=0}^{\infty}\frac{(-1)^l}{l!}\int \hat{X}'_{I_{l}}T^{\mu\nu}(t,\boldsymbol{x}')\text{d}^{3}x'\partial_{I_{l}}\left(\frac{1}{r}\right),
\end{eqnarray}
where the formula~(\ref{equ2.14}) is used, and $X'_{I_{l}}=r'^{l}N_{I_{l}}(\theta',\varphi')= x'_{i_{1}}x'_{i_{2}}\cdots x'_{i_{l}}$. For convenience, if we define
\begin{eqnarray}
\label{equ4.4}\mathcal{A}^{(\tilde{h})}_{\langle I_{l}\rangle}(t)=\frac{(-1)^l}{l!}\int \hat{X}'_{I_{l}}\frac{T^{00}(t,\boldsymbol{x}')}{c^2}\text{d}^{3}x',\\
\label{equ4.5}\mathcal{U}^{(\tilde{h})}_{i\langle I_{l}\rangle}(t)=\frac{(-1)^l}{l!}\int \hat{X}'_{I_{l}}\frac{T^{0i}(t,\boldsymbol{x}')}{c}\text{d}^{3}x',\\
\label{equ4.6}\mathcal{V}^{(\tilde{h})}_{ij\langle I_{l}\rangle}(t)=\frac{(-1)^l}{l!}\int \hat{X}'_{I_{l}}T^{ij}(t,\boldsymbol{x}')\text{d}^{3}x',
\end{eqnarray}
the components of $\tilde{h}^{\mu\nu}$ up to $1/c^3$ order can be written as
\begin{eqnarray}
\label{equ4.7}\tilde{h}^{00}(t,\boldsymbol{x})
&=&-\frac{4G}{c^{2}}\sum_{l=0}^{\infty}\mathcal{A}^{(\tilde{h})}_{\langle I_{l}\rangle}(t)\partial_{I_{l}}\left(\frac{1}{r}\right),\qquad \\
\label{equ4.8}\tilde{h}^{0i}(t,\boldsymbol{x})
&=&-\frac{4G}{c^{3}}\sum_{l=0}^{\infty}\mathcal{U}^{(\tilde{h})}_{i\langle I_{l}\rangle}(t)\partial_{I_{l}}\left(\frac{1}{r}\right),\qquad \\
\label{equ4.9}\tilde{h}^{ij}(t,\boldsymbol{x})&=&0.
\end{eqnarray}

Next, we will utilize eqs.~(\ref{equ2.7})--(\ref{equ2.10}) to decompose the reducible tensor $\mathcal{U}^{(\tilde{h})}_{i\langle I_{l}\rangle}$ into irreducible pieces. By virtue of these formulas, there is
\begin{eqnarray}
\label{equ4.10}\mathcal{U}^{(\tilde{h})}_{i\langle I_{l}\rangle}&=&\hat{R}^{(\tilde{h}+)}_{iI_{l}}+\frac{l}{l+1}\epsilon_{ai\langle i_{l}}\hat{R}^{(\tilde{h}0)}_{i_{1}\cdots i_{l-1}\rangle a}+\frac{2l-1}{2l+1}\delta_{i\langle i_{l}}\hat{R}^{(\tilde{h}-)}_{i_{1}\cdots i_{l-1}\rangle}
\end{eqnarray}
with
\begin{eqnarray}
\label{equ4.11}&&\hat{R}^{(\tilde{h}+)}_{I_{l+1}}=\hat{\mathcal{U}}^{(\tilde{h})}_{I_{l+1}},\\
\label{equ4.12}&&\hat{R}^{(\tilde{h}0)}_{I_{l}}=\mathcal{U}^{(\tilde{h})}_{pq\langle i_{1}\cdots i_{l-1}}\epsilon_{i_{l}\rangle pq},\\
\label{equ4.13}&&\hat{R}^{(\tilde{h}-)}_{I_{l-1}}=\mathcal{U}^{(\tilde{h})}_{aaI_{l-1}}.
\end{eqnarray}
Substituting the decomposition~(\ref{equ4.10}) in the expansion~(\ref{equ4.8}) outside the source, we derive, after suitable changes of the summation index,
\begin{eqnarray}
\label{equ4.14}\tilde{h}^{0i}&=&-\frac{4G}{c^{3}}\left[\sum_{l=0}^{\infty}\hat{\mathcal{B}}^{(\tilde{h})}_{I_{l}}\partial_{iI_{l}}\bigg(\frac{1}{r}\bigg)
+\sum_{l=1}^{\infty}\hat{\mathcal{C}}^{(\tilde{h})}_{iI_{l-1}}\partial_{I_{l-1}}\bigg(\frac{1}{r}\bigg)+\sum_{l=1}^{\infty}\epsilon_{iab}\hat{\mathcal{D}}^{(\tilde{h})}_{bI_{l-1}}\partial_{aI_{l-1}}\bigg(\frac{1}{r}\bigg)\right]\quad
\end{eqnarray}
with
\begin{eqnarray}
\label{equ4.15}&&\hat{\mathcal{B}}^{(\tilde{h})}_{I_{l}}:=\frac{2l+1}{2l+3}\hat{R}^{(\tilde{h}-)}_{I_{l}},\\
\label{equ4.16}&&\hat{\mathcal{C}}^{(\tilde{h})}_{I_{l}}:=\hat{R}^{(\tilde{h}+)}_{I_{l}},\\
\label{equ4.17}&&\hat{\mathcal{D}}^{(\tilde{h})}_{I_{l}}:=\frac{l}{l+1}\hat{R}^{(\tilde{h}0)}_{I_{l}}.
\end{eqnarray}
From expansions~(\ref{equ4.7}) and (\ref{equ4.14}), we recognize that the tensor field $\tilde{h}^{\mu\nu}$ up to $1/c^3$ order is expressed in terms of four multipole tensor coefficients $\mathcal{A}^{(\tilde{h})}_{\langle I_{l}\rangle}$, $\mathcal{B}^{(\tilde{h})}_{\langle I_{l}\rangle}$, $\mathcal{C}^{(\tilde{h})}_{\langle I_{l}\rangle}$, and $\mathcal{D}^{(\tilde{h})}_{\langle I_{l}\rangle}$. Since the components of $\tilde{h}^{\mu\nu}$ ought to satisfy the identity
\begin{eqnarray}
\label{equ4.18}&&\frac{1}{c}\partial_{t}\tilde{h}^{00}+\partial_{i}\tilde{h}^{i0}=0
\end{eqnarray}
supplied by the gauge condition~(\ref{equ3.6}) up to $1/c^3$ order, these tensor coefficients have the following relation,
\begin{eqnarray}
\label{equ4.19}\partial_{t}\hat{\mathcal{A}}^{(\tilde{h})}
+\sum_{l=1}^{\infty}\left(\partial_{t}\hat{\mathcal{A}}^{(\tilde{h})}_{I_{l}}+\hat{\mathcal{C}}^{(\tilde{h})}_{I_{l}}\right)\partial_{I_{l}}\bigg(\frac{1}{r}\bigg)=0,
\end{eqnarray}
from which, by virtue of eqs.~(\ref{equ2.12})--(\ref{equ2.14}), we obtain
\begin{eqnarray}
\label{equ4.20}\partial_{t}\hat{\mathcal{A}}^{(\tilde{h})}=0,\qquad \partial_{t}\hat{\mathcal{A}}^{(\tilde{h})}_{I_{l}}+\hat{\mathcal{C}}^{(\tilde{h})}_{I_{l}}=0,\quad l\geqslant1.
\end{eqnarray}
With these identities, the expansion of $\tilde{h}^{0i}$ is reduced to
\begin{eqnarray}
\label{equ4.21}\tilde{h}^{0i}&=&-\frac{4G}{c^{3}}\left[
-\sum_{l=1}^{\infty}\partial_{t}\hat{\mathcal{A}}^{(\tilde{h})}_{iI_{l-1}}\partial_{I_{l-1}}\bigg(\frac{1}{r}\bigg)+\sum_{l=1}^{\infty}\epsilon_{iab}\hat{\mathcal{D}}^{(\tilde{h})}_{bI_{l-1}}\partial_{aI_{l-1}}\bigg(\frac{1}{r}\bigg)
+\sum_{l=0}^{\infty}\hat{\mathcal{B}}^{(\tilde{h})}_{I_{l}}\partial_{iI_{l}}\bigg(\frac{1}{r}\bigg)\right].\qquad\quad
\end{eqnarray}

The further simplification of the expansion of $\tilde{h}^{0i}$ relies on the residual gauge freedom in the gauge condition~(\ref{equ3.6}).  One can prove that outside the source, up to $1/c^{3}$ order,
\begin{equation}\label{equ4.22}
\left\{\begin{array}{ll}
\displaystyle\varepsilon^{0}(t,\boldsymbol{x})=&\displaystyle \frac{4G}{c^{3}}\sum_{l=0}^{\infty}\hat{\mathcal{B}}^{(\tilde{h})}_{I_{l}}(t)\partial_{I_{l}}\bigg(\frac{1}{r}\bigg)\sim O\left(\frac{1}{c^3}\right),\smallskip\\
\displaystyle\varepsilon^{i}(t,\boldsymbol{x})=&\displaystyle 0
\end{array}\right.
\end{equation}
satisfies that
\begin{equation}\label{equ4.23}
\square_{\eta}\varepsilon^{\mu}\approx\Delta\varepsilon^{\mu}=0,
\end{equation}
and hence, by use of  the gauge transformation~(\ref{equ3.18.4}),
\begin{eqnarray}
\label{equ4.24}\tilde{h}'^{00}
&=&\tilde{h}^{00},\qquad \\
\label{equ4.25}\tilde{h}'^{0i}
&=&\tilde{h}^{0i}+\partial_{i}\varepsilon^{0}=\tilde{h}^{0i}+\frac{4G}{c^{3}}\sum_{l=0}^{\infty}\hat{\mathcal{B}}^{(\tilde{h})}_{I_{l}}(t)\partial_{iI_{l}}\bigg(\frac{1}{r}\bigg), \\
\label{equ4.26}\tilde{h}'^{ij}&=&0
\end{eqnarray}
are acquired. Combining the above results with expansions~(\ref{equ4.7}), (\ref{equ4.9}), and (\ref{equ4.14}),  we finally get
\begin{eqnarray}
\label{equ4.27}\tilde{h}^{00}
&=&-\frac{4G}{c^{2}}\sum_{l=0}^{\infty}\hat{\mathcal{A}}^{(\tilde{h})}_{ I_{l}}\partial_{I_{l}}\left(\frac{1}{r}\right),\qquad \\
\label{equ4.28}\tilde{h}^{0i}
&=&-\frac{4G}{c^{3}}\left[
-\sum_{l=1}^{\infty}\partial_{t}\hat{\mathcal{A}}^{(\tilde{h})}_{iI_{l-1}}\partial_{I_{l-1}}\bigg(\frac{1}{r}\bigg)+\sum_{l=1}^{\infty}\epsilon_{iab}\hat{\mathcal{D}}^{(\tilde{h})}_{bI_{l-1}}\partial_{aI_{l-1}}\bigg(\frac{1}{r}\bigg)\right], \\
\label{equ4.29}\tilde{h}^{ij}&=&0,
\end{eqnarray}
where the primes `` $'$ '' in $\tilde{h}'^{\mu\nu}$ have been omitted.
As a massless tensor field, eqs.~(\ref{equ4.1}) and the gauge transformation~(\ref{equ3.18.4}) clearly indicate that $\tilde{h}^{\mu\nu}$ is similar to the gravitational field amplitude $h_{\text{GR}}^{\mu\nu}$  in GR. In order to make the above expansion of $\tilde{h}^{\mu\nu}$ compatible with the
truncation of the relativistic time-dependent multipole expansion of $h_{\text{GR}}^{\mu\nu}$~\cite{Damour:1990gj} to $1/c^3$ order,  we need to define
\begin{eqnarray}
\label{equ4.30}
\hat{M}_{I_{l}}^{(\tilde{h})}
&=&(-1)^{l}l!\hat{\mathcal{A}}^{(\tilde{h})}_{I_{l}},\\
\label{equ4.31}\hat{S}_{I_{l}}^{(\tilde{h})}
&=&\frac{(-1)^{l+1}(l+1)!}{l}\hat{\mathcal{D}}^{(\tilde{h})}_{I_{l}},
\end{eqnarray}
from which, the multipole expansion of $\tilde{h} ^{\mu\nu}$ should be recast as
\begin{eqnarray}
\label{equ4.32}\tilde{h}^{00}(t,\boldsymbol{x})&=& -\frac{4G}{c^{2}}\sum_{l=0}^{\infty}\frac{(-1)^{l}}{l!}\hat{M}_{I_{l}}^{(\tilde{h})}(t)\partial_{I_{l}}\left( \frac{1}{r}\right),\qquad \\
\label{equ4.33}\tilde{h}^{0i}(t,\boldsymbol{x})&=& -\frac{4G}{c^{3}}\left[-\sum_{l=1}^{\infty}\frac{(-1)^{l}}{l!}\left(\partial_{t}\hat{M}_{iI_{l-1}}^{(\tilde{h})}(t)\right)\partial_{I_{l-1}}\left(\frac{1}{r}\right)\right. \notag\\ &&\left.+\sum_{l=1}^{\infty}\frac{(-1)^{l}l}{(l+1)!}\epsilon_{iab}\hat{S}_{aI_{l-1}}^{(\tilde{h})}(t)\partial_{bI_{l-1}}\left( \frac{1}{r}\right)\right],\\
\label{equ4.34}\tilde{h}^{ij}(t,\boldsymbol{x})&=&0,
\end{eqnarray}
where by means of eqs.~(\ref{equ4.4}), (\ref{equ4.5}), (\ref{equ4.12}) and (\ref{equ4.17}), the two sets of source multipole moments are
\begin{eqnarray}
\label{equ4.35}
\hat{M}_{I_{l}}^{(\tilde{h})}(t)&=&\int\hat{X'}_{I_{l}}\frac{T^{00}(t,\boldsymbol{x}')}{c^2}\text{d}^{3}x',\quad l\geqslant0,\\
\label{equ4.36}\hat{S}_{I_{l}}^{(\tilde{h})}(t)&=&-(-1)^{l}l!\hat{R}^{(\tilde{h}0)}_{I_{l}}=
-(-1)^{l}l!\mathcal{U}^{(\tilde{h})}_{pq\langle i_{1}\cdots i_{l-1}}\epsilon_{i_{l}\rangle pq}\notag\\
&=&\int \epsilon_{pq\langle i_{1}}\hat{X'}_{i_{2}\cdots i_{l}\rangle p}\frac{T^{0q}(t,\boldsymbol{x}')}{c}\text{d}^{3}x',\quad l\geqslant1.
\end{eqnarray}

The above two equations indicate that
\begin{eqnarray}
\label{equ4.37}
M:=\hat{M}_{I_{0}}^{(\tilde{h})}(t)=\int\frac{T^{00}(t,\boldsymbol{x}')}{c^2}\text{d}^{3}x'
\end{eqnarray}
is the total mass of the source, and
\begin{eqnarray}
\label{equ4.38}J_{b}:=\hat{S}_{b}^{(\tilde{h})}(t)=\int\epsilon_{bij}x'_{i}\frac{T^{0j}(t,\boldsymbol{x}')}{c}\text{d}^{3}x'
\end{eqnarray}
is the conserved angular momentum of the source, so $\hat{M}_{I_{l}}^{(\tilde{h})}(t)$ and $\hat{S}_{I_{l}}^{(\tilde{h})}(t)$ are referred to as the mass and spin multipole moments, respectively.
When the leading pole moments are considered in the stationary spacetime, those terms associated with $M$ and $J_{b}$ contribute to $\tilde{h}^{\mu\nu}$, namely,
\begin{eqnarray}
\label{equ4.39}\tilde{h}^{00}&=&-\frac{4GM}{c^{2}r},\qquad \\
\label{equ4.40}\tilde{h}^{0i}&=& \frac{2G\epsilon_{iab}x_{a}J_{b}}{c^{3}r^3},\\
\label{equ4.41}\tilde{h}^{ij}&=&0.
\end{eqnarray}
Under the case that  $F(X,Y,Z)$ gravity reduces to GR and the source is rotating around the $z$-axis, one can prove that the metric derived from $h^{\mu\nu}_{\text{GR}}=\tilde{h}^{\mu\nu}$ recovers the Lense-Thirring metric in the isotropic coordinate. The neat derivation in this subsection showcases that the fundamental feature of the multipole expansion of $\tilde{h} ^{\mu\nu}$ up to $1/c^3$ order is the Coulomb-like dependence, which is completely different from the Yukawa-like dependence for the massive fields $X^{(1)}$ and $P^{\mu\nu}$. In addition, it is worth noting that the residual gauge transformation plays an inevitable role in the process of eliminating redundant source multipole moments, which does not apply to the cases of $X^{(1)}$ and $P^{\mu\nu}$ because both of them have nothing to do with the gauge transformation.
\subsection{Multipole expansion of  $X^{(1)}$ outside a spatially compact source up to $1/c^3$ order}
Compared with the result of the massless field $\tilde{h} ^{\mu\nu}$, the multipole expansions of the massive field $X^{(1)}$ outside a spatially compact source up to $1/c^3$ order display new features.
It satisfies the screened Poisson equation~(\ref{equ3.28}) whose
Green's function is $\mathcal{G}_{1}(\boldsymbol{x};\boldsymbol{x}')$. The multipole expansion of $\mathcal{G}_{1}(\boldsymbol{x};\boldsymbol{x}')$ has been obtained in ref.~\cite{Wu:2022akq}, namely,
\begin{eqnarray}
\label{equ4.42}\mathcal{G}_{1}(\boldsymbol{x};\boldsymbol{x}')=\sum_{l=0}^{\infty}\frac{(2l+1)!!}{4\pi l!}m_{1}\text{i}_{l}(m_{1}r_<)\text{k}_{l}(m_{1}r_> )\hat{N}_{I_{l}}(\theta',\varphi')\hat{N}_{I_{l}}(\theta,\varphi),
\end{eqnarray}
where
\begin{eqnarray}
\label{equ4.43}\text{i}_{l}(z):=\sqrt{\frac{\pi}{2z}}\text{I}_{l+\frac{1}{2}}(z),\qquad
\text{k}_{l}(z):=\sqrt{\frac{2}{\pi z}}\text{K}_{l+\frac{1}{2}}(z)
\end{eqnarray}
are the spherical modified Bessel functions of $l$-order~\cite{Arfken1985} with
$\text{I}_{l+1/2}(z)$, $\text{K}_{l+1/2}(z)$ as the modified Bessel functions of $(l+1/2)$-order. From eqs.~(\ref{equ3.25}) and (\ref{equ3.28}), outside the source, the initial multipole expansion of $X^{(1)}$ up to $1/c^{3}$ order is
\begin{eqnarray}\label{equ4.44}
&&X^{(1)}(t,\boldsymbol{x})=\int\mathcal{G}_{1}(\boldsymbol{x};\boldsymbol{x}')\left(-m_{1}^2\kappa T(t,\boldsymbol{x}')\right)\text{d}^{3}x'=\int\mathcal{G}_{1}(\boldsymbol{x};\boldsymbol{x}')\left(m_{1}^2\kappa T^{00}(t,\boldsymbol{x}')\right)\text{d}^{3}x'\notag\\
&=&\frac{2Gm_{1}^{3}}{c^2}\sum_{l=0}^{\infty}\frac{(-1)^l}{l!}
\int\left((2l+1)!!\text{i}_{l}(m_{1}r')\hat{N}_{I_{l}}(\theta',\varphi')\frac{T^{00}(t,\boldsymbol{x}')}{c^2}(-1)^l\text{k}_{l}(m_{1}r)\hat{N}_{I_{l}}(\theta,\varphi)\right)\text{d}^{3}x'.\notag\\
\end{eqnarray}
This expansion needs to be rewritten in a neat form. According to ref.~\cite{Arfken1985}, the spherical modified Bessel functions of $l$-order have the following expressions,
\begin{eqnarray}
\label{equ4.45}\text{i}_{l}(z)=z^l\left(\frac{\text{d}}{z\text{d}z}\right)^{l}\left(\frac{\sinh{z}}{z}\right),\qquad
\text{k}_{l}(z)=\frac{\text{e}^{-z}}{z}\sum_{k=0}^{l}\frac{(l+k)!}{k!(l-k)!}\frac{1}{(2z)^{k}}.
\end{eqnarray}
If we define the function
\begin{eqnarray}
\label{equ4.46}&&\updelta_{l}(z):=(2l+1)!!\bigg(\frac{\text{d}}{z\text{d}z}\bigg)^{l}\bigg(\frac{\sinh{z}}{z}\bigg),
\end{eqnarray}
there is
\begin{eqnarray}
\label{equ4.47}&&\text{i}_{l}(m_{1}r')=\frac{(m_{1}r')^{l}}{(2l+1)!!}\updelta_{l}(m_{1}r').
\end{eqnarray}
In addition, with formulas~(\ref{equ2.14}) and
\begin{eqnarray}
\label{equ4.48}\text{e}^{-z}=(-1)^{l-k}\frac{\text{d}^{l-k}}{\text{d}z^{l-k}}\text{e}^{-z},
\end{eqnarray}
it can be inferred that
\begin{eqnarray}
\label{equ4.49}&&\text{k}_{l}(m_{1}r)\hat{N}_{I_{l}}(\theta,\varphi)=\frac{(-1)^{l}}{m_{1}^{l+1}}\hat\partial_{I_{l}}\left(\frac{\text{e}^{-m_{1}r}}{r}\right).
\end{eqnarray}
The substitution of eqs.~(\ref{equ4.47}) and (\ref{equ4.49}) in the expansion~(\ref{equ4.44}) gives rise to the final multipole expansion of $X^{(1)}$ up to $1/c^{3}$,
\begin{eqnarray}\label{equ4.50}
X^{(1)}(t,\boldsymbol{x})&=&
\frac{2Gm_{1}^{2}}{c^2}\sum_{l=0}^{\infty}\frac{(-1)^l}{l!}
\int\updelta_{l}(m_{1}r')r'^{l}\hat{N}_{I_{l}}(\theta',\varphi')\frac{T^{00}(t,\boldsymbol{x}')}{c^2}\text{d}^{3}x'\hat\partial_{I_{l}}\left(\frac{\text{e}^{-m_{1}r}}{r}\right)\notag\\
&=&\frac{2Gm_{1}^{2}}{c^2}\sum_{l=0}^{\infty}\frac{(-1)^l}{l!}
\int\updelta_{l}(m_{1}r')\hat{X}'_{I_{l}}\frac{T^{00}(t,\boldsymbol{x}')}{c^2}\text{d}^{3}x'\partial_{I_{l}}\left(\frac{\text{e}^{-m_{1}r}}{r}\right)\notag\\
&=&\frac{2Gm_{1}^{2}}{c^2}\sum_{l=0}^{\infty}\frac{(-1)^l}{l!}\hat{Q}_{I_{l}}(t)\partial_{I_{l}}\left(\frac{\text{e}^{-m_{1}r}}{r}\right),
\end{eqnarray}
where the $l$-pole scalar multipole moments $\hat{Q}_{I_{l}}$ are defined by
\begin{eqnarray}
\label{equ4.51}\hat{Q}_{I_{l}}(t)&=&\int\updelta_{l}(m_{1}r') \hat{X'}_{I_{l}}\frac{T^{00}(t,\boldsymbol{x}')}{c^2}\text{d}^{3}x'.
\end{eqnarray}

Compared with eq.~(\ref{equ4.35}), one will find that the expressions of the scalar multipole moments $\hat{Q}_{I_{l}}$ seem to be similar to those of the mass multipole moments $\hat{M}_{I_{l}}^{(\tilde{h})}$, but it can be proved that even at the leading pole order,
\begin{eqnarray}
\label{equ4.52}Q&:=&\hat{Q}_{I_{0}}(t)=\int\frac{\sinh{(m_{1}r')}}{m_{1}r'}\frac{T^{00}(t,\boldsymbol{x}')}{c^2}\text{d}^{3}x'\neq M,
\end{eqnarray}
and therefore, they are quite distinct.  Obviously,  all the scalar multipole moments are the spatial STF moments of the modulated source by a $l$-dependent radial function $\updelta_{l}(m_{1}r')$. As shown in the derivation, the radial function originates from the spherical modified Bessel function $\text{i}_{l}(m_{1}r')$  in the multipole expansion of the Green's function in eq.~(\ref{equ4.42}). According to ref.~\cite{Arfken1985},
\begin{eqnarray}
\label{equ4.53}\lim_{m_{1}\rightarrow0}\updelta_{l}(m_{1}r')=1
\end{eqnarray}
holds, and thus, it is understood that it is the massive parameter $m_{1}$ that leads to the appearance of the radial function in the scalar multipole moments. Additionally, from the result~(\ref{equ4.50}), we also find that because of the existence of the massive parameter $m_{1}$, the multipole expansion of the massive field $X^{(1)}$ bears the Yukawa-like dependence.  If one intends to derive the multipole expansion of a massive vector or tensor field up to $1/c^{3}$ order,  the above expansion for $X^{(1)}$ should be the starting point of the derivation, and therefore, it is concluded that ``the Yukawa-like dependence on massive parameter'' and ``source multipole moments are the spatial STF moments of the modulated source by the radial function'' are the two essential features of the multipole expansions of massive fields up to $1/c^3$ order.
\subsection{Multipole expansions of  $P^{\mu\nu}$ outside a spatially compact source up to $1/c^3$ order}

In this subsection, we will deal with $P^{\mu\nu}$, and from eqs.~(\ref{equ3.29}), (\ref{equ3.34}), and (\ref{equ3.35}), it satisfies
\begin{equation}\label{equ4.54}
\left\{\begin{array}{ll}
\displaystyle &\Delta P^{\mu\nu}-m_{2}^2 P^{\mu\nu}=-m_{2}^2\kappa S^{\mu\nu},\smallskip\\
\displaystyle &\partial_{\mu}P^{\mu\nu}=0,\smallskip\\
\displaystyle &\eta_{\mu\nu}P^{\mu\nu}=0
\end{array}\right.
\end{equation}
outside the source.  As a massive symmetric tensor field, one can expect that its multipole expansion up to $1/c^{3}$ order should possess some features of the expansions of $\tilde{h}^{\mu\nu}$ and $X^{(1)}$.
In view that the derivation is lengthy,  here we give a bullet point list to point out the individual steps that are taken in what follows.
\begin{itemize}
\item Providing the initial expansions of the components of $P^{\mu\nu}$;
\item Expressing $P^{0i}$ and $P^{ij}$ in the form of the STF-tensor spherical harmonics expansions;
\item Deducing the relations among the multipole tensor coefficients appearing in the STF-tensor spherical harmonics expansions of $P^{0i}$ and $P^{ij}$;
\item Deriving the expressions of $P^{0i}$ and $P^{ij}$ in terms of independent tensor coefficients;
\item Presenting the multipole expansion of  $P^{\mu\nu}$ outside a spatially compact source up to $1/c^3$ order.
\end{itemize}

As stated earlier,  the Green's function of the above screened Poisson equation is $\mathcal{G}_{2}(\boldsymbol{x};\boldsymbol{x}')$, and comparing it with $\mathcal{G}_{1}(\boldsymbol{x};\boldsymbol{x}')$, it is shown that the multipole expansion of  $\mathcal{G}_{2}(\boldsymbol{x};\boldsymbol{x}')$ is just that of $\mathcal{G}_{1}(\boldsymbol{x};\boldsymbol{x}')$ in eq.~(\ref{equ4.42}) with the replacement
of the massive parameter $m_{1}$ by $m_{2}$. As a consequence, the initial expansions of the components of $P^{\mu\nu}$ can be obtained by following the process from eq.~(\ref{equ4.44}) to eq.~(\ref{equ4.51}),
\begin{eqnarray}
\label{equ4.55}P^{00}(t,\boldsymbol{x})
&=&\frac{2Gm^{2}_{2}}{c^2}\sum_{l=0}^{\infty}\frac{(-1)^l}{l!}F^{00}_{\langle I_{l}\rangle}(t)\partial_{I_{l}}\left(\frac{\text{e}^{-m_{2}r}}{r}\right),\qquad \\
\label{equ4.56}P^{0i}(t,\boldsymbol{x})
&=&\frac{2Gm^{2}_{2}}{c^3}\sum_{l=0}^{\infty}\frac{(-1)^l}{l!}F^{0i}_{\langle I_{l}\rangle}(t)\partial_{I_{l}}\left(\frac{\text{e}^{-m_{2}r}}{r}\right),\qquad \\
\label{equ4.57}P^{ij}(t,\boldsymbol{x})
&=&\frac{2Gm^{2}_{2}}{c^2}\sum_{l=0}^{\infty}\frac{(-1)^l}{l!}F^{ij}_{\langle I_{l}\rangle}(t)\partial_{I_{l}}\left(\frac{\text{e}^{-m_{2}r}}{r}\right)\qquad
\end{eqnarray}
with
\begin{eqnarray}
\label{equ4.58}F^{00}_{\langle I_{l}\rangle}(t)&=&\int \updelta_{l}(m_{2}r')\hat{X'}_{I_{l}}\frac{S^{00}(t,\boldsymbol{x}')}{c^2}\text{d}^{3}x', \\
\label{equ4.59}F^{0i}_{\langle I_{l}\rangle}(t)&=&\int \updelta_{l}(m_{2}r')\hat{X'}_{I_{l}}\frac{S^{0i}(t,\boldsymbol{x}')}{c}\text{d}^{3}x', \\
\label{equ4.60}F^{ij}_{\langle I_{l}\rangle}(t)&=&\int \updelta_{l}(m_{2}r')\hat{X'}_{I_{l}}\frac{S^{ij}(t,\boldsymbol{x}')}{c^2}\text{d}^{3}x'.
\end{eqnarray}
The complete multipole expansion of $P^{\mu\nu}$ demands that the reducible tensors $F^{0i}_{\langle I_{l}\rangle}$ and $F^{ij}_{\langle I_{l}\rangle}$ should be decomposed into irreducible pieces by applying the STF formalism.
For convenience, let us define
\begin{eqnarray}
\label{equ4.61}\mathcal{A}^{(P)}_{\langle I_{l}\rangle}&:=&\frac{(-1)^l}{l!}F^{00}_{\langle I_{l}\rangle},\\
\label{equ4.62}\mathcal{U}^{(P)}_{i\langle I_{l}\rangle}&:=&\frac{(-1)^l}{l!}F^{0i}_{\langle I_{l}\rangle},\\
\label{equ4.63}\mathcal{V}^{(P)}_{ij\langle I_{l}\rangle}&:=&\frac{(-1)^l}{l!}F^{ij}_{\langle I_{l}\rangle},
\end{eqnarray}
and then, the initial expansions of the components of $P^{\mu\nu}$ are rewritten as
\begin{eqnarray}
\label{equ4.64}P^{00}
&=&\frac{2Gm^{2}_{2}}{c^2}\sum_{l=0}^{\infty}\mathcal{A}^{(P)}_{\langle I_{l}\rangle}\partial_{I_{l}}\bigg(\frac{\text{e}^{-m_{2}r}}{r}\bigg),\\
\label{equ4.65}P^{0i}
&=&\frac{2Gm^{2}_{2}}{c^3}\sum_{l=0}^{\infty}\mathcal{U}^{(P)}_{i\langle I_{l}\rangle}\partial_{I_{l}}\bigg(\frac{\text{e}^{-m_{2}r}}{r}\bigg),\\
\label{equ4.66}P^{ij}
&=&\frac{2Gm^{2}_{2}}{c^2}\sum_{l=0}^{\infty}\mathcal{V}^{(P)}_{ij\langle I_{l}\rangle}\partial_{I_{l}}\bigg(\frac{\text{e}^{-m_{2}r}}{r}\bigg).
\end{eqnarray}
In Appendix A, by repeatedly using formulas.~(\ref{equ2.7})--(\ref{equ2.10}) to decompose the reducible tensors $\mathcal{U}^{(P)}_{i\langle I_{l}\rangle}$ and $\mathcal{V}^{(P)}_{ij\langle I_{l}\rangle}$,
$P^{0i}$ and $P^{ij}$ are expressed in the form of  the STF-tensor spherical harmonics expansions,
\begin{eqnarray}
\label{equ4.67}P^{0i}&=&\frac{2Gm^{2}_{2}}{c^3}\left[\sum_{l=0}^{\infty}\hat{\mathcal{B}}^{(P)}_{I_{l}}\partial_{iI_{l}}\bigg(\frac{\text{e}^{-m_{2}r}}{r}\bigg)+\sum_{l=1}^{\infty}\hat{\mathcal{C}}^{(P)}_{iI_{l-1}}\partial_{I_{l-1}}\bigg(\frac{\text{e}^{-m_{2}r}}{r}\bigg)\right.\notag\\
&&\left.+\sum_{l=1}^{\infty}\epsilon_{iab}\hat{\mathcal{D}}^{(P)}_{bI_{l-1}}\partial_{aI_{l-1}}\bigg(\frac{\text{e}^{-m_{2}r}}{r}\bigg)\right],\\
\label{equ4.68}P^{ij}&=&\frac{2Gm^{2}_{2}}{c^2}\left[\sum_{l=0}^{\infty}\hat{\mathcal{E}}^{(P)}_{I_{l}}\partial_{ijI_{l}}\bigg(\frac{\text{e}^{-m_{2}r}}{r}\bigg)+\sum_{l=0}^{\infty}\hat{\mathcal{F}}^{(P)}_{I_{l}}\delta_{ij}\partial_{I_{l}}\bigg(\frac{\text{e}^{-m_{2}r}}{r}\bigg)\right.\notag\\
&&\left.+\sum_{l=1}^{\infty}\hat{\mathcal{G}}^{(P)}_{I_{l-1}(i}\partial_{j)I_{l-1}}\bigg(\frac{\text{e}^{-m_{2}r}}{r}\bigg)+\sum_{l=1}^{\infty}\hat{\mathcal{H}}^{(P)}_{bI_{l-1}}\epsilon_{ab(i}\partial_{j)aI_{l-1}}\bigg(\frac{\text{e}^{-m_{2}r}}{r}\bigg)\right.\notag\\
&&\left.+\sum_{l=2}^{\infty}\hat{\mathcal{I}}^{(P)}_{ijI_{l-2}}\partial_{I_{l-2}}\bigg(\frac{\text{e}^{-m_{2}r}}{r}\bigg)+\sum_{l=2}^{\infty}\epsilon_{ab(i}\hat{\mathcal{J}}^{(P)}_{j)bI_{l-2}}\partial_{aI_{l-2}}\bigg(\frac{\text{e}^{-m_{2}r}}{r}\bigg)\right],\qquad
\end{eqnarray}
where the multipole tensor coefficients $\hat{\mathcal{B}}^{(P)}_{I_{l}},\hat{\mathcal{C}}^{(P)}_{I_{l}},\cdots, \hat{\mathcal{J}}^{(P)}_{I_{l}}$, are defined by
\begin{eqnarray}
\label{equ4.69}\hat{\mathcal{B}}^{(P)}_{I_{l}}:&=&\frac{2l+1}{2l+3}\hat{R}^{(P-)}_{I_{l}},\\
\label{equ4.70}\hat{\mathcal{C}}^{(P)}_{I_{l}}:&=&\hat{R}^{(P+)}_{I_{l}}-\frac{m_{2}^2l}{2l+3}\hat{R}^{(P-)}_{I_{l}},\\
\label{equ4.71}\hat{\mathcal{D}}^{(P)}_{I_{l}}:&=&\frac{l}{l+1}\hat{R}^{(P0)}_{I_{l}},\\
\label{equ4.72}\hat{\mathcal{E}}^{(P)}_{I_{l}}:&=&\frac{2l+1}{2l+5}\hat{\mathcal{R}}^{(P--)}_{I_{l}},\\
\label{equ4.73}\hat{\mathcal{F}}^{(P)}_{I_{l}}:&=&\frac{2l+1}{(2l+3)(l+1)}\hat{\mathcal{R}}^{(P+-)}_{I_{l}}-\frac{l^2}{(l+1)^2}\hat{\mathcal{R}}^{(P00)}_{I_{l}}-\frac{(2l+1)m_{2}^2}{(2l+5)(2l+3)}\hat{\mathcal{R}}^{(P--)}_{I_{l}},\\
\label{equ4.74}\hat{\mathcal{G}}^{(P)}_{I_{l}}:&=&\frac{(2l-1)l}{(2l+3)(l+1)}\hat{\mathcal{R}}^{(P+-)}_{I_{l}}+\frac{l(2l-1)}{(l+1)^2}\hat{\mathcal{R}}^{(P00)}_{I_{l}}+\frac{2l-1}{2l+1}\hat{\mathcal{R}}^{(P-+)}_{I_{l}}\notag\\
&&-\frac{2l(2l+1)m_{2}^2}{(2l+5)(2l+3)}\hat{\mathcal{R}}^{(P--)}_{I_{l}},\\
\label{equ4.75}\hat{\mathcal{H}}^{(P)}_{I_{l}}:&=&\frac{(2l+1)l}{(2l+3)(l+2)}\hat{\mathcal{R}}^{(P0-)}_{I_{l}}+\frac{l(2l+1)}{(2l+3)(l+1)}\hat{\mathcal{R}}^{(P-0)}_{I_{l}},\\
\label{equ4.76}\hat{\mathcal{I}}^{(P)}_{I_{l}}:&=&\hat{\mathcal{R}}^{(P++)}_{I_{l}}-\frac{l(l-1)m_{2}^{2}}{(2l+3)(l+1)}\hat{\mathcal{R}}^{(P+-)}_{I_{l}}-\frac{l(l-1)m_{2}^{2}}{(l+1)^2}\hat{\mathcal{R}}^{(P00)}_{I_{l}}\notag\\
&&-\frac{(l-1)m_{2}^{2}}{2l+1}\hat{\mathcal{R}}^{(P-+)}_{I_{l}}+\frac{l(l-1)m_{2}^4}{(2l+5)(2l+3)}\hat{\mathcal{R}}^{(P--)}_{I_{l}},\qquad\\
\label{equ4.77}\hat{\mathcal{J}}^{(P)}_{I_{l}}:&=&\frac{l-1}{l+1}\hat{\mathcal{R}}^{(P+0)}_{I_{l}}+\frac{l-1}{l}\hat{\mathcal{R}}^{(P0+)}_{I_{l}}-\frac{l(l-1)m_{2}^{2}}{(l+2)(2l+3)}\hat{\mathcal{R}}^{(P0-)}_{I_{l}}\notag\\
&&-\frac{l(l-1)m_{2}^2}{(2l+3)(l+1)}\hat{\mathcal{R}}^{(P-0)}_{I_{l}}.
\end{eqnarray}
The expressions of the STF tensors $\hat{R}^{(P+)}_{I_{l}}$, $\hat{R}^{(P0)}_{I_{l}}$, $\hat{R}^{(P-)}_{I_{l}}$,
$\hat{\mathcal{R}}^{(P++)}_{I_{l}}$, $\hat{\mathcal{R}}^{(P+0)}_{I_{l}}$, $\hat{\mathcal{R}}^{(P+-)}_{I_{l}}$,
$\hat{\mathcal{R}}^{(P0+)}_{I_{l}}$, $\hat{\mathcal{R}}^{(P00)}_{I_{l}}$, $\hat{\mathcal{R}}^{(P0-)}_{I_{l}}$,
$\hat{\mathcal{R}}^{(P-+)}_{I_{l}}$, $\hat{\mathcal{R}}^{(P-0)}_{I_{l}}$, and $\hat{\mathcal{R}}^{(P--)}_{I_{l}}$, given in terms of $\mathcal{U}^{(P)}_{i\langle I_{l}\rangle}$ and $\mathcal{V}^{(P)}_{ij\langle I_{l}\rangle}$, can be deduced from eqs.~(\ref{equA2})--(\ref{equA4}), (\ref{equA6})--(\ref{equA8}), and (\ref{equA12})--(\ref{equA20}). A obvious distinction between $P^{\mu\nu}$ and $\tilde{h}^{\mu\nu}$ is that up to $1/c^3$ order, $P^{ij}\neq0$ but $\tilde{h}^{ij}=0$, which originates from the fact that  $\kappa T^{ij}\sim O\big(1/c^4\big)$ and $\kappa S^{ij}\sim O\big(1/c^2\big)$. In addition, a glance at eqs.~(\ref{equ4.69})--(\ref{equ4.77}) reveals that the massive parameter $m_{2}$ appears in the definitions of the multipole tensor coefficients. These two conclusions imply that compared with the case of $\tilde{h}^{\mu\nu}$, more complicated multipole moments will emerge in the multipole expansion of $P^{\mu\nu}$.

We are now in a position to handle last two equations in~(\ref{equ4.54}), from which, based on eqs.~(\ref{equ3.22}), (\ref{equ3.24}), (\ref{equ4.64}), (\ref{equ4.67}), and (\ref{equ4.68}),  three identities fulfilled by the components of $P^{\mu\nu}$ up to $1/c^3$ order are supplied,
\begin{eqnarray}
\label{equ4.78}&&\frac{1}{c}\partial_{t}P^{00}+\partial_{i}P^{i0}=0, \\
\label{equ4.79}&&\partial_{i}P^{ij}=0,\\
\label{equ4.80}&&-P^{00}+P^{ii}=0.
\end{eqnarray}
Due to them, not all of the multipole tensor coefficients $\hat{\mathcal{A}}^{(P)}_{I_{l}},\hat{\mathcal{B}}^{(P)}_{I_{l}},\cdots, \hat{\mathcal{J}}^{(P)}_{I_{l}}$ are independent, and in what follows, we will deduce
the relations among them.  Plugging eqs.~(\ref{equ4.64}),  (\ref{equ4.67}), and (\ref{equ4.68})  into the above three identities yields
\begin{eqnarray}
\label{equ4.81}&&\partial_{t}\hat{\mathcal{A}}^{(P)}+m_{2}^2\hat{\mathcal{B}}^{(P)}
+\sum_{l=1}^{\infty}\left(\partial_{t}\hat{\mathcal{A}}^{(P)}_{I_{l}}+m_{2}^2\hat{\mathcal{B}}^{(P)}_{I_{l}}+\hat{\mathcal{C}}^{(P)}_{I_{l}}\right)\partial_{I_{l}}\bigg(\frac{\text{e}^{-m_{2}r}}{r}\bigg)=0,\\
\label{equ4.82}&&m_{2}^{2}\frac{\hat{\mathcal{G}}^{(P)}_{j}}{2}\frac{\text{e}^{-m_{2}r}}{r}+\frac{m_{2}^{2}}{3}\left(m_{2}^{2}\hat{\mathcal{E}}^{(P)}_{j}+\hat{\mathcal{F}}^{(P)}_{j}+\frac{\hat{\mathcal{G}}^{(P)}_{j}}{2}\right)\frac{\text{e}^{-m_{2}r}}{r}+\left(m_{2}^{2}\hat{\mathcal{E}}^{(P)}+\hat{\mathcal{F}}^{(P)}\right)\partial_{j}\bigg(\frac{\text{e}^{-m_{2}r}}{r}\bigg)\notag\\
&&+m_{2}^{2}\frac{\hat{\mathcal{H}}^{(P)}_{b}}{2}\epsilon_{abj}\partial_{a}\bigg(\frac{\text{e}^{-m_{2}r}}{r}\bigg)+\left(m_{2}^{2}\frac{\hat{\mathcal{G}}^{(P)}_{ji}}{2}+\hat{\mathcal{I}}^{(P)}_{ji}\right)\partial_{i}\bigg(\frac{\text{e}^{-m_{2}r}}{r}\bigg)\notag\\
&&+\frac{2m_{2}^{2}}{5}\left(m_{2}^{2}\hat{\mathcal{E}}^{(P)}_{ji}+\hat{\mathcal{F}}^{(P)}_{ji}+\frac{\hat{\mathcal{G}}^{(P)}_{ji}}{2}\right)\partial_{i}\bigg(\frac{\text{e}^{-m_{2}r}}{r}\bigg)+\sum_{l=2}^{\infty}\left(m_{2}^{2}\frac{\hat{\mathcal{G}}^{(P)}_{jI_{l}}}{2}+\hat{\mathcal{I}}^{(P)}_{jI_{l}}\right)\hat{\partial}_{I_{l}}\bigg(\frac{\text{e}^{-m_{2}r}}{r}\bigg)\notag\\
&&+\sum_{l=2}^{\infty}\left(m_{2}^{2}\hat{\mathcal{E}}^{(P)}_{I_{l-1}}+\hat{\mathcal{F}}^{(P)}_{I_{l-1}}+\frac{\hat{\mathcal{G}}^{(P)}_{I_{l-1}}}{2}\right)\hat{\partial}_{jI_{l-1}}\bigg(\frac{\text{e}^{-m_{2}r}}{r}\bigg)\notag\\
&&+\sum_{l=2}^{\infty}\frac{m_{2}^2(l+1)}{2l+3}\left(m_{2}^{2}\hat{\mathcal{E}}^{(P)}_{jI_{l}}+\hat{\mathcal{F}}^{(P)}_{jI_{l}}+\frac{\hat{\mathcal{G}}^{(P)}_{jI_{l}}}{2}\right)\hat{\partial}_{I_{l}}\bigg(\frac{\text{e}^{-m_{2}r}}{r}\bigg)\notag\\
&&+\sum_{l=2}^{\infty}\frac{1}{2}\left(m_{2}^{2}\hat{\mathcal{H}}^{(P)}_{bI_{l-1}}+\hat{\mathcal{J}}^{(P)}_{bI_{l-1}}\right)\epsilon_{abj}\partial_{aI_{l-1}}\bigg(\frac{\text{e}^{-m_{2}r}}{r}\bigg)=0,\\
\label{equ4.83}&&\left(-\hat{\mathcal{A}}^{(P)}+m_{2}^2\hat{\mathcal{E}}^{(P)}+3\hat{\mathcal{F}}^{(P)}\right)
+\sum_{l=1}^{\infty}\left(-\hat{\mathcal{A}}^{(P)}_{I_{l}}+m_{2}^{2}\hat{\mathcal{E}}^{(P)}_{I_{l}}+3\hat{\mathcal{F}}^{(P)}_{I_{l}}+\hat{\mathcal{G}}^{(P)}_{I_{l}}\right)\partial_{I_{l}}\bigg(\frac{\text{e}^{-m_{2}r}}{r}\bigg)=0,\notag\\
\end{eqnarray}
where in the derivation of eq.~(\ref{equ4.82}), we have used
\begin{eqnarray}
\label{equ4.84}&&\partial_{i\langle I_{l}\rangle}:=\partial_{i}\hat{\partial}_{I_{l}}=\hat{\partial}_{iI_{l}}+\frac{l}{2l+1}\delta_{i\langle i_{l}}\hat{\partial}_{i_{1}\cdots i_{l-1}\rangle}\Delta
\end{eqnarray}
whose proof is the same as that of the formula~(\ref{equ2.11}) in the STF formalism. Furthermore, with the help of formulas~(\ref{equ2.12})--(\ref{equ2.14}),  the formal relations among $\hat{\mathcal{A}}^{(P)}_{I_{l}},\hat{\mathcal{B}}^{(P)}_{I_{l}},\cdots, \hat{\mathcal{J}}^{(P)}_{I_{l}}$ are inferred from eqs.~(\ref{equ4.81})--(\ref{equ4.83}),
\begin{eqnarray}
\label{equ4.85}&&\hat{\mathcal{B}}^{(P)}=-\frac{1}{m_{2}^2}\partial_{t}\hat{\mathcal{A}}^{(P)},\\
\label{equ4.86}&&\hat{\mathcal{C}}^{(P)}_{I_{l}}=-\partial_{t}\hat{\mathcal{A}}^{(P)}_{I_{l}}-m_{2}^2\hat{\mathcal{B}}^{(P)}_{I_{l}},\quad l\geqslant1,\\
\label{equ4.87}&&\hat{\mathcal{E}}^{(P)}=-\frac{1}{2m_{2}^{2}}\hat{\mathcal{A}}^{(P)},\quad \hat{\mathcal{F}}^{(P)}=\frac{1}{2}\hat{\mathcal{A}}^{(P)},\quad  \hat{\mathcal{G}}^{(P)}_{i}=0,\quad \hat{\mathcal{H}}^{(P)}_{i}=0,\\
\label{equ4.88}&&\hat{\mathcal{E}}^{(P)}_{I_{l}}=-\frac{1}{2m_{2}^{2}}\hat{\mathcal{A}}^{(P)}_{I_{l}}-\frac{1}{4m_{2}^2}\hat{\mathcal{G}}^{(P)}_{I_{l}},\quad l\geqslant1,\\
\label{equ4.89}&&\hat{\mathcal{F}}^{(P)}_{I_{l}}=\frac{1}{2}\hat{\mathcal{A}}^{(P)}_{I_{l}}-\frac{1}{4}\hat{\mathcal{G}}^{(P)}_{I_{l}},\quad l\geqslant1,\\
\label{equ4.90}&&\hat{\mathcal{I}}^{(P)}_{I_{l}}=-\frac{m_{2}^{2}}{2}\hat{\mathcal{G}}^{(P)}_{I_{l}},\quad l\geqslant 2,\\
\label{equ4.91}&&\hat{\mathcal{J}}^{(P)}_{I_{l}}=-m_{2}^{2}\hat{\mathcal{H}}^{(P)}_{I_{l}},\quad l\geqslant 2.
\end{eqnarray}
After replacing these relations into eqs.~(\ref{equ4.67}) and (\ref{equ4.68}), explicit expressions of $P^{0i}$ and $P^{ij}$  in terms of independent tensor coefficients $\hat{\mathcal{A}}^{(P)}_{I_{l}},\hat{\mathcal{B}}^{(P)}_{I_{l}}, \hat{\mathcal{D}}^{(P)}_{I_{l}},  \hat{\mathcal{G}}^{(P)}_{I_{l}}$, and $ \hat{\mathcal{H}}^{(P)}_{I_{l}}$ are achieved,
\begin{eqnarray}
\label{equ4.92}P^{0i}&=&\frac{2Gm^{2}_{2}}{c^3}\left[-\sum_{l=1}^{\infty}\partial_{t}\hat{\mathcal{A}}^{(P)}_{iI_{l-1}}\partial_{I_{l-1}}\left(\frac{\text{e}^{-m_{2}r}}{r}\right)+\sum_{l=1}^{\infty}\epsilon_{iab}\hat{\mathcal{D}}^{(P)}_{bI_{l-1}}\partial_{aI_{l-1}}\bigg(\frac{\text{e}^{-m_{2}r}}{r}\bigg)\right.\nonumber\\
&&\left.+\sum_{l=0}^{\infty}\hat{\mathcal{B}}^{(P)}_{I_{l}}\partial_{iI_{l}}\bigg(\frac{\text{e}^{-m_{2}r}}{r}\bigg)
-m^{2}_{2}\sum_{l=1}^{\infty}\hat{\mathcal{B}}^{(P)}_{iI_{l-1}}\partial_{I_{l-1}}\bigg(\frac{\text{e}^{-m_{2}r}}{r}\bigg)\right],\\
P^{ij}
&=&\frac{2Gm^{2}_{2}}{c^2}\bigg\{\frac{1}{2}\sum_{l=0}^{\infty}\hat{\mathcal{A}}^{(P)}_{I_{l}}\left[\delta_{ij}\partial_{I_{l}}\left(\frac{\text{e}^{-m_{2}r}}{r}\right)-\frac{1}{m^{2}_{2}}\partial_{ij I_{l}}\bigg(\frac{\text{e}^{-m_{2}r}}{r}\bigg)\right]\notag\\
&&-\frac{1}{4}\sum_{l=2}^{\infty}\hat{\mathcal{G}}^{(P)}_{I_{l} }\left[\delta_{ij}\partial_{I_{l}}\bigg(\frac{\text{e}^{-m_{2}r}}{r}\bigg)+\frac{1}{m^{2}_{2}}\partial_{ij I_{l}}\bigg(\frac{\text{e}^{-m_{2}r}}{r}\bigg)\right]\notag\\
\label{equ4.93}
&&+\sum_{l=2}^{\infty}\left[\hat{\mathcal{G}}^{(P)}_{I_{l-1}(i}\partial_{j)I_{l-1}}\bigg(\frac{\text{e}^{-m_{2}r}}{r}\bigg)
-\frac{m_{2}^{2}}{2}\hat{\mathcal{G}}^{(P)}_{ijI_{l-2}}\partial_{I_{l-2}}\bigg(\frac{\text{e}^{-m_{2}r}}{r}\bigg)\right]\notag\\
&&+\sum_{l=2}^{\infty}\left[\hat{\mathcal{H}}^{(P)}_{bI_{l-1}}\epsilon_{ab(i}\partial_{j)aI_{l-1}}\bigg(\frac{\text{e}^{-m_{2}r}}{r}\bigg)-m_{2}^2\epsilon_{ab(i}\hat{\mathcal{H}}^{(P)}_{j)bI_{l-2}}\partial_{aI_{l-2}}\bigg(\frac{\text{e}^{-m_{2}r}}{r}\bigg)\right]\bigg\}.\qquad
\end{eqnarray}
Similarly to the situations of $\tilde{h}^{\mu\nu}$ and $X^{(1)}$, in order to keep the source multipole moments of $P^{\mu\nu}$ invariant under the parity transformation $x'_{i}=-x_{i}$, we define them as
\begin{eqnarray}
\label{equ4.94}&&\hat{M}^{(P)}_{I_{l}}=(-1)^{l}l!\hat{\mathcal{A}}^{(P)}_{I_{l}},\\
\label{equ4.95}&&\hat{S}^{(P)}_{I_{l}}=\frac{(-1)^{l+1}(l+1)!}{l}\hat{\mathcal{D}}^{(P)}_{I_{l}},\\
\label{equ4.96}&&\hat{B}^{(P)}_{I_{l}}=(-1)^{l+1}(l+1)!\hat{\mathcal{B}}^{(P)}_{I_{l}},\\
\label{equ4.97}&&\hat{G}^{(P)}_{I_{l}}=(-1)^{l+1}l!\hat{\mathcal{G}}^{(P)}_{I_{l}},\\
\label{equ4.98}&&\hat{H}^{(P)}_{I_{l}}=(-1)^{l+1}(l+1)!\hat{\mathcal{H}}^{(P)}_{I_{l}},
\end{eqnarray}
and then, the substitution of eqs.~(\ref{equ4.94})--(\ref{equ4.98}) in eqs.~(\ref{equ4.64}), (\ref{equ4.92}), and (\ref{equ4.93}) leads to the multipole expansion of $P^{\mu\nu}$ up to $1/c^3$ order outside the source, namely,
\begin{eqnarray}
\label{equ4.99}P^{00}(t,\boldsymbol{x})&=& \frac{2Gm^{2}_{2}}{c^2}\sum_{l=0}^{\infty}\frac{(-1)^l}{l!}\hat{M}^{(P)}_{I_{l}}(t)\partial_{I_{l}}\left(\frac{\text{e}^{-m_{2}r}}{r}\right),\\
\label{equ4.100}P^{0i}(t,\boldsymbol{x})&=& \frac{2Gm^{2}_{2}}{c^3}\left[-\sum_{l=1}^{\infty}\frac{(-1)^l}{l!}\left(\partial_{t}\hat{M}^{(P)}_{iI_{l-1}}(t)\right)\partial_{I_{l-1}}\left(\frac{\text{e}^{-m_{2}r}}{r}\right)\right.\notag\\
&&+\sum_{l=1}^{\infty}\frac{(-1)^{l}l}{(l+1)!}\epsilon_{iab}\hat{S}^{(P)}_{aI_{l-1}}(t)\partial_{bI_{l-1}}\bigg(\frac{\text{e}^{-m_{2}r}}{r}\bigg) \notag\\
&&\left.+\sum_{l=0}^{\infty}\frac{(-1)^{l+1}}{(l+1)!}\hat{B}^{(P)}_{I_{l}}(t)\partial_{iI_{l}}\bigg(\frac{\text{e}^{-m_{2}r}}{r}\bigg)\right.\notag\\
&&\left.-m^{2}_{2}\sum_{l=1}^{\infty}\frac{(-1)^{l+1}}{(l+1)!}\hat{B}^{(P)}_{iI_{l-1}}(t)\partial_{I_{l-1}}\bigg(\frac{\text{e}^{-m_{2}r}}{r}\bigg)\right],\\
P^{ij}(t,\boldsymbol{x})&=& \frac{2Gm^{2}_{2}}{c^2}\bigg\{\frac{1}{2}\sum_{l=0}^{\infty}\frac{(-1)^{l}}{l!}\hat{M}^{(P)}_{I_{l}}(t)\left[\delta_{ij}\partial_{I_{l}}\left(\frac{\text{e}^{-m_{2}r}}{r}\right)-\frac{1}{m^{2}_{2}}\partial_{ij I_{l}}\bigg(\frac{\text{e}^{-m_{2}r}}{r}\bigg)\right]\notag\\
&&+\frac{1}{4}\sum_{l=2}^{\infty}\frac{(-1)^{l}}{l!}\hat{G}^{(P)}_{I_{l} }(t)\left[\delta_{ij}\partial_{I_{l}}\bigg(\frac{\text{e}^{-m_{2}r}}{r}\bigg)+\frac{1}{m^{2}_{2}}\partial_{ij I_{l}}\bigg(\frac{\text{e}^{-m_{2}r}}{r}\bigg)\right]\notag\\
&& -\sum_{l=2}^{\infty}\frac{(-1)^{l}}{l!}\left[\hat{G}^{(P)}_{I_{l-1}(i}(t)\partial_{j)I_{l-1}}\bigg(\frac{\text{e}^{-m_{2}r}}{r}\bigg)
-\frac{m_{2}^{2}}{2}\hat{G}^{(P)}_{ijI_{l-2}}(t)\partial_{I_{l-2}}\bigg(\frac{\text{e}^{-m_{2}r}}{r}\bigg)\right]\notag\\
&& +\sum_{l=2}^{\infty}\frac{(-1)^{l+1}}{(l+1)!}\left[\hat{H}^{(P)}_{bI_{l-1}}(t)\epsilon_{ab(i}\partial_{j)aI_{l-1}}\bigg(\frac{\text{e}^{-m_{2}r}}{r}\bigg)\right.\notag\\
\label{equ4.101}&&\phantom{\sum_{l=2}^{\infty}\frac{(-1)^{l+1}}{(l+1)!}[\ \ \ }\left.-m_{2}^2\epsilon_{ab(i}\hat{H}^{(P)}_{j)bI_{l-2}}(t)\partial_{aI_{l-2}}\bigg(\frac{\text{e}^{-m_{2}r}}{r}\bigg)\right]\bigg\},
\end{eqnarray}
where after tedious calculations, the expressions of the source multipole moments in terms of the source $S^{\mu\nu}$ are derived from eqs.~(\ref{equ4.58})--(\ref{equ4.63}), (\ref{equ4.69})--(\ref{equ4.77}), (\ref{equA2})--(\ref{equA20}), and (\ref{equ4.94})--(\ref{equ4.98}),
\begin{eqnarray}
\label{equ4.102}&&\hat{M}^{(P)}_{I_{l}}(t)=\int \updelta_{l}(m_{2}r')\hat{X'}_{I_{l}}\frac{S^{00}(t,\boldsymbol{x}')}{c^2}\text{d}^{3}x',\quad l\geqslant0,\\
\label{equ4.103}&&\hat{S}^{(P)}_{I_{l}}(t)=\int\updelta_{l}(m_{2}r')\epsilon_{pq\langle i_{1}}\hat{X'}_{i_{2}\cdots i_{l}\rangle p}\frac{S^{0q}(t,\boldsymbol{x}')}{c}\text{d}^{3}x',\quad l\geqslant1,\\
\label{equ4.104}&&\hat{B}^{(P)}_{I_{l}}(t)=\frac{2l+1}{2l+3}\int \updelta_{l+1}(m_{2}r')\hat{X'}_{aI_{l}}\frac{S^{0a}(t,\boldsymbol{x}')}{c}\text{d}^{3}x',\quad l\geqslant0,\\
\label{equ4.105}&&\hat{G}^{(P)}_{I_{l}}(t)=2\int \updelta_{l}(m_{2}r')\hat{X'}_{I_{l}}\frac{S^{aa}(t,\boldsymbol{x}')}{c^2}\text{d}^{3}x'\notag\\
&&\phantom{\hat{G}^{(P)}_{I_{l}}(t)=}+\frac{4(2l+1)m_{2}^{2}}{(l+1)(l+2)(2l+5)}\int \updelta_{l+2}(m_{2}r')\hat{X'}_{abI_{l}}\frac{S^{ab}(t,\boldsymbol{x}')}{c^2}\text{d}^{3}x',\quad l\geqslant2,\qquad \\
\label{equ4.106}&&\hat{H}^{(P)}_{I_{l}}(t)=\frac{2l(2l+1)}{(l+2)(2l+3)}\int\updelta_{l+1}(m_{2}r')\epsilon_{ab\langle i_{1}}\hat{X'}_{i_{2}\cdots i_{l}\rangle bc}\frac{S^{ac}(t,\boldsymbol{x}')}{c^2}\text{d}^{3}x',\quad l\geqslant2.
\end{eqnarray}
The above results display that the expansion of $P^{\mu\nu}$ presents the Yukawa-like dependence on massive parameter $m_{2}$ and that all the source multipole moments are the spatial STF moments of the modulated source by the radial function $\updelta_{l}(m_{2}r')$, which is consistent with the two essential features of the multipole expansions of massive fields up to $1/c^3$ order mentioned earlier. In particular, the fact that the expansion of $P^{00}$ is almost identical to that of $X^{(1)}$ indicates that these two fields share the same massive nature. In addition, as a tensor field, the expansion of $P^{\mu\nu}$ also has some similarity to that of $\tilde{h}^{\mu\nu}$. For them, if the space-space components are not taken into account, their expansions are similar. Comparing eqs.~(\ref{equ4.64}) and (\ref{equ4.67}) with eqs.~(\ref{equ4.7}) and (\ref{equ4.14}), respectively, we find that the reason is that at the beginning of the derivation, the time-time and time-space components of $P^{\mu\nu}$ and $\tilde{h}^{\mu\nu}$ are expressed in terms of four multipole tensor coefficients. However, since these two fields possess different physical nature, $\hat{\mathcal{B}}^{(\tilde{h})}_{I_{l}}$ in the expansion of $\tilde{h}^{0i}$ can be transformed away by employing the residual gauge freedom in the gauge condition~(\ref{equ3.6}), whereas $\hat{\mathcal{B}}^{(P)}_{I_{l}}$ in the expansion of $P^{0i}$ can not be eliminated in the same way. Thus, in the above expansions of $P^{00}$ and $P^{0i}$, it could be recognized that $\hat{M}^{(P)}_{I_{l}}$ and $\hat{S}^{(P)}_{I_{l}}$ should be the counterparts of the mass and spin multipole moments for $\tilde{h}^{\mu\nu}$, and $\hat{B}^{(P)}_{I_{l}}$  constitute the additional set of multipole moments encoding the expansion of $P^{0i}$. Finally, it needs to be emphasized that because another two sets of  multipole moments $\hat{G}^{(P)}_{I_{l}}$ and $\hat{H}^{(P)}_{I_{l}}$ appear in the expansion of $P^{ij}$, compared with the massless tensor field $\tilde{h}^{\mu\nu}$, there are a total of five sets of multipole moments encoding the expansion of the massive tensor field $P^{\mu\nu}$.
\subsection{Multipole expansion of the metric outside a spatially compact source up to $1/c^3$ order}
Now, the problem posed at the beginning of this section has been solved outside the source. That is to say, since the
conditions~(\ref{equ3.6}), (\ref{equ3.34}), and (\ref{equ3.35}) have been used to reduce the redundant tensor coefficients in the multipole expansions of $\tilde{h} ^{\mu\nu}$ and $P^{\mu\nu}$ outside the source, they do not need to be considered again when the external metric is derived based on the multipole expansions of $\tilde{h} ^{\mu\nu}$, $X^{(1)}$, and $P^{\mu\nu}$. As the final task of this section, we will deduce the multipole expansion of the metric for the external gravitational field of a spatially compact source up to $1/c^3$ order.

According to eqs.~(\ref{equ3.16})--(\ref{equ3.18}) in the weak-field approximation framework of $F(X,Y,Z)$ gravity,
the multipole expansion of the metric outside a spatially compact source up to $1/c^3$ order is obtained only after
the corresponding expansion for the gravitational field amplitude $h^{\mu\nu}$ is given.
With eqs.~(\ref{equ4.32})--(\ref{equ4.34}), (\ref{equ4.50}), and (\ref{equ4.99})--(\ref{equ4.101}),  the expansion of the gravitational field amplitude $h^{\mu\nu}$ can immediately be obtained from eq.~(\ref{equ3.16}),
\begin{eqnarray}
\label{equ4.107}h^{00}(t,\boldsymbol{x})&=& -\frac{4G}{c^2}\sum_{l=0}^{\infty}\frac{(-1)^{l}}{l!}\hat{M}_{I_{l}}^{(\tilde{h})}(t)\partial_{I_{l}}\left( \frac{1}{r}\right)\notag\\
&&+\frac{2G}{3c^2}\left(1+\frac{m_{1}^2}{m_{2}^{2}}\right)\sum_{l=0}^{\infty}\frac{(-1)^l}{l!}\hat{Q}_{I_{l}}(t)\partial_{I_{l}}\left(\frac{\text{e}^{-m_{1}r}}{r}\right)\notag\\
&&+\frac{4G}{c^2}\sum_{l=0}^{\infty}\frac{(-1)^l}{l!}\hat{M}^{(P)}_{I_{l}}(t)\partial_{I_{l}}\left(\frac{\text{e}^{-m_{2}r}}{r}\right),\\
\label{equ4.108}h^{0i}(t,\boldsymbol{x})&=& \frac{4G}{c^3}\sum_{l=1}^{\infty}\frac{(-1)^{l}}{l!}\left(\partial_{t}\hat{M}_{iI_{l-1}}^{(\tilde{h})}(t)\right)\partial_{I_{l-1}}\left(\frac{1}{r}\right)\notag
\end{eqnarray}
\begin{eqnarray}
&&-\frac{4G}{c^3}\sum_{l=1}^{\infty}\frac{(-1)^{l}l}{(l+1)!}\epsilon_{iab}\hat{S}_{aI_{l-1}}^{(\tilde{h})}(t)\partial_{bI_{l-1}}\left( \frac{1}{r}\right)\notag\\ &&-\frac{4G}{3m_{2}^{2}c^{3}}\sum_{l=0}^{\infty}\frac{(-1)^l}{l!}\left(\partial_{t}\hat{Q}_{I_{l}}(t)\right)\partial_{iI_{l}}\left(\frac{\text{e}^{-m_{1}r}}{r}\right)\notag\\
&&-\frac{4G}{c^3}\sum_{l=1}^{\infty}\frac{(-1)^l}{l!}\left(\partial_{t}\hat{M}^{(P)}_{iI_{l-1}}(t)\right)\partial_{I_{l-1}}\left(\frac{\text{e}^{-m_{2}r}}{r}\right)\notag\\
&&+\frac{4G}{c^3}\sum_{l=1}^{\infty}\frac{(-1)^{l}l}{(l+1)!}\epsilon_{iab}\hat{S}^{(P)}_{aI_{l-1}}(t)\partial_{bI_{l-1}}\bigg(\frac{\text{e}^{-m_{2}r}}{r}\bigg)\notag\\
&&+\frac{4G}{c^3}\sum_{l=0}^{\infty}\frac{(-1)^{l+1}}{(l+1)!}\hat{B}^{(P)}_{I_{l}}(t)\partial_{iI_{l}}\bigg(\frac{\text{e}^{-m_{2}r}}{r}\bigg)\notag\\
&&\left.
-\frac{4Gm^{2}_{2}}{c^3}\sum_{l=1}^{\infty}\frac{(-1)^{l+1}}{(l+1)!}\hat{B}^{(P)}_{iI_{l-1}}(t)\partial_{I_{l-1}}\bigg(\frac{\text{e}^{-m_{2}r}}{r}\bigg)\right],\\
\label{equ4.109}h^{ij}(t,\boldsymbol{x})&=&
-\delta_{ij}\frac{2G}{3c^{2}}\left(1+\frac{m_{1}^2}{m_{2}^{2}}\right)\sum_{l=0}^{\infty}\frac{(-1)^l}{l!}\hat{Q}_{I_{l}}(t)\partial_{I_{l}}\left(\frac{\text{e}^{-m_{1}r}}{r}\right)\notag\\
&&+\frac{4G}{3m_{2}^{2}c^{2}}\sum_{l=0}^{\infty}\frac{(-1)^l}{l!}\hat{Q}_{I_{l}}(t)\partial_{ijI_{l}}\left(\frac{\text{e}^{-m_{1}r}}{r}\right)+\frac{2}{m_{2}^{2}}P^{ij}.
\end{eqnarray}
Then, if we let $h$ and $\tilde{h}$ be the traces of $h^{\mu\nu}$ and $\tilde{h}^{\mu\nu}$ in Minkowski spacetime, respectively, there is
\begin{eqnarray}\label{equ4.110}
h(t,\boldsymbol{x})&=&\tilde{h}(t,\boldsymbol{x})-\frac{2}{3}\left(\frac{2}{m_{1}^{2}}+\frac{1}{m_{2}^{2}}\right)X^{(1)}(t,\boldsymbol{x})\notag\\
&=&\frac{4G}{c^2}\sum_{l=0}^{\infty}\frac{(-1)^{l}}{l!}\hat{M}_{I_{l}}^{(\tilde{h})}(t)\partial_{I_{l}}\left( \frac{1}{r}\right)\notag\\
&&-\frac{4G}{3c^2}\left(2+\frac{m_{1}^2}{m_{2}^{2}}\right)\sum_{l=0}^{\infty}\frac{(-1)^l}{l!}\hat{Q}_{I_{l}}(t)\partial_{I_{l}}\left(\frac{\text{e}^{-m_{1}r}}{r}\right).
\end{eqnarray}
By substituting the above results in eqs.~(\ref{equ3.17}) and (\ref{equ3.18}), the metric, presented in the form of eq.~(\ref{equ3.22}),  for the external gravitational field of the source is gained, where the multipole expansions of the potentials $U(t,\boldsymbol{x})$, $U_{i}(t,\boldsymbol{x})$, and $U_{ij}(t,\boldsymbol{x})$ are, respectively,
\begin{eqnarray}
\label{equ4.111}U(t,\boldsymbol{x})&=& G\sum_{l=0}^{\infty}\frac{(-1)^{l}}{l!}\hat{M}_{I_{l}}^{(\tilde{h})}(t)\partial_{I_{l}}\left( \frac{1}{r}\right)+\frac{G}{3}\sum_{l=0}^{\infty}\frac{(-1)^l}{l!}\hat{Q}_{I_{l}}(t)\partial_{I_{l}}\left(\frac{\text{e}^{-m_{1}r}}{r}\right)\notag\\
&&-2G\sum_{l=0}^{\infty}\frac{(-1)^l}{l!}\hat{M}^{(P)}_{I_{l}}(t)\partial_{I_{l}}\left(\frac{\text{e}^{-m_{2}r}}{r}\right),\\
\label{equ4.112}U_{i}(t,\boldsymbol{x})&=& -G\sum_{l=1}^{\infty}\frac{(-1)^{l}}{l!}\left(\partial_{t}\hat{M}_{iI_{l-1}}^{(\tilde{h})}(t)\right)\partial_{I_{l-1}}\left(\frac{1}{r}\right) \notag\\ &&+G\sum_{l=1}^{\infty}\frac{(-1)^{l}l}{(l+1)!}\epsilon_{iab}\hat{S}_{aI_{l-1}}^{(\tilde{h})}(t)\partial_{bI_{l-1}}\left( \frac{1}{r}\right)\notag
\end{eqnarray}
\begin{eqnarray}
&&+\frac{G}{3m_{2}^{2}}\sum_{l=0}^{\infty}\frac{(-1)^l}{l!}\left(\partial_{t}\hat{Q}_{I_{l}}(t)\right)\partial_{iI_{l}}\left(\frac{\text{e}^{-m_{1}r}}{r}\right)\notag\\
&&+G\sum_{l=1}^{\infty}\frac{(-1)^l}{l!}\left(\partial_{t}\hat{M}^{(P)}_{iI_{l-1}}(t)\right)\partial_{I_{l-1}}\left(\frac{\text{e}^{-m_{2}r}}{r}\right)\notag\\
&&-G\sum_{l=1}^{\infty}\frac{(-1)^{l}l}{(l+1)!}\epsilon_{iab}\hat{S}^{(P)}_{aI_{l-1}}(t)\partial_{bI_{l-1}}\bigg(\frac{\text{e}^{-m_{2}r}}{r}\bigg)\notag\\
&&-G\sum_{l=0}^{\infty}\frac{(-1)^{l+1}}{(l+1)!}\hat{B}^{(P)}_{I_{l}}(t)\partial_{iI_{l}}\bigg(\frac{\text{e}^{-m_{2}r}}{r}\bigg)\notag\\
&&\left.
+m^{2}_{2}G\sum_{l=1}^{\infty}\frac{(-1)^{l+1}}{(l+1)!}\hat{B}^{(P)}_{iI_{l-1}}(t)\partial_{I_{l-1}}\bigg(\frac{\text{e}^{-m_{2}r}}{r}\bigg)\right],\\
\label{equ4.113}U_{ij}(t,\boldsymbol{x})&=& \delta_{ij}G\sum_{l=0}^{\infty}\frac{(-1)^{l}}{l!}\hat{M}_{I_{l}}^{(\tilde{h})}(t)\partial_{I_{l}}\left( \frac{1}{r}\right)-\delta_{ij}\frac{G}{3}\sum_{l=0}^{\infty}\frac{(-1)^l}{l!}\hat{Q}_{I_{l}}(t)\partial_{I_{l}}\left(\frac{\text{e}^{-m_{1}r}}{r}\right)\notag\\
&&-\frac{2G}{3m_{2}^{2}}\sum_{l=0}^{\infty}\frac{(-1)^l}{l!}\hat{Q}_{I_{l}}(t)\partial_{ijI_{l}}\left(\frac{\text{e}^{-m_{1}r}}{r}\right)\notag\\
&& -G\bigg\{\sum_{l=0}^{\infty}\frac{(-1)^{l}}{l!}\hat{M}^{(P)}_{I_{l}}(t)\left[\delta_{ij}\partial_{I_{l}}\left(\frac{\text{e}^{-m_{2}r}}{r}\right)-\frac{1}{m^{2}_{2}}\partial_{ij I_{l}}\bigg(\frac{\text{e}^{-m_{2}r}}{r}\bigg)\right]\notag\\
&& +\frac{1}{2}\sum_{l=2}^{\infty}\frac{(-1)^{l}}{l!}\hat{G}^{(P)}_{I_{l} }(t)\left[\delta_{ij}\partial_{I_{l}}\bigg(\frac{\text{e}^{-m_{2}r}}{r}\bigg)+\frac{1}{m^{2}_{2}}\partial_{ij I_{l}}\bigg(\frac{\text{e}^{-m_{2}r}}{r}\bigg)\right]\notag\\
&&-\sum_{l=2}^{\infty}\frac{(-1)^{l}}{l!}\left[2\hat{G}^{(P)}_{I_{l-1}(i}(t)\partial_{j)I_{l-1}}\bigg(\frac{\text{e}^{-m_{2}r}}{r}\bigg)
-m_{2}^{2}\hat{G}^{(P)}_{ijI_{l-2}}(t)\partial_{I_{l-2}}\bigg(\frac{\text{e}^{-m_{2}r}}{r}\bigg)\right]\notag\\
&& +2\sum_{l=2}^{\infty}\frac{(-1)^{l+1}}{(l+1)!}\left[\hat{H}^{(P)}_{bI_{l-1}}(t)\epsilon_{ab(i}\partial_{j)aI_{l-1}}\bigg(\frac{\text{e}^{-m_{2}r}}{r}\bigg)\right.\notag\\
&&\left.-m_{2}^2\epsilon_{ab(i}\hat{H}^{(P)}_{j)bI_{l-2}}(t)\partial_{aI_{l-2}}\bigg(\frac{\text{e}^{-m_{2}r}}{r}\bigg)\right]\bigg\}.
\end{eqnarray}
These expansions indicate that if $m_{i}\ (i=1,2)$ approach positive infinity, all the terms related to $m_{i}$ go to zero. As a consequence, it can be easily verified that when $F(X,Y,Z)=R$, since eqs.~(\ref{equ3.14}) and (\ref{equ3.15}) imply that $m_{1}\rightarrow+\infty$ and $m_{2}\rightarrow+\infty$ the metric potentials reduce to
\begin{eqnarray}
\label{equ4.114}U^{\text{GR}}(t,\boldsymbol{x})&=& G\sum_{l=0}^{\infty}\frac{(-1)^{l}}{l!}\hat{M}_{I_{l}}^{(\tilde{h})}(t)\partial_{I_{l}}\left( \frac{1}{r}\right),\\
\label{equ4.115}U^{\text{GR}}_{i}(t,\boldsymbol{x})&=& -G\sum_{l=1}^{\infty}\frac{(-1)^{l}}{l!}\left(\partial_{t}\hat{M}_{iI_{l-1}}^{(\tilde{h})}(t)\right)\partial_{I_{l-1}}\left(\frac{1}{r}\right)\notag\\ &&+G\sum_{l=1}^{\infty}\frac{(-1)^{l}l}{(l+1)!}\epsilon_{iab}\hat{S}_{aI_{l-1}}^{(\tilde{h})}(t)\partial_{bI_{l-1}}\left( \frac{1}{r}\right),\\
\label{equ4.116}U^{\text{GR}}_{ij}(t,\boldsymbol{x})&=& \delta_{ij}G\sum_{l=0}^{\infty}\frac{(-1)^{l}}{l!}\hat{M}_{I_{l}}^{(\tilde{h})}(t)\partial_{I_{l}}\left( \frac{1}{r}\right),
\end{eqnarray}
which means the metric in this case is exactly the result in GR~\cite{Wu:2021uws}. Therefore, in the metric~(\ref{equ3.22}) within $F(X,Y,Z)$ gravity, those terms, not related to the two massive parameters, constitute the massless tensor part. As for the remaining terms, those  related to $\text{e}^{-m_{1}r}$ and $\text{e}^{-m_{2}r}$
constitute the massive scalar part and the massive tensor part, respectively. But despite all that, it is also important to note that since the following two terms
\begin{eqnarray}
\label{equ4.117}&&\frac{G}{3m_{2}^{2}}\sum_{l=0}^{\infty}\frac{(-1)^l}{l!}\left(\partial_{t}\hat{Q}_{I_{l}}(t)\right)\partial_{iI_{l}}\left(\frac{\text{e}^{-m_{1}r}}{r}\right),\\
\label{equ4.118}&&-\frac{2G}{3m_{2}^{2}}\sum_{l=0}^{\infty}\frac{(-1)^l}{l!}\hat{Q}_{I_{l}}(t)\partial_{ijI_{l}}\left(\frac{\text{e}^{-m_{1}r}}{r}\right)
\end{eqnarray}
are associated with both $m_{1}$ and $m_{2}$, they should originate from the coupling of the two massive propagations in $F(X,Y,Z)$ gravity. When $F(X,Y,Z)\rightarrow f(R)$ or $f(R,\mathcal{G})$,  it can be proved that because of $m_{2}\rightarrow\infty$~\cite{Wu:2022mna}, the metric potentials are
\begin{eqnarray}
\label{equ4.119}U^{\text{f}}(t,\boldsymbol{x})&=& G\sum_{l=0}^{\infty}\frac{(-1)^{l}}{l!}\hat{M}_{I_{l}}^{(\tilde{h})}(t)\partial_{I_{l}}\left( \frac{1}{r}\right)\notag\\
&&+\frac{G}{3}\sum_{l=0}^{\infty}\frac{(-1)^l}{l!}\hat{Q}_{I_{l}}(t)\partial_{I_{l}}\left(\frac{\text{e}^{-m_{1}r}}{r}\right),\\
\label{equ4.120}U^{\text{f}}_{i}(t,\boldsymbol{x})&=& -G\sum_{l=1}^{\infty}\frac{(-1)^{l}}{l!}\left(\partial_{t}\hat{M}_{iI_{l-1}}^{(\tilde{h})}(t)\right)\partial_{I_{l-1}}\left(\frac{1}{r}\right) \notag\\ &&+G\sum_{l=1}^{\infty}\frac{(-1)^{l}l}{(l+1)!}\epsilon_{iab}\hat{S}_{aI_{l-1}}^{(\tilde{h})}(t)\partial_{bI_{l-1}}\left( \frac{1}{r}\right),\\
\label{equ4.121}U^{\text{f}}_{ij}(t,\boldsymbol{x})&=& \delta_{ij}\left[G\sum_{l=0}^{\infty}\frac{(-1)^{l}}{l!}\hat{M}_{I_{l}}^{(\tilde{h})}(t)\partial_{I_{l}}\left( \frac{1}{r}\right)\right.\notag\\
&&\left.-\frac{G}{3}\sum_{l=0}^{\infty}\frac{(-1)^l}{l!}\hat{Q}_{I_{l}}(t)\partial_{I_{l}}\left(\frac{\text{e}^{-m_{1}r}}{r}\right)\right],
\end{eqnarray}
which shows that the metric under this circumstance indeed recovers the one in $f(R)$ or $f(R,\mathcal{G})$ gravity presented in ref.~\cite{Wu:2021uws}. Due to the fact that the Gauss-Bonnet scalar $\mathcal{G}$, being a topological invariant, has no contribution to the gravitational field dynamics, the metrics in $f(R)$ gravity and $f(R,\mathcal{G})$ gravity are identical~\cite{Wu:2021uws}.

It is clear that the massless tensor part, as GR-like metric, shows the Coulomb-like dependence and is characterized by the mass and spin multipole moments, whereas the massive scalar and tensor parts bear the Yukawa-like dependence on the two massive parameters and predict the appearance of six additional sets of source multipole moments. Five of these additional sets of multipole moments come from the massive tensor part, and among them besides the counterparts of the mass and spin multipole moments, the remaining three, unlike the situation encountered in the massless tensor part, can not be transformed away by the gauge symmetry. The massive scalar and tensor parts actually provide the corrections to GR-like metric in $F(X,Y,Z)$ gravity, and six additional sets of multipole moments characterize the corrections. This fact implies that up to $1/c^3$ order, the number of the degrees of freedom beyond GR in $F(X,Y,Z)$ gravity is six. In addition, from all the expressions of the multipole moments,  it is observed that the integrals involved are carried out only over the actual source distributions, which means that the metric obtained in this section can avoid infinite integrals appearing in previous expressions of the metric in refs.~\cite{stabile2015,Capozziello:2009ss,Stabile:2010mz}, and therefore, it can be conveniently employed in practical application. As an attempt, in the next section,  for a gyroscope moving around the source without experiencing any torque, the metric will be utilized to deduce the precessional angular velocity of its spin.
\section{Multipole expansion of the precessional angular velocity of the gyroscopic spin in $F(X,Y,Z)$ gravity~\label{Sec:fifth}}
From the discussions in the previous section, it is shown that up to $1/c^3$ order, there exist a total of six degrees of freedom beyond GR in $F(X,Y,Z)$ gravity, whereas in $f(R)$ or $f(R,\mathcal{G})$ gravity the number of such degrees of freedom is only one. Therefore, compared with these two models, $F(X,Y,Z)$ gravity is able to predict more effects beyond GR. As a typical application example, in this section, we intend to make use of the metric obtained in this paper to deal with gyroscopic precession so as to study the effects of the degrees of freedom beyond GR appearing in $F(X,Y,Z)$ gravity.

The basic method on how to evaluate the precessional angular velocity of the spin of a gyroscope in the stationary spacetime is presented in ref.~\cite{MTW1973},  and by following it, we can begin the task. However, it should be noted that since the metric~(\ref{equ3.22}) is time-dependent in general, this basic method needs to be extended in the derivation. Consider a gyroscope moving around the spatially compact source without experiencing any torque, and let  $x^{\mu}(\tau)$ be its world line with $\tau$ as the proper time. The spin $S^{\alpha}$ (i.e., the angular momentum vector) of the gyroscope is always orthogonal to its four-velocity $u^{\alpha}$~\cite{Naf:2010zy} and also satisfies the Fermi-Walker transport equation along its world line $x^{\mu}(\tau)$~\cite{MTW1973}. Thus, there are
\begin{eqnarray}
\label{equ5.1}u^{\alpha}S_{\alpha}&=&0,\\
\label{equ5.2}u^{\alpha}\nabla_{\alpha}S^{\beta}&=&\frac{\text{d}S^{\beta}}{\text{d}\tau}+u^{\alpha}S^{\lambda}\varGamma^{\beta}_{\lambda\alpha}=\frac{1}{c^2}a^{\rho}S_{\rho}u^{\beta},
\end{eqnarray}
where $\varGamma^{\beta}_{\lambda\alpha}$ is the Christoffel symbol and $a^{\rho}$ is the gyroscopic four-acceleration. With eqs.~(\ref{equ5.1}) and (\ref{equ5.2}), it is concluded that $S^{\beta}S_{\beta}$ remains fixed along $x^{\mu}(\tau)$ because the equality
\begin{eqnarray}
\label{equ5.3}\frac{\text{d}(S^{\beta}S_{\beta})}{\text{d}\tau}=u^{\alpha}\nabla_{\alpha}(S^{\beta}S_{\beta})=2S_{\beta}u^{\alpha}\nabla_{\alpha}S^{\beta}=0
\end{eqnarray}
holds. The above three equations constitute the starting point for dealing with gyroscopic precession.

According to ref.~\cite{MTW1973}, a local orthonormal tetrad $\boldsymbol{e}_{[\sigma]}$, at rest in the coordinate frame, needs to be defined, and when the gyroscope moves, it carries with itself an orthonormal frame $\boldsymbol{e}_{(\alpha)}$, which is related to the tetrad $\boldsymbol{e}_{[\sigma]}$ by a pure Lorentz boost.
Here, in view that the vectors of  $\boldsymbol{e}_{[\sigma]}$ and $\boldsymbol{e}_{(\alpha)}$ are denoted by
the Greek indices within square and round brackets, the components of a tensor with respect to $\boldsymbol{e}_{[\sigma]}$ and $\boldsymbol{e}_{(\alpha)}$ will also be denoted by such indices. The construction of the tetrad $\boldsymbol{e}_{[\sigma]}$ is a bit subtle in $F(X,Y,Z)$ gravity.
As required, the three spatial vectors of the tetrad $\boldsymbol{e}_{[\sigma]}$ should be put on an equal footing, so that it can reduce to a Minkowskian coordinate tetrad when the spacetime is flat. If one applies the Gram-Schmidt orthogonalization to the coordinate frame
\begin{eqnarray}
\label{equ5.4}\boldsymbol{g}_{\rho}:=\frac{\partial}{\partial x^{\rho}}
\end{eqnarray}
like the case in $f(R)$ gravity~\cite{Wu:2021uws}, the tetrad meeting the above requirement will not be gained because in $F(X,Y,Z)$ gravity, $g_{ij}=2U_{ij}/c^2\ (i\neq j)$ are nonvanishing in general. In this paper, by defining
\begin{eqnarray}
\label{equ5.5}\mathcal{W}^{0}_{\phantom{0}[0]}&=&1+\frac{1}{c^{2}}U(t,\boldsymbol{x}),\\
\label{equ5.6}\mathcal{W}^{0}_{\phantom{0}[i]}&=&-\frac{4}{c^{3}}U_{i}(t,\boldsymbol{x}),\\
\label{equ5.7}\mathcal{W}^{j}_{\phantom{0}[0]}&=&0,\\
\label{equ5.8}\mathcal{W}^{j}_{\phantom{0}[i]}&=&\delta_{ji}-\frac{1}{c^{2}}U_{ji}(t,\boldsymbol{x}),
\end{eqnarray}
the tetrad is constructed as
\begin{eqnarray}
\label{equ5.9}\boldsymbol{e}_{[\sigma]}:=\boldsymbol{g}_{\rho}\mathcal{W}^{\rho}_{\phantom{\rho}[\sigma]},
\end{eqnarray}
where eq.~(\ref{equ5.8}) implies that its three spatial vectors $\boldsymbol{e}_{[i]}$ are on an equal footing, and
it can easily be proved that up to $1/c^3$ order, $\boldsymbol{e}_{[\sigma]}\cdot\boldsymbol{e}_{[\rho]}=\eta_{\sigma\rho}$. As mentioned above, the orthonormal frame $\boldsymbol{e}_{(\alpha)}$ comoving with the gyroscope is related to the tetrad $\boldsymbol{e}_{[\sigma]}$ by a pure Lorentz boost. Next, we will employ the gyroscopic four-velocity $u^{\alpha}$ to derive the pure Lorentz boost $\mathcal{K}^{[\sigma]}_{\phantom{\sigma}(\alpha)}$ so as to determine the frame $\boldsymbol{e}_{(\alpha)}$ by
\begin{eqnarray}
\label{equ5.10}\boldsymbol{e}_{(\alpha)}:=\boldsymbol{e}_{[\sigma]}\mathcal{K}^{[\sigma]}_{\phantom{[\sigma]}(\alpha)}.
\end{eqnarray}
In the coordinate frame, if the gyroscopic velocity is represented by $v^{i}$,  the components of the gyroscopic four-velocity are
\begin{equation}\label{equ5.11}
\left\{\begin{array}{l}
\displaystyle u^{0}=c\frac{\text{d}t}{\text{d}\tau},\smallskip\\
\displaystyle u^{i}=v^{i}\frac{\text{d}t}{\text{d}\tau}.
\end{array}\right.
\end{equation}
Then, by means of the identity
\begin{eqnarray}
\label{equ5.12}u_{\beta}u^{\beta}=-c^2,
\end{eqnarray}
and the metric~(\ref{equ3.22}), one can acquire the expression of $\text{d}t/\text{d}\tau$ up to $1/c^3$ order,
\begin{eqnarray}
\label{equ5.13}\frac{\text{d}t}{\text{d}\tau}=1+\frac{1}{c^{2}}U(t,\boldsymbol{x})+\frac{1}{2c^{2}}v^{k}v^{k}.
\end{eqnarray}
In the local tetrad $\boldsymbol{e}_{[\sigma]}$, if the proper time of the observer at rest is $t_{0}$, and the gyroscopic velocity is labeled by $v^{[i]}$, the components of the gyroscopic four-velocity in this case are
\begin{equation}\label{equ5.14}
\left\{\begin{array}{l}
\displaystyle u^{[0]}=c\frac{\text{d}t_{0}}{\text{d}\tau},\smallskip\\
\displaystyle u^{[i]}=v^{[i]}\frac{\text{d}t_{0}}{\text{d}\tau}.
\end{array}\right.
\end{equation}
By use of eq.~(\ref{equ5.9}), the relation between $u^{[\sigma]}$ and $u^{\rho}$ is given as
\begin{eqnarray}\label{equ5.15}
u^{[\sigma]}=\big(\mathcal{W}^{-1}\big)^{[\sigma]}_{\phantom{[\sigma]}\rho}u^{\rho},
\end{eqnarray}
and then, together with eqs.~(\ref{equ5.5})--(\ref{equ5.8}), (\ref{equ5.11}), (\ref{equ5.13}), and (\ref{equ5.14}), we get
\begin{eqnarray}
\label{equ5.16}\frac{\text{d}t_{0}}{\text{d}\tau}&=&1+\frac{1}{2c^2}v^{k}v^{k},\\
\label{equ5.17}v^{[i]}&=&\left(1+\frac{1}{c^{2}}U(t,\boldsymbol{x})\right)v^{i}+\frac{1}{c^2}U_{ij}(t,\boldsymbol{x})v^{j}.
\end{eqnarray}
Remember that $v^{[i]}$ is the gyroscopic velocity in the tetrad $\boldsymbol{e}_{[\sigma]}$, and thus, by definition, from eqs.~(\ref{equ5.16}) and (\ref{equ5.17}), the expressions of the components of the Lorentz boost $\mathcal{K}^{[\sigma]}_{\phantom{\sigma}(\alpha)}$ up to $1/c^3$ order are
\begin{eqnarray}
\label{equ5.18}\mathcal{K}^{0}_{\phantom{0}[0]}&=&1+\frac{1}{2c^2}v^{k}v^{k},\\
\label{equ5.19}\mathcal{K}^{0}_{\phantom{0}[i]}&=&\frac{1}{c}v^{i}+\frac{1}{2c^3}v^{i}v^{k}v^{k}+\frac{1}{c^3}U(t,\boldsymbol{x})v^{i}+\frac{1}{c^3}U_{ik}(t,\boldsymbol{x})v^{k},\\
\label{equ5.20}\mathcal{K}^{j}_{\phantom{0}[0]}&=&\frac{1}{c}v^{j}+\frac{1}{2c^3}v^{j}v^{k}v^{k}+\frac{1}{c^3}U(t,\boldsymbol{x})v^{j}+\frac{1}{c^3}U_{jk}(t,\boldsymbol{x})v^{k},\\
\label{equ5.21}\mathcal{K}^{j}_{\phantom{0}[i]}&=&\delta_{ji}+\frac{1}{2c^2}v^{j}v^{i}.
\end{eqnarray}
Equations~(\ref{equ5.9}) and (\ref{equ5.10}) indicate that
\begin{eqnarray}\label{equ5.22}
\boldsymbol{e}_{(\alpha)}=\boldsymbol{g}_{\rho}\mathcal{W}^{\rho}_{\phantom{\rho}[\sigma]}\mathcal{K}^{[\sigma]}_{\phantom{[\sigma]}(\alpha)}
=:\boldsymbol{g}_{\rho}\mathcal{L}^{\rho}_{\phantom{\rho}(\alpha)},
\end{eqnarray}
where  $\mathcal{L}^{\rho}_{\phantom{\rho}(\alpha)}=\mathcal{W}^{\rho}_{\phantom{\rho}[\sigma]}\mathcal{K}^{[\sigma]}_{\phantom{[\sigma]}(\alpha)}$
is the compound transformation, and with it, the frame $\boldsymbol{e}_{(\alpha)}$ can directly be determined from the coordinate frame $\boldsymbol{g}_{\rho}$.
Moreover, the above equation also yields
\begin{eqnarray}
\label{equ5.23}S^{\beta}=\mathcal{L}^{\beta}_{\phantom{\beta}(\alpha)}S^{(\alpha)},
\end{eqnarray}
from which, the components of the gyroscopic spin in the frame $\boldsymbol{e}_{(\alpha)}$ could be given,
\begin{eqnarray}
\label{equ5.24}S^{(\alpha)}=\big(\mathcal{L}^{-1}\big)^{(\alpha)}_{\phantom{(\alpha)}\beta}S^{\beta}.
\end{eqnarray}
Since $\boldsymbol{e}_{(\alpha)}$ is an orthonormal frame comoving with the gyroscope, the compound transformation $\mathcal{L}^{\rho}_{\phantom{\rho}(\alpha)}$ ought to satisfy
\begin{eqnarray}
\label{equ5.25}\mathcal{L}^{\rho}_{\phantom{\rho}(0)}&=& \frac{u^{\rho}}{c},\\
\label{equ5.26}\boldsymbol{e}_{(\alpha)}\cdot\boldsymbol{e}_{(\beta)}&=&\eta_{\alpha\beta}=g_{\rho\sigma}\mathcal{L}^{\rho}_{\phantom{\rho}(\alpha)}\mathcal{L}^{\sigma}_{\phantom{\sigma}(\beta)},
\end{eqnarray}
and then, combining eqs.~(\ref{equ5.1}) and (\ref{equ5.23}), we arrive at a conclusion that
\begin{eqnarray}
\label{equ5.27}S^{(0)}=0.
\end{eqnarray}
This conclusion implies that the spin of the gyroscope is a purely spatial vector in its comoving frame, and after substituting it in eq.~(\ref{equ5.3}), the equation
\begin{eqnarray}
\label{equ5.28}\frac{\text{d}\left(S^{(i)}S_{(i)}\right)}{\text{d}\tau}=\frac{\text{d}\left(S^{(i)}S^{(j)}\delta_{ij}\right)}{\text{d}\tau}=0
\end{eqnarray}
is obtained, which means that $S^{(i)}$ always precesses relative to this comoving frame. Thus, $S^{(i)}$ should fulfill the following precession equation
\begin{eqnarray}
\label{equ5.29}\frac{\text{d}S^{(i)}}{\text{d}\tau}=\epsilon^{ijk}\omega^{(j)}S^{(k)}
\end{eqnarray}
with $\omega^{(j)}$ as the angular velocity. The main purpose of this section is to derive the precessional angular velocity $\omega^{(j)}$ of the gyroscopic spin up to $1/c^3$ order under the WFSM approximation.

Now, let us calculate the expression of $\text{d}S^{(i)}/\text{d}\tau$ up to $1/c^3$ order. By virtue of eq.~(\ref{equ5.22}), we have
\begin{eqnarray}
\label{equ5.30}\big(\mathcal{L}^{-1}\big)^{(\alpha)}_{\phantom{(\alpha)}\beta}=
\big(\mathcal{K}^{-1}\big)^{(\alpha)}_{\phantom{(\alpha)}[\lambda]}\big(\mathcal{W}^{-1}\big)^{[\lambda]}_{\phantom{(\alpha)}\beta},
\end{eqnarray}
and then inserting the expressions of
$\big(\mathcal{K}^{-1}\big)^{(\alpha)}_{\phantom{(\alpha)}[\lambda]}$ and $\big(\mathcal{W}^{-1}\big)^{[\lambda]}_{\phantom{(\alpha)}\beta}$ up to $1/c^3$ order gives rise to
\begin{eqnarray}
\label{equ5.31}\big(\mathcal{L}^{-1}\big)^{0}_{\phantom{0}[0]}&=&1-\frac{1}{c^{2}}U(t,\boldsymbol{x})+\frac{1}{2c^2}v^{k}v^{k},\\
\label{equ5.32}\big(\mathcal{L}^{-1}\big)^{0}_{\phantom{0}[i]}&=&-\frac{1}{c}v^{i}-\frac{1}{2c^3}v^{i}v^{k}v^{k}-\frac{1}{c^3}U(t,\boldsymbol{x})v^{i}-\frac{2}{c^3}U_{ik}(t,\boldsymbol{x})v^{k}+\frac{4}{c^3}U_{i}(t,\boldsymbol{x}),\\
\label{equ5.33}\big(\mathcal{L}^{-1}\big)^{j}_{\phantom{0}[0]}&=&-\frac{1}{c}v^{j}-\frac{1}{2c^3}v^{j}v^{k}v^{k}-\frac{1}{c^3}U_{jk}(t,\boldsymbol{x})v^{k},\\
\label{equ5.34}\big(\mathcal{L}^{-1}\big)^{j}_{\phantom{0}[i]}&=&\delta_{ji}+\frac{1}{c^{2}}U_{ji}(t,\boldsymbol{x})+\frac{1}{2c^2}v^{j}v^{i}.
\end{eqnarray}
With these results, eqs.~(\ref{equ5.24}) and (\ref{equ5.27}) yield
\begin{eqnarray}
\label{equ5.35}S^{0}&=&\left(\frac{1}{c}v^{j}+\frac{2}{c^3}U(t,\boldsymbol{x})v^{j}+\frac{2}{c^3}U_{jk}(t,\boldsymbol{x})v^{k}-\frac{4}{c^3}U_{j}(t,\boldsymbol{x})\right)S^{j},\\
\label{equ5.36}S^{(i)}&=&S^{i}+\frac{1}{c^{2}}U_{ij}(t,\boldsymbol{x})S^{j}-\frac{1}{2c^2}v^{i}v^{j}S^{j},
\end{eqnarray}
where in the derivation of eq.~(\ref{equ5.36}), eq.~(\ref{equ5.35}) has been used, and thus,
by eqs.~(\ref{equ5.11}) and (\ref{equ5.13}), the derivative of $S^{(i)}$ with respect to the gyroscopic proper time $\tau$ is acquired,
\begin{eqnarray}
\label{equ5.37}\frac{\text{d}S^{(i)}}{\text{d}\tau}&=&\frac{\text{d}S^{i}}{\text{d}\tau}
+\frac{1}{c^{2}}U_{ij}(t,\boldsymbol{x})\frac{\text{d}S^{j}}{\text{d}\tau}
+\frac{1}{c^{2}}S^{j}\partial_{t}U_{ij}(t,\boldsymbol{x})
+\frac{1}{c^{2}}S^{j}v^{k}\partial_{k}U_{ij}(t,\boldsymbol{x})
-\frac{1}{2c^2}v^{j}S^{j}\frac{\text{d}v^{i}}{\text{d}\tau}\notag\\
&&-\frac{1}{2c^2}v^{i}S^{j}\frac{\text{d}v^{j}}{\text{d}\tau}-\frac{1}{2c^2}v^{i}v^{j}\frac{\text{d}S^{j}}{\text{d}\tau}.
\end{eqnarray}
In order to further handle the right-hand side of the above equation, the expression of $\text{d}S^{i}/\text{d}\tau$ up to $1/c^3$ order needs to be evaluated on the basis of eq.~(\ref{equ5.2}), namely,
\begin{eqnarray}
\label{equ5.38}\frac{\text{d}S^{i}}{\text{d}\tau}=-u^{\alpha}S^{\lambda}\varGamma^{i}_{\lambda\alpha}+\frac{1}{c^2}a^{\rho}S^{\sigma}g_{\rho\sigma}u^{i}.
\end{eqnarray}
The detailed calculation is a bit tedious and is put it in Appendix B. The result is
\begin{eqnarray}
\label{equ5.39}\frac{\text{d}S^{i}}{\text{d}\tau}&=&\frac{1}{c^2}v^{j}S^{j}\partial_{i}U(t,\boldsymbol{x})
-\frac{2}{c^2}\left[\partial_{i}U_{j}(t,\boldsymbol{x})-\partial_{j}U_{i}(t,\boldsymbol{x})\right]S^{j}
-\frac{1}{c^2}\partial_{t}U_{ij}(t,\boldsymbol{x})S^{j}\notag\\
&&-\frac{1}{c^2}v^{k}S^{j}\left[\partial_{j}U_{ik}(t,\boldsymbol{x})+\partial_{k}U_{ij}(t,\boldsymbol{x})-\partial_{i}U_{jk}(t,\boldsymbol{x})\right]
+\frac{1}{c^2}a^{j}v^{i}S^{j}.
\end{eqnarray}
From eq.~(\ref{equ5.36}), it is shown that up to $1/c^3$ order, the equation
\begin{eqnarray}
\label{equ5.40}\frac{1}{c^2}S^{i}=\frac{1}{c^2}S^{(i)}
\end{eqnarray}
holds, and with it, the substitution of the result~(\ref{equ5.39}) in eq.~(\ref{equ5.37}) brings about
\begin{eqnarray}
\label{equ5.41}\frac{\text{d}S^{(i)}}{\text{d}\tau}&=&\frac{1}{2c^2}\left[v^{j}\partial_{i}U(t,\boldsymbol{x})-v^{i}\partial_{j}U(t,\boldsymbol{x})\right]S^{(j)}
-\frac{2}{c^2}\left[\partial_{i}U_{j}(t,\boldsymbol{x})-\partial_{j}U_{i}(t,\boldsymbol{x})\right]S^{(j)}\notag\\
&+&\frac{1}{c^2}\left[\partial_{i}U_{jk}(t,\boldsymbol{x})-\partial_{j}U_{ik}(t,\boldsymbol{x})\right]v^{k}S^{(j)}
+\frac{1}{2c^2}\left(a^{j}v^{i}-a^{i}v^{j}\right)S^{(j)},\qquad
\end{eqnarray}
where eq.~(\ref{equB12}) is used. Finally, after the right-hand side of the above equation is rewritten as
the form of $\epsilon^{ijk}\omega^{(j)}S^{(k)}$, the precessional angular velocity
$\omega^{(j)}$ of the gyroscopic spin up to $1/c^3$ order is achieved,
\begin{eqnarray}
\label{equ5.42}\omega^{(j)}&=&\frac{1}{2c^2}\epsilon^{jpq}v^{p}\partial_{q}U(t,\boldsymbol{x})+\frac{2}{c^2}\epsilon^{jpq}\partial_{p}U_{q}(t,\boldsymbol{x})
-\frac{1}{c^2}\epsilon^{jpq}\partial_{p}U_{ql}(t,\boldsymbol{x})v^{l}-\frac{1}{2c^2}\epsilon^{jpq}v^{p}a^{q}.\qquad
\end{eqnarray}
Up to $1/c^3$ order under the WFSM approximation, eqs.~(\ref{equ5.29}) and (\ref{equ5.42}) describe in complete generality the precession of the gyroscopic spin relative to the comoving orthonormal frame $\boldsymbol{e}_{(\alpha)}$ in $F(X,Y,Z)$ gravity. According to the conventional definitions~\cite{MTW1973},
there should be three types of precession. In the precessional angular velocity $\omega^{(j)}$, the term
\begin{eqnarray}
\label{equ5.43}\omega^{(j)}_{\text{Tho}}&:=&-\frac{1}{2c^2}\epsilon^{jpq}v^{p}a^{q}
\end{eqnarray}
is associated with the gyroscopic four-acceleration, so it represents the Thomas precession, which plays a significant role in the fine structure of atomic spectra~\cite{MTW1973}. It is observed that $\omega^{(j)}_{\text{Tho}}$ in $F(X,Y,Z)$ gravity is presented in the same form as that in GR, but even so, they are completely distinct because with the help of eqs.~(\ref{equ5.13}) and (\ref{equB12}), $\omega^{(j)}_{\text{Tho}}$ in eq.~(\ref{equ5.43}) can be recast as
\begin{eqnarray}
\label{equ5.44}\omega^{(j)}_{\text{Tho}}&=&-\frac{1}{2c^2}\epsilon^{jpq}v^{p}\frac{\text{d}v^{q}}{\text{d}t}+\frac{1}{2c^2}\epsilon^{jpq}v^{p}\partial_{q}U(t,\boldsymbol{x}),
\end{eqnarray}
and the potential $U(t,\boldsymbol{x})$ in $F(X,Y,Z)$ gravity is different from that in GR. Let
\begin{eqnarray}
\label{equ5.45}\omega^{(j)}_{\text{Geo}}&:=&\frac{1}{2c^2}\epsilon^{jpq}v^{p}\partial_{q}U(t,\boldsymbol{x})-\frac{1}{c^2}\epsilon^{jpq}\partial_{p}U_{ql}(t,\boldsymbol{x})v^{l},\\
\label{equ5.46}\omega^{(j)}_{\text{L-T}}&:=&\frac{2}{c^2}\epsilon^{jpq}\partial_{p}U_{q}(t,\boldsymbol{x}),
\end{eqnarray}
and in the case that the gyroscope moves along a geodesic, the precessional angular velocity of its spin should be
\begin{eqnarray}
\label{equ5.47}\omega^{(j)}=\omega^{(j)}_{\text{Geo}}+\omega^{(j)}_{\text{L-T}}.
\end{eqnarray}
Since the angular velocities $\omega^{(j)}_{\text{Geo}}$ and $\omega^{(j)}_{\text{L-T}}$ are associated with the gyroscopic velocity and the time-space components of the metric (the vector potential), respectively, they denote the geodetic precession and the Lense-Thirring precession. According to ref.~\cite{MTW1973}, the former is caused by the motion of the gyroscope through the source's curved and static spacetime geometry, and the latter originates from the rotation of the source.

In the previous section, the multipole expansions of the three potentials $U(t,\boldsymbol{x})$, $U_{i}(t,\boldsymbol{x})$, and $U_{ij}(t,\boldsymbol{x})$ have been obtained in eqs.~(\ref{equ4.111}), (\ref{equ4.112}), and (\ref{equ4.113}), and by substituting these results in eqs.~(\ref{equ5.44})--(\ref{equ5.46}), the multipole expansions of the gyroscopic spin's angular velocities of  the Thomas precession, the geodetic precession, and
the Lense-Thirring precession up to $1/c^3$ order are all achieved,
\begin{eqnarray}
\label{equ5.48}\omega^{(j)}_{\text{Tho}}&=&-\frac{1}{2c^2}\epsilon^{jpq}v^{p}\frac{\text{d}v^{q}}{\text{d}t}+\frac{G}{2c^2}\sum_{l=0}^{\infty}\frac{(-1)^{l}}{l!}\hat{M}_{I_{l}}^{(\tilde{h})}(t)\epsilon^{jpq}v^{p}\partial_{qI_{l}}\left( \frac{1}{r}\right)\notag\\
&&+\frac{G}{6c^2}\sum_{l=0}^{\infty}\frac{(-1)^l}{l!}\hat{Q}_{I_{l}}(t)\epsilon^{jpq}v^{p}\partial_{qI_{l}}\left(\frac{\text{e}^{-m_{1}r}}{r}\right)\notag\\
&&-\frac{G}{c^2}\sum_{l=0}^{\infty}\frac{(-1)^l}{l!}\hat{M}^{(P)}_{I_{l}}(t)\epsilon^{jpq}v^{p}\partial_{qI_{l}}\left(\frac{\text{e}^{-m_{2}r}}{r}\right),\\
\label{equ5.49}\omega^{(j)}_{\text{Geo}}&=& \frac{3G}{2c^2}\sum_{l=0}^{\infty}\frac{(-1)^{l}}{l!}\hat{M}_{I_{l}}^{(\tilde{h})}(t)\epsilon^{jpq}v^{p}\partial_{qI_{l}}\left( \frac{1}{r}\right)\notag\\
&&-\frac{G}{6c^2}\sum_{l=0}^{\infty}\frac{(-1)^l}{l!}\hat{Q}_{I_{l}}(t)\epsilon^{jpq}v^{p}\partial_{qI_{l}}\left(\frac{\text{e}^{-m_{1}r}}{r}\right)\notag\\
&& -\frac{2G}{c^2}\sum_{l=0}^{\infty}\frac{(-1)^{l}}{l!}\hat{M}^{(P)}_{I_{l}}(t)\epsilon^{jpq}v^{p}\partial_{qI_{l}}\left(\frac{\text{e}^{-m_{2}r}}{r}\right)\notag\\
&&-\frac{G}{2c^2}\sum_{l=2}^{\infty}\frac{(-1)^{l}}{l!}\hat{G}^{(P)}_{I_{l} }(t)\epsilon^{jpq}v^{p}\partial_{qI_{l}}\bigg(\frac{\text{e}^{-m_{2}r}}{r}\bigg)\notag\\
&&+\frac{G}{c^2}\sum_{l=2}^{\infty}\frac{(-1)^{l}}{l!}\hat{G}^{(P)}_{I_{l-1}p}(t)\epsilon^{jpq}v^{s}\partial_{qsI_{l-1}}\bigg(\frac{\text{e}^{-m_{2}r}}{r}\bigg)
\notag\\
&&-\frac{Gm_{2}^{2}}{c^2}\sum_{l=2}^{\infty}\frac{(-1)^{l}}{l!}
\hat{G}^{(P)}_{psI_{l-2}}(t)\epsilon^{jpq}v^{s}\partial_{qI_{l-2}}\bigg(\frac{\text{e}^{-m_{2}r}}{r}\bigg)\notag\\
&& +\frac{G}{c^2}\sum_{l=2}^{\infty}\frac{(-1)^{l+1}}{(l+1)!}\hat{H}^{(P)}_{I_{l}}(t)v^{s}\partial_{jsI_{l}}\bigg(\frac{\text{e}^{-m_{2}r}}{r}\bigg)\notag\\
&& +\frac{Gm_{2}^{2}}{c^2}\sum_{l=2}^{\infty}\frac{(-1)^{l+1}}{(l+1)!}\hat{H}^{(P)}_{I_{l}}(t)v^{j}\partial_{I_{l}}\bigg(\frac{\text{e}^{-m_{2}r}}{r}\bigg)\notag\\
&& -\frac{2Gm_{2}^{2}}{c^2}\sum_{l=2}^{\infty}\frac{(-1)^{l+1}}{(l+1)!}\hat{H}^{(P)}_{jI_{l-1}}(t)v^{s}\partial_{sI_{l-1}}\bigg(\frac{\text{e}^{-m_{2}r}}{r}\bigg)\notag\\
&& -\frac{2Gm_{2}^{2}}{c^2}\sum_{l=2}^{\infty}\frac{(-1)^{l+1}}{(l+1)!}\hat{H}^{(P)}_{sI_{l-1}}(t)v^{s}\partial_{jI_{l-1}}\bigg(\frac{\text{e}^{-m_{2}r}}{r}\bigg)\notag\\
&& +\frac{2Gm_{2}^{4}}{c^2}\sum_{l=2}^{\infty}\frac{(-1)^{l+1}}{(l+1)!}\hat{H}^{(P)}_{jsI_{l-2}}(t)v^{s}\partial_{I_{l-2}}\bigg(\frac{\text{e}^{-m_{2}r}}{r}\bigg),\\
\label{equ5.50}\omega^{(j)}_{\text{L-T}}&=& \frac{2G}{c^2}\sum_{l=1}^{\infty}\frac{(-1)^{l}}{l!}\left(\partial_{t}\hat{M}_{pI_{l-1}}^{(\tilde{h})}(t)\right)\epsilon^{jpq}\partial_{qI_{l-1}}\left(\frac{1}{r}\right) \notag
\end{eqnarray}
\newpage
\begin{eqnarray}
&&-\frac{2G}{c^2}\sum_{l=1}^{\infty}\frac{(-1)^{l}l}{(l+1)!}\hat{S}_{I_{l}}^{(\tilde{h})}(t)\partial_{jI_{l}}\left( \frac{1}{r}\right)\notag\\
&&-\frac{2G}{c^2}\sum_{l=1}^{\infty}\frac{(-1)^l}{l!}\left(\partial_{t}\hat{M}^{(P)}_{pI_{l-1}}(t)\right)\epsilon^{jpq}\partial_{qI_{l-1}}\left(\frac{\text{e}^{-m_{2}r}}{r}\right)\notag\\
&&-\frac{2Gm_{2}^{2}}{c^2}\sum_{l=1}^{\infty}\frac{(-1)^{l}l}{(l+1)!}\hat{S}^{(P)}_{jI_{l-1}}(t)\partial_{I_{l-1}}\bigg(\frac{\text{e}^{-m_{2}r}}{r}\bigg)\notag\\
&&+\frac{2G}{c^2}\sum_{l=1}^{\infty}\frac{(-1)^{l}l}{(l+1)!}\hat{S}^{(P)}_{I_{l}}(t)\partial_{jI_{l}}\bigg(\frac{\text{e}^{-m_{2}r}}{r}\bigg)\notag\\
&&
-\frac{2Gm^{2}_{2}}{c^2}\sum_{l=1}^{\infty}\frac{(-1)^{l+1}}{(l+1)!}\hat{B}^{(P)}_{pI_{l-1}}(t)\epsilon^{jpq}\partial_{qI_{l-1}}\bigg(\frac{\text{e}^{-m_{2}r}}{r}\bigg),
\end{eqnarray}
where in the derivation, $\Delta(1/r)=0\ (r\neq0)$ and eq.~(\ref{equA21})  have been used. The reduction method on the metric can also be applicable to the above three precessional angular velocities. When $F(X,Y,Z)=R$, the results in GR~\cite{Wu:2021uws} are given by $m_{1}\rightarrow+\infty$ and $m_{2}\rightarrow+\infty$,
\begin{eqnarray}
\label{equ5.51}\omega^{\text{GR}(j)}_{\text{Tho}}&=&-\frac{1}{2c^2}\epsilon^{jpq}v^{p}\frac{\text{d}v^{q}}{\text{d}t}+\frac{G}{2c^2}\sum_{l=0}^{\infty}\frac{(-1)^{l}}{l!}\hat{M}_{I_{l}}^{(\tilde{h})}(t)\epsilon^{jpq}v^{p}\partial_{qI_{l}}\left( \frac{1}{r}\right),\\
\label{equ5.52}\omega^{\text{GR}(j)}_{\text{Geo}}&=& \frac{3G}{2c^2}\sum_{l=0}^{\infty}\frac{(-1)^{l}}{l!}\hat{M}_{I_{l}}^{(\tilde{h})}(t)\epsilon^{jpq}v^{p}\partial_{qI_{l}}\left( \frac{1}{r}\right),\\
\label{equ5.53}\omega^{\text{GR}(j)}_{\text{L-T}}&=& \frac{2G}{c^2}\sum_{l=1}^{\infty}\frac{(-1)^{l}}{l!}\left(\partial_{t}\hat{M}_{pI_{l-1}}^{(\tilde{h})}(t)\right)\epsilon^{jpq}\partial_{qI_{l-1}}\left(\frac{1}{r}\right)\notag\\
&&-\frac{2G}{c^2}\sum_{l=1}^{\infty}\frac{(-1)^{l}l}{(l+1)!}\hat{S}_{I_{l}}^{(\tilde{h})}(t)\partial_{jI_{l}}\left( \frac{1}{r}\right),
\end{eqnarray}
where $\hat{M}_{I_{l}}^{(\tilde{h})}(t)$ and $\hat{S}_{I_{l}}^{(\tilde{h})}(t)$ are the mass and spin multipole moments characterizing the massless tensor part of the metric. It can be seen that if eqs~(\ref{equ5.51})--(\ref{equ5.53})
are considered at the leading pole order in the stationary spacetime, they reduce to the classical angular velocities of the three types of precession in GR~\cite{MTW1973}, namely,
\begin{eqnarray}
\label{equ5.54}\omega^{\text{GR}(j)}_{\text{Tho}}|_{l=0}&=&-\frac{1}{2c^2}\epsilon^{jpq}v^{p}\frac{\text{d}v^{q}}{\text{d}t}+\frac{GM}{2c^2r^3}\epsilon^{jpq}x^{p}v^{q},\\
\label{equ5.55}\omega^{\text{GR}(j)}_{\text{Geo}}|_{l=0}&=&\frac{3GM}{2c^2r^3}\epsilon^{jpq}x^{p}v^{q},\\
\label{equ5.56}\omega^{\text{GR}(j)}_{\text{L-T}}|_{l=1}&=&\frac{GJ_{i}}{c^2r^5}(3x_{i}x_{j}-\delta_{ij}r^2)
\end{eqnarray}
with $M$ and $J_{i}$ as the mass and the conserved angular momentum of the source. Furthermore, when $F(X,Y,Z)\rightarrow f(R)$ or $f(R,\mathcal{G})$,  the results in $f(R)$ gravity~\cite{Wu:2021uws} or $f(R,\mathcal{G})$ gravity are given by $m_{2}\rightarrow+\infty$,\newpage
\begin{eqnarray}
\label{equ5.57}\omega^{\text{f}(j)}_{\text{Tho}}&=&-\frac{1}{2c^2}\epsilon^{jpq}v^{p}\frac{\text{d}v^{q}}{\text{d}t}+\frac{G}{2c^2}\sum_{l=0}^{\infty}\frac{(-1)^{l}}{l!}\hat{M}_{I_{l}}^{(\tilde{h})}(t)\epsilon^{jpq}v^{p}\partial_{qI_{l}}\left( \frac{1}{r}\right)\notag\\
&&+\frac{G}{6c^2}\sum_{l=0}^{\infty}\frac{(-1)^l}{l!}\hat{Q}_{I_{l}}(t)\epsilon^{jpq}v^{p}\partial_{qI_{l}}\left(\frac{\text{e}^{-m_{1}r}}{r}\right),\\
\label{equ5.58}\omega^{\text{f}(j)}_{\text{Geo}}&=& \frac{3G}{2c^2}\sum_{l=0}^{\infty}\frac{(-1)^{l}}{l!}\hat{M}_{I_{l}}^{(\tilde{h})}(t)\epsilon^{jpq}v^{p}\partial_{qI_{l}}\left( \frac{1}{r}\right)\notag\\
&&-\frac{G}{6c^2}\sum_{l=0}^{\infty}\frac{(-1)^l}{l!}\hat{Q}_{I_{l}}(t)\epsilon^{jpq}v^{p}\partial_{qI_{l}}\left(\frac{\text{e}^{-m_{1}r}}{r}\right),\\
\label{equ5.59}\omega^{\text{f}(j)}_{\text{L-T}}&=& \frac{2G}{c^2}\sum_{l=1}^{\infty}\frac{(-1)^{l}}{l!}\left(\partial_{t}\hat{M}_{pI_{l-1}}^{(\tilde{h})}(t)\right)\epsilon^{jpq}\partial_{qI_{l-1}}\left(\frac{1}{r}\right)\notag\\
&&-\frac{2G}{c^2}\sum_{l=1}^{\infty}\frac{(-1)^{l}l}{(l+1)!}\hat{S}_{I_{l}}^{(\tilde{h})}(t)\partial_{jI_{l}}\left( \frac{1}{r}\right),
\end{eqnarray}
where besides the mass and spin multipole moments, the scalar multipole moments $\hat{Q}_{I_{l}}(t)$ characterizing the massive scalar part of the metric also appear, and those terms related to $\hat{Q}_{I_{l}}(t)$ correct $\omega^{\text{GR}(j)}_{\text{Tho}}$ and $\omega^{\text{GR}(j)}_{\text{Geo}}$ in these two models. Under this case, eqs~(\ref{equ5.57})--(\ref{equ5.59}) at the leading pole order in the stationary spacetime reduce to
\begin{eqnarray}
\label{equ5.60}\omega^{\text{f}(j)}_{\text{Tho}}|_{l=0}&=&-\frac{1}{2c^2}\epsilon^{jpq}v^{p}\frac{\text{d}v^{q}}{\text{d}t}+\frac{GM}{2c^2r^3}\epsilon^{jpq}x^{p}v^{q}
+\frac{GQ(1+m_{1}r)\text{e}^{-m_{1}r}}{6c^2r^3}\epsilon^{jpq}x^{p}v^{q},\\
\label{equ5.61}\omega^{\text{f}(j)}_{\text{Geo}}|_{l=0}&=&\frac{3GM}{2c^2r^3}\epsilon^{jpq}x^{p}v^{q}-\frac{GQ(1+m_{1}r)\text{e}^{-m_{1}r}}{6c^2r^3}\epsilon^{jpq}x^{p}v^{q},\\
\label{equ5.62}\omega^{\text{f}(j)}_{\text{L-T}}|_{l=1}&=&\frac{GJ_{i}}{c^2r^5}(3x_{i}x_{j}-\delta_{ij}r^2)
\end{eqnarray}
with $Q$ defined in eq.~(\ref{equ4.52}), and one can verify that the correction to $\omega^{\text{GR}(j)}_{\text{Geo}}|_{l=0}$ in $\omega^{\text{f}(j)}_{\text{Geo}}|_{l=0}$ recovers that for the gyroscope moving around a point-like~\cite{Naf:2010zy,Dass:2019hnb} or a ball-like source~\cite{Castel-Branco:2014exa}. It is worth noting that eqs.~(\ref{equ5.53}) and (\ref{equ5.59}) indicate that
$f(R)$ gravity and $f(R,\mathcal{G})$ gravity do not give the correction to the gyroscopic spin's angular velocity $\omega^{\text{GR}(j)}_{\text{L-T}}$ of the Lense-Thirring precession in GR, which means that the scalar multipole moments $\hat{Q}_{I_{l}}(t)$ have no contribution to the Lense-Thirring precession in these two model.

In $F(X,Y,Z)$ gravity, the case is different due to the existence of more degrees of freedom beyond GR. Obviously, eqs.~(\ref{equ5.48})--(\ref{equ5.50}) show the gyroscopic spin's angular velocities of the three types of precession in GR are all corrected by those terms related to the additional source multipole moments.
For the Thomas precession, only the multipole moments $\hat{M}^{(\tilde{h})}_{I_{l}}(t)$, $\hat{Q}_{I_{l}}(t)$, and  $\hat{M}^{(P)}_{I_{l}}(t)$ encoding the multipole expansion of the scalar potential $U(t,\boldsymbol{x})$ contribute it, and since the first term in $\omega^{(j)}_{\text{Tho}}$ is the result in Special Relativity, those terms related to these three sets of multipole moments actually provide the correction to this result brought about by the curved spacetime in $F(X,Y,Z)$ gravity. In terms of the geodesic precession, the expression of $\omega^{(j)}_{\text{Geo}}$ is lengthy because of the contributions from the multipole moments $\hat{G}^{(P)}_{I_{l}}(t)$ and $\hat{H}^{(P)}_{I_{l}}(t)$.
Equation~(\ref{equ5.49}) showcases that the multipole moments $\hat{G}^{(P)}_{I_{l}}(t)$ and $\hat{H}^{(P)}_{I_{l}}(t)$ play roles only at the quadrupole or higher-pole order, so their effects could be omitted in general applications. As a consequence, like the Thomas precession, in the expression of $\omega^{(j)}_{\text{Geo}}$, we only need to consider those terms related to the multipole moments $\hat{M}^{(\tilde{h})}_{I_{l}}(t)$, $\hat{Q}_{I_{l}}(t)$, and $\hat{M}^{(P)}_{I_{l}}(t)$ in general. It should be pointed out that even in this case,
the tensor potential $U_{ij}(t,\boldsymbol{x})$ also contributes to $\omega^{(j)}_{\text{Geo}}$, which can be seen from the first four terms in its multipole expansion~(\ref{equ4.113}). As to the Lense-Thirring precession, from eqs.~(\ref{equ4.112}) and (\ref{equ5.50}), it is found that except the scalar multipole moments $\hat{Q}_{I_{l}}(t)$,
all the other multipole moments $\hat{M}^{(\tilde{h})}_{I_{l}}(t)$, $\hat{S}^{(\tilde{h})}_{I_{l}}(t)$, $\hat{M}^{(P)}_{I_{l}}(t)$, $\hat{S}^{(P)}_{I_{l}}(t)$, and $\hat{B}^{(P)}_{I_{l}}(t)$ encoding the multipole expansion of the vector potential $U_{i}(t,\boldsymbol{x})$ contribute it, which implies that differently from the case in
$f(R)$ gravity or $f(R,\mathcal{G})$ gravity, the gyroscopic spin's angular velocity $\omega^{\text{GR}(j)}_{\text{L-T}}$ of Lense-Thirring precession in GR is corrected by those new degrees of freedom in $F(X,Y,Z)$ gravity. One interesting fact is that the angular velocity $\omega^{(j)}_{\text{L-T}}$ of Lense-Thirring precession in $F(X,Y,Z)$ gravity is independent of the nonvanishing monopole term in the vector potential $U_{i}(t,\boldsymbol{x})$, which results in that $\omega^{(j)}_{\text{L-T}}$ at the monopole order disappears.

The multipole expansions of the gyroscopic spin's angular velocities of the three types of precession in eqs.~(\ref{equ5.48})--(\ref{equ5.50}) describe all the effects of the external gravitational field of the source up to $1/c^3$ order on the gyroscopic spin. If one wants to apply them to gyroscope experiments, e.g., GP-B,  the Thomas-monopole term, the geodetic-monopole term, and the Lense-Thirring-dipole term in the stationary spacetime
are important because from them, the most main corrections to the classical angular velocities of the three types of precession in GR by the new degrees of freedom in $F(X,Y,Z)$ gravity can be read off. These terms can directly be gained by truncating the expansions~(\ref{equ5.48})--(\ref{equ5.50}),
\begin{eqnarray}
\label{equ5.63}\omega^{(j)}_{\text{Tho}}|_{l=0}&=&-\frac{1}{2c^2}\epsilon^{jpq}v^{p}\frac{\text{d}v^{q}}{\text{d}t}
+\frac{GM}{2c^2}\epsilon^{jpq}v^{p}\partial_{q}\left( \frac{1}{r}\right)\notag\\
&&+\frac{GQ}{6c^2}\epsilon^{jpq}v^{p}\partial_{q}\left(\frac{\text{e}^{-m_{1}r}}{r}\right)-\frac{G\hat{M}^{(P)}}{c^2}\epsilon^{jpq}v^{p}\partial_{q}\left(\frac{\text{e}^{-m_{2}r}}{r}\right),\\
\label{equ5.64}\omega^{(j)}_{\text{Geo}}|_{l=0}&=& \frac{3GM}{2c^2}\epsilon^{jpq}v^{p}\partial_{q}\left( \frac{1}{r}\right)-\frac{GQ}{6c^2}\epsilon^{jpq}v^{p}\partial_{q}\left(\frac{\text{e}^{-m_{1}r}}{r}\right)\notag\\
&& -\frac{2G\hat{M}^{(P)}}{c^2}\epsilon^{jpq}v^{p}\partial_{q}\left(\frac{\text{e}^{-m_{2}r}}{r}\right),\\
\label{equ5.65}\omega^{(j)}_{\text{L-T}}|_{l=1}&=& \frac{GJ_{i}}{c^2}\partial_{ji}\left( \frac{1}{r}\right)-\frac{G\hat{S}^{(P)}_{i}}{c^2}\partial_{ji}\bigg(\frac{\text{e}^{-m_{2}r}}{r}\bigg)\notag\\
&&+\frac{Gm_{2}^{2}}{c^2}\hat{S}^{(P)}_{j}\frac{\text{e}^{-m_{2}r}}{r}
-\frac{Gm^{2}_{2}}{c^2}\epsilon^{jpq}\hat{B}^{(P)}_{p}\partial_{q}\bigg(\frac{\text{e}^{-m_{2}r}}{r}\bigg).
\end{eqnarray}
As stated earlier, differently from $f(R)$ gravity, the most salient feature of the general $F(X,Y,Z)$ gravity is that it yields the correction to the gyroscopic spin's angular velocity of Lense-Thirring precession in GR (cf.~(\ref{equ5.56}) and (\ref{equ5.62})). Therefore, if one employs the Lense-Thirring-dipole term in eq.~(\ref{equ5.65}) to explain the corresponding data in gyroscope experiment GP-B, the constraint on the massive parameter
$m_{2}$ will be obtained, and in this manner, one can identify whether the degrees of freedom beyond $f(R)$ gravity exist or not in principle. After the massive parameter $m_{2}$ is determined, by further comparing
the geodetic-monopole term~(\ref{equ5.64}) with the measurement in
gyroscope experiment GP-B, the constraint on the massive parameter $m_{1}$ will also be obtained.
If both the parameters $m_{1}$ and $m_{2}$ can be fixed with the help of the gyroscope experiment GP-B, the effects of the degrees of freedom beyond GR in $F(X,Y,Z)$ gravity are clear in gyroscopic precession. According to eqs.~(\ref{equ3.4}), (\ref{equ3.14}) and (\ref{equ3.15}), both $m_{1}$ and $m_{2}$ are dependent on the coefficients
$F_{2}$, $F_{3}$, and $F_{11}$ appearing in the Lagrangian density of $F(X,Y,Z)$ gravity, so once the constraints on $m_{1}$ and $m_{2}$ are given the further constraints on these coefficients could also be acquired. Although the leading-pole terms of the angular velocities of the three types of precession are the most significant, if one intends to probe the influence of the scale and shape of the source on gyroscopic spin those terms at the next-leading and higher pole order need to be taken into account. In that case, more effects brought about by the degrees of freedom beyond GR will emerge.

In this section, the derivations of the precessional angular velocities of the gyroscopic spin could be treated as a typical application example of the metric derived in this paper. The whole process displays that one significant advantage of the metric is that all the integrals in the source multipole moments only need to be performed over the actual source distributions, so the metric can be conveniently employed to explore astrophysical phenomena happening  in the gravitational field outside a \emph{realistic} source. Moreover, another significant advantage of the metric is that it is expressed in a structurally transparent way, which enables us to easily recognize the GR-like metric and the massive scalar and tensor corrections to it in $F(X,Y,Z)$ gravity. As a consequence, when the metric is applied to some phenomenon, the corresponding results can also be presented in such way, which will greatly facilitate the subsequent analysis on the effects of the new degrees of freedom in $F(X,Y,Z)$ gravity. Besides gyroscopic precession, one is also able to make use of the metric in this paper to study other astrophysical phenomena like the gravitational redshift of light and the light bending, and it is expected that more effects beyond GR will be predicted by those new degrees of freedom in $F(X,Y,Z)$ gravity.
\section{Summary and discussion~\label{Sec:sixth}}
$F(X,Y,Z)$ gravity could be regarded as a general theoretical framework for FOTGs, so it is very meaningful to deduce the metric for the gravitational field of a source in this comprehensive model.
The gravitational field equations of $F(X,Y,Z)$ gravity are still very complicated  even after simplified by adopting the WFSM approximation. In this paper, based on the weak-field approximation framework of $F(X,Y,Z)$ gravity developed in ref.~\cite{Wu:2022mna}, a viable WFSM approximation method within this model is constructed. By applying this method, the metric is decomposed into three separated parts relevant to the fields $\tilde{h} ^{\mu\nu}$, $X^{(1)}$, and $P^{\mu\nu}$, respectively, and the complicated linearized gravitational field equations of $F(X,Y,Z)$ gravity up to $1/c^{3}$ order are converted to an easy-to-handle system of equations consisting of a Poisson equation satisfied by
$\tilde{h} ^{\mu\nu}$ and two screened Poisson equations satisfied by both $X^{(1)}$ and $P^{\mu\nu}$. In principle, after solving these equations by use of the Green's function method, one can obtain the metric for a source. However, it should note that the tensor fields $\tilde{h} ^{\mu\nu}$ and $P^{\mu\nu}$ also satisfy some additional conditions (cf.~eqs.~(\ref{equ3.6}), (\ref{equ3.34}), and (\ref{equ3.35})), which results in that the metric obtained in the above manner is still inconvenient to be employed in practical application because these conditions need to be considered separately.

This problem can be resolved with the help of the STF formalism in terms of the irreducible Cartesian tensors. By applying the techniques in the STF formalism, starting from the three equations fulfilled by $\tilde{h} ^{\mu\nu}$, $X^{(1)}$, and $P^{\mu\nu}$, these fields are expressed  in the form of the STF-tensor spherical harmonics expansions encoded by some multipole tensor coefficients. For the tensor fields $\tilde{h}^{\mu\nu}$ and $P^{\mu\nu}$, those additional conditions fulfilled by them are used to eliminate the redundant coefficients in their expansions. As a consequence, by such approach, the multipole expansions of $\tilde{h} ^{\mu\nu}$, $X^{(1)}$, and $P^{\mu\nu}$ outside a spatially compact source up to $1/c^3$ order are acquired, and based on these results, the multipole expansion of the metric for the external gravitational field of the source up to $1/c^3$ order is also derived in this paper.  In the expansion of the metric, the closed-form expressions for the source multipole moments are all presented explicitly, and it is shown that all the integrals involved in the multipole moments are carried out only over the actual source distribution. Therefore, compared with previous expressions of the metric given in refs.~\cite{stabile2015,Capozziello:2009ss,Stabile:2010mz}, our result is not plagued by the infinite integrals over the whole space like those in eqs.~(\ref{equ1.2}) and thus yields a ready-to-use form of the metric.

In the metric, the part relevant to the massless tensor field $\tilde{h} ^{\mu\nu}$ is exactly the result in GR when $F(X,Y,Z)=R$, and as the GR-like metric, it shows the Coulomb-like dependence and is characterized by the mass and spin multipole moments. Furthermore, when $F(X,Y,Z)\rightarrow f(R)$ or $f(R,\mathcal{G})$,  it can be proved that the GR-like metric plus the part relevant to the massive scalar field $X^{(1)}$ recovers the metric in $f(R)$ or $f(R,\mathcal{G})$ gravity, so this part actually provides the correction to the GR-like metric in either of the two models. Here, as a well-known fact, one needs to note that the Gauss-Bonnet scalar $\mathcal{G}$ has no contribution to the gravitational field dynamics. Thus, from the above discussions, we know that the remaining part of the metric relevant to the massive tensor field $P^{\mu\nu}$ should be the further correction to the GR-like metric in $F(X,Y,Z)$ gravity. The two parts relevant to the massive fields $X^{(1)}$ and $P^{\mu\nu}$ bear the Yukawa-like dependence on the two massive parameters and predict the appearance of six additional sets of source multipole moments. Five of these additional sets of multipole moments come from the massive tensor part, and among them besides the counterparts of
the mass and spin multipole moments, the remaining three, unlike the situation encountered in the massless tensor part, can not be transformed away by the gauge symmetry.

The appearance of the six additional sets of multipole moments in the metric indicates that up to $1/c^3$ order, there exist a total of six degrees of freedom beyond GR in $F(X,Y,Z)$ gravity. In order to study the effects of these new degrees of freedom, the metric is employed to deal with gyroscopic precession. By reasonably extending the basic method analyzing gyroscopic precession in the stationary spacetime in ref.~\cite{MTW1973}, for a gyroscope moving around the source without experiencing any torque, the multipole expansions of its spin's angular velocities of the Thomas precession, the geodetic precession, and the Lense-Thirring precession in $F(X,Y,Z)$ gravity are derived in this paper. In these expansions, those terms associated with the massless tensor part of the metric are the corresponding angular velocities of the gyroscopic spin in GR~\cite{Wu:2021uws}, and the remaining terms yield the corrections to them.
The effects of the degrees of freedom beyond GR appearing in $F(X,Y,Z)$ gravity can be read off from these corrections,
because they stem from the massive scalar and tensor parts of the metric and are characterized by the additional multipole moments. When $F(X,Y,Z)\rightarrow f(R)$ or $f(R,\mathcal{G})$, the corrections in $f(R)$ gravity~\cite{Wu:2021uws} or $f(R,\mathcal{G})$ gravity are recovered, and compared with these two models, the most salient feature of the general $F(X,Y,Z)$ gravity is that it gives the nonvanishing correction to the  gyroscopic spin's angular velocity of the Lense-Thirring precession in GR. Moreover, one interesting point is that for the Lense-Thirring precession, the angular velocity of the gyroscopic spin at the monopole order does not exist, although the monopole terms in the time-space components of the metric are nonvanishing in general.

All the effects of the external gravitational field of the source up to $1/c^3$ order on the gyroscopic spin are described by
the multipole expansions of the gyroscopic spin's angular velocities of the three types of precession. By directly truncating these expansions, the Thomas-monopole term, the geodetic-monopole term, and the Lense-Thirring-dipole term in the stationary spacetime are gained, and from them, one can read off the most main corrections to the classical angular velocities of the three types of precession in GR by the new degrees of freedom in $F(X,Y,Z)$. With these results, by successively comparing the Lense-Thirring-dipole term and the geodetic-monopole term with the corresponding  measurements in gyroscope experiment GP-B, the constraints on the massive parameter $m_{2}$ and $m_{1}$ will be obtained, and by them, the further constraints on the coefficients
$F_{2}$, $F_{3}$, and $F_{11}$ appearing in the Lagrangian density of $F(X,Y,Z)$ gravity could also be acquired.
In this manner, one can identify whether the degrees of freedom beyond $f(R)$ gravity and GR exist or not in principle, and thus, the effects of the new degrees of freedom in $F(X,Y,Z)$ gravity are clear in gyroscopic precession.
In the above process, if those terms at the next-leading and higher pole order are taken into account, one is also able to
probe the influence of the scale and shape of the source on gyroscopic spin, and in those circumstances, more effects brought about by the degrees of freedom beyond GR will emerge.

As a typical application example of the metric obtained in this paper, the derivations of the precessional angular velocities of the gyroscopic spin suggest that the metric bears two advantages. The first is that all the integrals involved in the source multipole moments only need to be performed over the actual source distributions, so the metric can be conveniently employed to explore astrophysical phenomena outside a \emph{realistic} source. The second is that the structurally transparent presentation of the metric enables us to easily recognize the GR-like metric and the massive scalar and tensor corrections to it in $F(X,Y,Z)$ gravity. Thus, while the metric is utilized to study some phenomenon, the corresponding results can also be expressed in such way, which will greatly facilitate the analysis on the effects of the new degrees of freedom appearing in $F(X,Y,Z)$ gravity. This example also indicates that although only the metric up to $1/c^{3}$ order in $F(X,Y,Z)$ gravity is obtained,  it is sufficient for us to analyze some astrophysical phenomena happening in the external gravitational field of a realistic source. It should be noted that the metric obtained in this paper is only applicable to near-zone phenomena because it is derived under the WFSM approximation~\cite{eric2018}, and as a consequence, the metric can not be utilized to discuss gravitational waves in $F(X,Y,Z)$ gravity. If one wants to embark on this topic, the multipole expansion of the metric needs to be performed starting from eqs.~(\ref{equ3.8})--(\ref{equ3.10}) instead of eqs.~(\ref{equ3.27})--(\ref{equ3.29}) so that the retardation effects can be incorporated.
\appendix
\section{Derivations for eqs.~(\ref{equ4.67})--(\ref{equ4.77})}
In order to derive eqs.~(\ref{equ4.67})--(\ref{equ4.77}), the reducible tensors $\mathcal{U}^{(P)}_{i\langle I_{l}\rangle}$ and $\mathcal{V}^{(P)}_{ij\langle I_{l}\rangle}$ in eqs.~(\ref{equ4.65}) and (\ref{equ4.66}) need to be
decomposed  into irreducible pieces, so that the components of $P^{\mu\nu}$ can be expressed in the form of  the STF-tensor spherical harmonics expansions.
By directly using eqs.~(\ref{equ2.7})--(\ref{equ2.10}), we acquire
\begin{eqnarray}
\label{equA1}\mathcal{U}^{(P)}_{i\langle I_{l}\rangle}&=&\hat{R}^{(P+)}_{iI_{l}}+\frac{l}{l+1}\epsilon_{ai\langle i_{l}}\hat{R}^{(P0)}_{i_{1}\cdots i_{l-1}\rangle a}+\frac{2l-1}{2l+1}\delta_{i\langle i_{l}}\hat{R}^{(P-)}_{i_{1}\cdots i_{l-1}\rangle}
\end{eqnarray}
with\newpage
\begin{eqnarray}
\label{equA2}&&\hat{R}^{(P+)}_{I_{l+1}}=\hat{\mathcal{U}}^{(P)}_{I_{l+1}},\\
\label{equA3}&&\hat{R}^{(P0)}_{I_{l}}=\mathcal{U}^{(P)}_{pq\langle i_{1}\cdots i_{l-1}}\epsilon_{i_{l}\rangle pq},\\
\label{equA4}&&\hat{R}^{(P-)}_{I_{l-1}}=\mathcal{U}^{(P)}_{aaI_{l-1}}
\end{eqnarray}
and
\begin{eqnarray}
\label{equA5}\mathcal{V}^{(P)}_{ij\langle I_{l}\rangle}&=&\hat{\mathcal{R}}^{(P+)}_{|i|jI_{l}}+\frac{l}{l+1}\hat{\mathcal{R}}^{(P0)}_{|i|a\langle i_{1}\cdots i_{l-1}}\epsilon_{i_{l}\rangle aj}+\frac{2l-1}{2l+1}\hat{\mathcal{R}}^{(P-)}_{|i|\langle i_{1}\cdots i_{l-1}}\delta_{i_{l}\rangle j}
\end{eqnarray}
with
\begin{eqnarray}
\label{equA6}&&\hat{\mathcal{R}}^{(P+)}_{|i|I_{l+1}}:=\mathcal{R}^{(P+)}_{i\langle I_{l+1}\rangle}=\mathcal{V}^{(P)}_{i\langle I_{l+1}\rangle},\\
\label{equA7}&&\hat{\mathcal{R}}^{(P0)}_{|i|I_{l}}:=\mathcal{R}^{(P0)}_{i\langle I_{l}\rangle}=\mathcal{V}^{(P)}_{ipq\langle i_{1}\cdots i_{l-1}}\epsilon_{i_{l}\rangle pq},\\
\label{equA8}&&\hat{\mathcal{R}}^{(P-)}_{|i|I_{l-1}}:=\mathcal{R}^{(P-)}_{i\langle I_{l-1}\rangle}=\mathcal{V}^{(P)}_{iaaI_{l-1}}.\\
\notag
\end{eqnarray}
Equation~(\ref{equA1}) yields the decomposition of $\mathcal{U}^{(P)}_{i\langle I_{l}\rangle}$, but eq.~(\ref{equA5}) does not yield that of $\mathcal{V}^{(P)}_{ij\langle I_{l}\rangle}$ because the tensors $\mathcal{R}^{(P+)}_{i\langle I_{l+1}\rangle}$, $\mathcal{R}^{(P0)}_{i\langle I_{l}\rangle}$, and $\mathcal{R}^{(P-)}_{i\langle I_{l-1}\rangle}$ are still reducible. By employing eqs.~(\ref{equ2.7})--(\ref{equ2.10}) again,
the decompositions of $\mathcal{R}^{(P+)}_{i\langle I_{l+1}\rangle}$, $\mathcal{R}^{(P0)}_{i\langle I_{l}\rangle}$, and $\mathcal{R}^{(P-)}_{i\langle I_{l-1}\rangle}$ are gained,
\begin{eqnarray}
\label{equA9}\mathcal{R}^{(P+)}_{i\langle I_{l+1}\rangle}&=&\hat{\mathcal{R}}^{(P++)}_{iI_{l+1}}+\frac{l+1}{l+2}\epsilon_{ai\langle i_{l+1}}\hat{\mathcal{R}}^{(P+0)}_{i_{1}\cdots i_{l}\rangle a}+\frac{2l+1}{2l+3}\delta_{i\langle i_{l+1}}\hat{\mathcal{R}}^{(P+-)}_{i_{1}\cdots i_{l}\rangle},\\
\label{equA10}\mathcal{R}^{(P0)}_{i\langle I_{l}\rangle}&=&\hat{\mathcal{R}}^{(P0+)}_{iI_{l}}+\frac{l}{l+1}\epsilon_{ai\langle i_{l}}\hat{\mathcal{R}}^{(P00)}_{i_{1}\cdots i_{l-1}\rangle a}+\frac{2l-1}{2l+1}\delta_{i\langle i_{l}}\hat{\mathcal{R}}^{(P0-)}_{i_{1}\cdots i_{l-1}\rangle},\\
\label{equA11}\mathcal{R}^{(P-)}_{i\langle I_{l-1}\rangle}&=&\hat{\mathcal{R}}^{(P-+)}_{iI_{l-1}}+\frac{l-1}{l}\epsilon_{ai\langle i_{l-1}}\hat{\mathcal{R}}^{(P-0)}_{i_{1}\cdots i_{l-2}\rangle a}+\frac{2l-3}{2l-1}\delta_{i\langle i_{l-1}}\hat{\mathcal{R}}^{(P--)}_{i_{1}\cdots i_{l-2}\rangle}
\end{eqnarray}
with
\begin{eqnarray}
\label{equA12}&&\hat{\mathcal{R}}^{(P++)}_{I_{l+2}}=\hat{\mathcal{R}}^{(P+)}_{I_{l+2}},\\
\label{equA13}&&\hat{\mathcal{R}}^{(P+0)}_{I_{l+1}}=\mathcal{R}^{(P+)}_{pq\langle i_{1}\cdots i_{l}}\epsilon_{i_{l+1}\rangle pq},\\
\label{equA14}&&\hat{\mathcal{R}}^{(P+-)}_{I_{l}}=\mathcal{R}^{(P+)}_{aa\langle I_{l}\rangle},\\
\label{equA15}&&\hat{\mathcal{R}}^{(P0+)}_{I_{l+1}}=\hat{\mathcal{R}}^{(P0)}_{I_{l+1}},\\
\label{equA16}&&\hat{\mathcal{R}}^{(P00)}_{I_{l}}=\mathcal{R}^{(P0)}_{pq\langle i_{1}\cdots i_{l-1}}\epsilon_{i_{l}\rangle pq},\\
\label{equA17}&&\hat{\mathcal{R}}^{(P0-)}_{I_{l-1}}=\mathcal{R}^{(P0)}_{aa\langle I_{l-1}\rangle},\\
\label{equA18}&&\hat{\mathcal{R}}^{(P-+)}_{I_{l}}=\hat{\mathcal{R}}^{(P-)}_{I_{l}},\\
\label{equA19}&&\hat{\mathcal{R}}^{(P-0)}_{I_{l-1}}=\mathcal{R}^{(P-)}_{pq\langle i_{1}\cdots i_{i-2}}\epsilon_{i_{l-1}\rangle pq},\\
\label{equA20}&&\hat{\mathcal{R}}^{(P--)}_{I_{l-2}}=\mathcal{R}^{(P-)}_{aa\langle I_{l-2}\rangle}.
\end{eqnarray}
Then, inserting eqs.~(\ref{equA9})--(\ref{equA11}) into eq.~(\ref{equA5}) gives rise to the decomposition of $\mathcal{V}^{(P)}_{ij\langle I_{l}\rangle}$. With the help of the decompositions of $\mathcal{U}^{(P)}_{i\langle I_{l}\rangle}$ and $\mathcal{V}^{(P)}_{ij\langle I_{l}\rangle}$, from eqs.~(\ref{equ4.65}) and (\ref{equ4.66}) , we finally derive, after suitable changes of the summation index, eqs.~(\ref{equ4.67})--(\ref{equ4.77}), where in the derivation,  the identity
\begin{eqnarray}
\label{equA21}\left(\Delta-\lambda^{2}\right)\bigg(\frac{\text{e}^{-\lambda r}}{r}\bigg)=0
\end{eqnarray}
outside the source has been used.
\section{Derivation for eq.~(\ref{equ5.39})}
According to eq.~(\ref{equ5.38}), in order to obtain the expression of $\text{d}S^{i}/\text{d}\tau$ up to $1/c^3$ order, the Christoffel symbol $\varGamma^{i}_{\lambda\alpha}$ and the gyroscopic four-acceleration $a^{\rho}$ need to be first obtained. From eq.~(\ref{equ3.17}), the metric is expressed in the form of
$g_{\mu\nu}=\eta_{\mu\nu}-\overline{h}_{\mu\nu}$ under the weak-field approximation, and in this case, the Christoffel symbol should be evaluated by
\begin{eqnarray}
\label{equB1}\varGamma^{\beta}_{\lambda\alpha}=\frac{1}{2}\left[\partial_{\alpha}\big(-\overline{h}^{\beta}_{\phantom{\beta}\lambda}\big)
+\partial_{\lambda}\big(-\overline{h}^{\beta}_{\phantom{\beta}\alpha}\big)-\partial^{\beta}\big(-\overline{h}_{\lambda\alpha}\big)\right].
\end{eqnarray}
Up to $1/c^3$ order, as shown in eq.~(\ref{equ3.22}), the metric is further expressed in terms of the three potentials
$U(t,\boldsymbol{x})$, $U_{i}(t,\boldsymbol{x})$, and $U_{ij}(t,\boldsymbol{x})$, which leads to that
\begin{equation}\label{equB2}
\left\{\begin{array}{ll}
\displaystyle -\overline{h}_{00}(t,\boldsymbol{x})&=\displaystyle \frac{2}{c^{2}}U(t,\boldsymbol{x}),\smallskip\\
\displaystyle -\overline{h}_{0i}(t,\boldsymbol{x})&=\displaystyle -\frac{4}{c^{3}}U_{i}(t,\boldsymbol{x}),\smallskip\\
\displaystyle -\overline{h}_{ij}(t,\boldsymbol{x})&=\displaystyle \frac{2}{c^{2}}U_{ij}(t,\boldsymbol{x}).
\end{array}\right.
\end{equation}
After inserting eqs.~(\ref{equB2}) into eq.~(\ref{equB1}), the expressions of the Christoffel symbols up to $1/c^3$ order are given,
\begin{eqnarray}
\label{equB3}\varGamma^{0}_{00}&=&-\frac{1}{c^3}\partial_{t}U(t,\boldsymbol{x}),\\
\label{equB4}\varGamma^{0}_{0j}&=&-\frac{1}{c^2}\partial_{j}U(t,\boldsymbol{x}),\\
\label{equB5}\varGamma^{0}_{jk}&=&\frac{2}{c^3}\left[\partial_{j}U_{k}(t,\boldsymbol{x})+\partial_{k}U_{j}(t,\boldsymbol{x})\right]+\frac{1}{c^3}\partial_{t}U_{jk}(t,\boldsymbol{x}),\\
\label{equB6}\varGamma^{i}_{00}&=&-\frac{1}{c^2}\partial_{i}U(t,\boldsymbol{x}),\\
\label{equB7}\varGamma^{i}_{0j}&=&\frac{2}{c^3}\left[\partial_{i}U_{j}(t,\boldsymbol{x})-\partial_{j}U_{i}(t,\boldsymbol{x})\right]+\frac{1}{c^3}\partial_{t}U_{ij}(t,\boldsymbol{x}),\\
\label{equB8}\varGamma^{i}_{jk}&=&\frac{1}{c^2}\left[\partial_{j}U_{ik}(t,\boldsymbol{x})+\partial_{k}U_{ij}(t,\boldsymbol{x})-\partial_{i}U_{jk}(t,\boldsymbol{x})\right].
\end{eqnarray}
With these results, together with eqs.~(\ref{equ5.11}), (\ref{equ5.13}), and (\ref{equ5.35}), the first term in the right-hand side of eq.~(\ref{equ5.38}) up to $1/c^3$ order is
\begin{eqnarray}
\label{equB9}-u^{\alpha}S^{\lambda}\varGamma^{i}_{\lambda\alpha}&=&\frac{1}{c^2}v^{j}S^{j}\partial_{i}U(t,\boldsymbol{x})
-\frac{2}{c^2}\left[\partial_{i}U_{j}(t,\boldsymbol{x})-\partial_{j}U_{i}(t,\boldsymbol{x})\right]S^{j}
-\frac{1}{c^2}\partial_{t}U_{ij}(t,\boldsymbol{x})S^{j}\notag\\
&&-\frac{1}{c^2}v^{k}S^{j}\left[\partial_{j}U_{ik}(t,\boldsymbol{x})+\partial_{k}U_{ij}(t,\boldsymbol{x})-\partial_{i}U_{jk}(t,\boldsymbol{x})\right].
\end{eqnarray}
As to the gyroscopic four-acceleration, it is defined by
\begin{eqnarray}
\label{equB10}a^{\rho}=u^{\lambda}\nabla_{\lambda}u^{\rho}=\frac{\text{d}u^{\rho}}{\text{d}\tau}+u^{\lambda}u^{\sigma}\varGamma^{\rho}_{\lambda\sigma},
\end{eqnarray}
and with the aid of eqs.~(\ref{equ5.11}), (\ref{equ5.13}), and (\ref{equB3})--(\ref{equB8}), up to $1/c^3$ order, the following two identities are obtained,
\begin{eqnarray}
\label{equB11}\frac{1}{c^2}a^{0}&=&\frac{1}{2c^3}\frac{\text{d}(v^{k}v^{k})}{\text{d}\tau}-\frac{1}{c^3}v^{j}\partial_{j}U(t,\boldsymbol{x}),\\
\label{equB12}\frac{1}{c^2}a^{i}&=&\frac{1}{c^2}\frac{\text{d}v^{i}}{\text{d}\tau}-\frac{1}{c^2}\partial_{i}U(t,\boldsymbol{x}).
\end{eqnarray}
Then, plugging them into the second term in the right-hand side of eq.~(\ref{equ5.38}) and combining eqs.~(\ref{equ3.22}), (\ref{equ5.11}), and (\ref{equ5.35}), one is able to achieve its expression up to $1/c^3$ order
\begin{eqnarray}
\label{equB13}\frac{1}{c^2}a^{\rho}S^{\sigma}g_{\rho\sigma}u^{i}=\frac{1}{c^2}a^{j}v^{i}S^{j}.
\end{eqnarray}
Finally, based on eqs.~(\ref{equB9}) and (\ref{equB13}), the expression of $\text{d}S^{i}/\text{d}\tau$ up to $1/c^3$ order, namely eq.~(\ref{equ5.39}), is derived.
\acknowledgments
This work was supported by the National Natural Science Foundation of China (Grants Nos.~12105039 and 12133003).
We are very grateful to Prof. Chao-Guang Huang for the guidance to Bofeng Wu.


\end{document}